\documentclass{emulateapj-rtx4}
\usepackage{pdflscape,longtable}
\usepackage{lscape}
\usepackage{longtable}

\usepackage{graphicx}
\usepackage{color}
\usepackage{amsmath,amssymb}
\usepackage{natbib}

\newcommand{\salt}{{\tt SALT2}}

\newcommand{\OII}{[O~{\footnotesize II}]}
\newcommand{\OIII}{[O~{\footnotesize III}]}
\newcommand{\NII}{[N~{\footnotesize II}]}
\newcommand{\SII}{[S~{\footnotesize II}]}
\newcommand{\spec}[2]{#1~{\footnotesize #2}}
\newcommand{\Ha}{H$\alpha$}
\newcommand{\Hb}{H$\beta$}

\newcommand{\sdss}{SDSS}
\newcommand{\sdsssns}{SDSS-SNS}
\newcommand{\E}[1]{\ensuremath{\times 10^{#1}}}
\newcommand{\unt}[1]{\mbox{\;#1}}
\newcommand{\Msun}{\ensuremath{M_\odot}}
\newcommand{\NspecSN}{499}
\newcommand{\Ntrans}{10,258}
\newcommand{\numpm}{345}
\newcommand{\nummzs}{144}

\begin{document}

\title{SDSS-II Supernova Survey: An Analysis of the Largest Sample of Type Ia Supernovae and Correlations with Host-Galaxy Spectral Properties}

\shortauthors{Wolf et al.}

\author{Rachel C. Wolf\altaffilmark{1},
Chris B. D'Andrea\altaffilmark{2,3},
Ravi R. Gupta\altaffilmark{1,4},
Masao Sako\altaffilmark{1}, John A. Fischer\altaffilmark{1}, Rick Kessler\altaffilmark{5},
Saurabh W. Jha\altaffilmark{6}, Marisa C. March\altaffilmark{1}, Daniel M. Scolnic\altaffilmark{5}, Johanna-Laina Fischer\altaffilmark{1}, Heather Campbell\altaffilmark{3,7}, Robert C. Nichol\altaffilmark{3}, Matthew D. Olmstead\altaffilmark{8,9}, Michael Richmond\altaffilmark{10}, Donald P. Schneider\altaffilmark{11,12}, Mathew Smith\altaffilmark{2,13}}

\email{rcane@physics.upenn.edu}

\altaffiltext{1}{\label{penn}
  Department of Physics and Astronomy, University of Pennsylvania, 
  209 South 33rd Street, Philadelphia, PA 19104, USA
}
   \altaffiltext{2}{\label{sh}
School of Physics and Astronomy, University of Southampton, Southampton, SO17 1BJ, UK
  } 
\altaffiltext{3}{\label{icg}
  Institute of Cosmology and Gravitation, University of Portsmouth, 
  Dennis Sciama Building, Burnaby Road, Portsmouth, PO1 3FX, UK
}
\altaffiltext{4}{\label{anl}
  Argonne National Laboratory, 
  9700 South Cass Avenue, Lemont, IL 60439, USA
}
 \altaffiltext{5}{\label{chi}
  The University of Chicago, The Kavli Institute for Cosmological Physics, 
  933 East 56th Street, Chicago, IL 60637, USA
  } 
  \altaffiltext{6}{\label{rutgers}
  Department of Physics and Astronomy, Rutgers, the State University of New Jersey, 
  136 Frelinghuysen Road, Piscataway, NJ 08854, USA
  }
    \altaffiltext{7}{\label{cam}
Institute of Astronomy, University of Cambridge, Madingley
Road, Cambridge CB3 0HA, UK
  } 
     \altaffiltext{8}{\label{kc}
Department of Chemistry and Physics, KingÕs College, Wilkes-Barre, PA 18711, USA
  } 
 \altaffiltext{9}{\label{utah}
Department of Physics and Astronomy, University of Utah, Salt Lake City, UT 84112, USA
  } 
  \altaffiltext{10}{\label{rochester} 
   School of Physics and Astronomy, Rochester Institute of
Technology, Rochester, New York 14623, USA
}
   \altaffiltext{11}{\label{don1}
 Department of Astronomy and Astrophysics, The Pennsylvania State University,
   University Park, PA 16802, USA
  } 
 \altaffiltext{12}{\label{don2}
Institute for Gravitation and the Cosmos, The Pennsylvania State University,
   University Park, PA 16802, USA
  } 
   \altaffiltext{13}{\label{mat1}
Department of Physics, University of the Western Cape,
Cape Town, 7535, South Africa
  }

\begin{abstract}
Using the largest single-survey sample of Type Ia supernovae (SNe\,Ia) to date, we study the relationship between 
properties of SNe\,Ia and those of their host galaxies, focusing primarily on correlations with Hubble residuals (HR). 
Our sample consists of \numpm~photometrically-classified or spectroscopically-confirmed SNe\,Ia 
discovered as part of the SDSS-II Supernova Survey (\sdsssns).  This analysis utilizes host-galaxy spectroscopy obtained 
during the SDSS-I/II spectroscopic survey and from an ancillary program on the SDSS-III Baryon Oscillation 
Spectroscopic Survey (BOSS) that obtained spectra for nearly all host galaxies of SDSS-II SN candidates.  
In addition, we use photometric host-galaxy properties from the \sdsssns\ data release \citep{Sako14} such as host 
stellar mass and star-formation rate. 
We confirm the well-known relation between HR and host-galaxy mass and find a $3.6\sigma$ significance of a non-zero linear slope.  We also recover correlations between HR and host-galaxy gas-phase metallicity and specific star-formation rate as they are reported in the literature.  With our large dataset, we examine correlations between HR and multiple host-galaxy properties simultaneously and find no evidence of a significant correlation.  We also independently analyze our spectroscopically-confirmed and photometrically-classified SNe\,Ia and comment on the significance of similar combined datasets for future surveys.       
 
\end{abstract}



\section{INTRODUCTION}
\label{sec:intro}

Type Ia supernovae (SNe\,Ia) are crucial observational probes for investigating the history of our expanding universe.  The origin of these phenomena remains a mystery, although there is evidence for two distinct SN\,Ia progenitor systems (the so-called single degenerate and double degenerate scenarios) that result in a thermonuclear explosion occurring as the mass of a carbon-oxygen white dwarf approaches the Chandresekhar limit \citep{WI73,Nomoto,IT84,Webbink,HN}.  Observations of these incredibly bright explosions, visible out to high redshifts, have provided evidence for the accelerating expansion of the universe \citep{R98,Perlmutter} and are used to measure cosmological parameters with increasing precison \citep{Astier,Kessler09a,Conley11,B14,Scolnic14}.  Their efficacy as ``standard candles", however, relies on the ability to calibrate intrinsic luminosity with SNe light-curve width (`stretch') and optical color \citep{Phillips,Hamuy96,RPK96,Tripp}.  After applying these corrections using light-curve fitting techniques, there remains a $1\sigma$ dispersion in peak brightness of about 0.1 magnitudes, corresponding to about five percent in distance \citep{B14,Conley11}.  The origin of this scatter remains unknown; yet it is postulated that both the progenitor and its environment play a role \citep{Gallagher05,Gallagher08,Neill09,Howell09,Kelly10}.  

Standardization of SNe\,Ia luminosity can be improved by searching for additional parameters that  correlate with the Hubble Residual (HR), which quantifies the difference in the distance modulus, corrected for stretch and color, and what is predicted by the best-fit cosmology.  \citet{Gallagher05} studied the host-galaxy properties of nearby SNe\,Ia and found a tenuous correlation between the HR and host-galaxy gas-phase metallicity.  More recently, \citet{Kelly10}, \citet{Sullivan10}, and \citet{Lampeitl10}, using independent data sets, demonstrated that SNe\,Ia in more massive hosts are about $\sim$ 0.1 magnitudes brighter (after light-curve corrections) than those in lower mass hosts.  Understanding this relation is crucial for future precision cosmology experiments.
In the recent literature, there have been several studies indicating that rather than a continuous linear slope, 
the HR trend with host stellar mass behaves more like a ``step" function, which has a transition region 
connecting the two levels \citep{Childress13,Johansson13,Rigault13}.  This trend has become known 
as the ``mass step."

\citet[hereafter C13]{Childress13} combined their sample of SNe\,Ia from the Nearby Supernova Factory with SNe\,Ia from 
the literature (namely, \citealt{Kelly10}, \citealt{Sullivan10}, and \citealt{Gupta11}) to create a sample of 
601 SNe\,Ia spanning low and high redshift.  
They used this combined sample to analyze the trend between HR and host-galaxy mass 
and found that the structure of the trend is consistent with a plateau at low and high mass 
separated by a transition region from $\log(M/M_\odot)=9.8$ to $\log(M/M_\odot)=10.4$.  
Several physical models for this behavior were expounded and compared to the data, and the authors 
concluded that the cause of the trend may be due to a combination of the shape of the galaxy mass-metallicity 
relation, the evolution of SN\,Ia progenitor age along the galaxy mass sequence, and the uncertain 
effects of SN color and host galaxy dust.

\citet[hereafter J13]{Johansson13} analyzed a sample of 247 SDSS SNe\,Ia using only SDSS host-galaxy photometry.  
They found that, as in C13, the HR-mass 
relation behaves as a sloped step function, with essentially zero slope at the high- and low-mass ends and a 
non-zero slope in the region $9.5 < \log(M/M_\odot) < 10.2$.  They reported that the step in the HR-mass 
plane is close to the evolutionary transition mass of low-redshift galaxies first described by 
\citet{Kauffmann03a}.  
This transition mass occurs at $\log(M/M_\odot) \sim 10.5$ and signifies a change in galaxy morphology 
and stellar populations.  J13 concluded that differences between SN\,Ia progenitors in these 
populations could imply the existence of two samples of SNe\,Ia with high and low HR.

Following on the work of C13, \citet{Rigault13} used integral field spectroscopy for a 
sample of 89 SNe\,Ia from the Nearby Supernova Factory to measure \Ha\ emission within a 1 kpc 
radius around each SN.  This \Ha\ surface brightness was used to define SN environments as either ``locally 
star-forming" or ``locally passive" and they found that the mean standardized brightness for SNe\,Ia with local \Ha\ 
emission is on average 0.09 magnitudes fainter than for those without.  They found a bimodal structure in HR, and 
claim that the intrinsically brighter mode, exclusive to locally passive environments, is responsible 
for the mass step.  They argue that HRs are highly dependent on local environment, with local  
\Ha\ emission being more fundamental than global host properties.

There is no known mechanism by which the mass of the host galaxy can directly influence the explosion of a single white dwarf; therefore, other host properties that are correlated with galaxy mass must be invoked to explain the underlying physical mechanism of this relation.  For example, host galaxy gas-phase metallicity is widely assumed to be a proxy for progenitor metallicity, and there are models suggesting that SN\,Ia luminosities depend on the stellar metallicity of the progenitor \citep{Timmes, KRW}.  Therefore, correlations between host 
metallicity and SN properties have been of recent interest as well. 
\citet[hereafter D11]{D11} used a complete sample of all 34 SNe\,Ia with $ z < 0.15 $ detected by the Sloan Digital Sky Survey-II SN Survey \citep[hereafter \sdsssns;][]{Frieman08} and corresponding host-galaxy spectra and found significant correlations between gas-phase metallicity and specific star formation rate with HR.  Similar trends were observed by C13 and \citet[hereafter P14]{Pan14} using data from the SNFactory and PTF, respectively.
\citet{Konishi} also analyzed host spectra of SDSS SNe and concluded that SNe\,Ia in metal-rich galaxies are 0.13 magnitudes brighter after correcting for light-curve width and color.  Given that broadband photometry of galaxies is more readily available than galaxy spectra, several studies have estimated host galaxy physical properties from photometry. 
\citet{Gupta11} used 206 SNe\,Ia from the \sdsssns\ and host-galaxy multi-wavelength photometry and found that while the relation of HR with host stellar mass was highly significant, the relation with mass-weighted age of the host was not.  Building on this work, \citet{Hayden} calibrated the fundamental metallicity relation (FMR) of \citet{Mannucci10} to better estimate host metallicity from photometry, and found that using the FMR improves HR correlation beyond the stellar mass alone. 
More recently, using empirical models of galaxy star formation histories and theoretical SN delay time distribution models, \citet{Childress14} have argued that the mean ages of SNe\,Ia progenitors are responsible for driving the HR correlation with host mass.

Many recent studies (D11, C13, P14), utilize host-galaxy spectroscopy to study these relations.  Campbell et al. (2015, submitted) use SNe\,Ia from SDSS to explore correlations with spectroscopic host-galaxy properties, using published BOSS data products from the SDSS DR10 catalog \citep{Ahn14} and focusing on cosmological constraints.  Using spectroscopy rather than photometry provides direct access to the galaxy SED and a better estimate of dust extinction.  It also allows for derivations of the gas-phase metallicity and star-formation rates via narrow emission lines.  In this work, we study the relationship between SN\,Ia HRs and properties of their host galaxies, including metallicity and star-formation rate, using SN data from the full three-year SDSS-SNS \citep[hereafter S14]{Sako14} and a combination of host-galaxy spectra from an ancillary program of the SDSS-III Baryon Oscillation Spectroscopic Survey (BOSS) \citep{Olmstead14,Dawson13} and from the SDSS I/II spectroscopic survey \citep{DR7,Strauss}.  In comparison to recent literature, this is the largest single-survey sample of spectroscopically-confirmed or photometrically-classified SN\,Ia light curves and host-galaxy spectroscopic data.  As newer, larger surveys, such as the Dark Energy Survey \citep{Bernstein12}, Pan-STARRS \citep{Kaiser}, and LSST \citep{LSST} will also heavily rely on photometrically-classifed samples of SNe\,Ia, the biases and selection effects discussed in this work will be critical for future host-galaxy studies.  

In this paper we adopt the best-fit flat, $\Lambda$CDM cosmology for SNe\,Ia alone as determined by \citet[hereafter B14]{B14}, a joint analysis
of 740 spectroscopically-confirmed SNe\,Ia from a compilation of surveys of low, intermediate, and high redshift ranges ($\Omega_{M}=0.295$).  
The B14 sample combines 242 high-$z$ SNe\,Ia from the first 3 years of the Supernova Legacy Survey (SNLS) and 374 SNe\,Ia from the full 3-year data release of SDSS-SNS ($0.05 < z < 0.4$).  The remainder of the SN\,Ia sample are from a collection of low-$z$ surveys, with most from the third release \citep{Hicken09} of photometric data acquired at the F. L. Whipple Observatory of the Harvard-Smithsonian Center for Astrophysics (CfA3).  Since the value of the Hubble constant is degenerate with the absolute magnitude 
of SNe\,Ia, we adopt $H_0 = 70\unt{km s}^{-1}\unt{Mpc}^{-1}$.  
We use this cosmology to compute HR, defined as HR~$ \equiv \mu_{\mathrm{SN}} - \mu_z$, where $\mu_{\mathrm{SN}}$ is the distance modulus estimated from fitting SN\,Ia light curves and $\mu_z$ is the distance modulus computed using the redshift and our assumed cosmology.  The HR quantifies whether our SNe\,Ia are over-luminous (negative HR) or under-luminous (positive HR) after light-curve correction. 

The general structure of this work is as follows: In Section~\ref{sec:data} we describe our SN and galaxy data.  Section~\ref{sec:SN} highlights light-curve quality requirements for our SNe\,Ia sample and describes the treatment of effects such as Malmquist bias.  Section~\ref{sec:GalSpec} details our methods for extracting galaxy spectroscopy and the selection cuts we impose on our host galaxy sample. Section~\ref{sec:derivprop} outlines how we derive host galaxy properties from emission-line fluxes.  The sample selection requirements discussed in Sections \ref{sec:SN}, \ref{sec:GalSpec}, and \ref{sec:derivprop} ultimately yield our two final samples for analysis, which contain \numpm~and \nummzs~SNe\,Ia, respectively.  In Section~\ref{sec:results} we present our findings and we discuss our results in Section~\ref{sec:Disc}.

\section{OBSERVATIONAL DATA}
\label{sec:data}

Observations from the SDSS-SNS were used for our SN\,Ia sample and a combination of spectra from SDSS and BOSS were utilized for host-galaxy spectroscopy.  
Spectra of host galaxies are important not only for securing redshifts of their SNe, but also as probes 
of the physical properties of galaxies themselves.  As summarized in the previous section, these properties 
can influence the SN\,Ia progenitor and the subsequent explosion.
We describe how we obtain our SN and host galaxy data in Sections \ref{sec:data-sn} and \ref{sec:data-gal}, 
respectively.

\subsection{Supernovae}
\label{sec:data-sn}

All SNe in this work were discovered and observed by the \sdsssns.  Data were collected over a three month observing season (September - November) in 2005, 2006, and 2007 using the wide-field \sdss\ CCD camera \citep{SDSS-Camera} on the 2.5 meter SDSS telescope at the Apache Point Observatory in New Mexico \citep{SDSS-Telescope}.   The survey observed Stripe 82, a 300 deg$^{2}$ equatorial region of the Southern sky, in drift-scan mode, obtaining nearly simultaneous 55 second exposures in each of the \emph{ugriz} SDSS filters \citep{SDSS-Filters}.  Descriptions of the SDSS absolute photometric calibration are found in \citet{Ivezic}, \citet{Pad08}, and \citet{B13}.  The average cadence of the survey, including losses due to weather and sky brightness, was $\sim 4$ days.  Selection of the supernova candidates is described in \citet{Sako08}. High quality light curves were obtained \citep{JH08} with optical photometry that is internally consistent to $\sim1\%$ \citep{Ivezic}.  For a technical summary of the \sdss, see \citet{York}.

A full description of data acquisition and reduction from the SDSS-SNS can be found in the final Data Release paper (S14).  Over its three-year run, the SDSS-SNS discovered \Ntrans\ new variable objects in the redshift range $0.01 < z < 0.55$.  Of these, \NspecSN\ were spectroscopically-classified as SNe\,Ia (`Spec-Ia').  In S14, these SNe\,Ia are typed ``SNIa".  

Analyses that use spectroscopically-identified samples of SNe\,Ia \citep[e.g.,][]{Kessler09a,B14} are highly pure, as they contain, to high confidence, only SNe\,Ia.  However such samples, as in the case of the SDSS-SNS, can be biased, as the likelihood of an SN\,Ia being spectroscopically-classified is a function of many factors:  its location within the host galaxy, its relative brightness compared to the surface brightness of the host galaxy, and its color (but not the intrinsic brightness; see Figure 10 of S14).  Additionally, the expense of spectroscopy is a limiting factor in rolling supernova surveys such as the SDSS-SNS: resources are typically unavailable (or observing conditions disadvantageous) for a complete spectroscopic program.  For these reasons, we also use in this paper SDSS-SNS transients that have been \textit{photometrically-classified}, using the host-galaxy spectroscopic redshift as a prior, as SNe\,Ia (`Phot-Ia').  In S14, these SNe\,Ia are typed ``zSNIa".  We describe the classification and data-quality cuts applied to this catalog of transients in Section \ref{sec:SN}.

\subsection{Host Galaxies}
\label{sec:data-gal}
Our primary source of SN host-galaxy spectroscopy is the BOSS survey of SDSS-III \citep{Eisenstein}.  BOSS, which ran from 2008 to 2014, was designed to measure the scale of baryon acoustic oscillations (BAO) by observing 1.5 million galaxies to redshift $z < 0.7$ and 150,000 quasars at redshifts $2.15 < z  < 3.5$ over an area of $10,000\unt{deg}^{2}$.  To accommodate this survey, the original SDSS spectrograph was rebuilt with smaller fibers (2\arcsec\ diameter, allowing a larger number of targets per pointing), more sensitive detectors in both the blue and red channels, and a wider wavelength range ($361-1014\unt{nm}$).  These improvements allowed the survey to reach higher galaxy redshifts and observe about one magnitude deeper than SDSS.  A detailed description of the BOSS spectrograph (as the upgraded instrument is now known) can be found in \citet{Smee}. 

Approximately 5\% of the BOSS fibers were allocated to ancillary science programs, one of which was the systematic targeting of host galaxies of SN candidates from the SDSS-SNS.  Targets for this program were prioritized based on the probability of the observed transient being a type Ia or core-collapse SN using the photometric--classification software PSNID (see Section \ref{sec:SN}), as well as on the \textit{r}-band fiber-magnitude of the host galaxy (\textit{r}$_{\textrm{fiber}}<21.25$).  A total of 3,761 of the 4,777 requested targets were observed, with non-observations primarily due to the finite availability of fibers and clashes with higher priority targets.  The SDSS-SNS target selection for this ancillary program is detailed in \citet{Olmstead14} and \citet{Campbell}. 

We use in this analysis the host-galaxy matching done in S14.  Here each detected transient is matched to the SDSS Data Release 8 \citep{Aihara} catalog using an algorithm that identifies the ``nearest" 
galaxy in a parameter space that accounts for the apparent size and surface brightness profile of each galaxy 
within a 30\arcsec\ radius of the transient coordinates.
It is estimated that this method is able to match host galaxies with 97\% accuracy (S14).

The host-galaxy matching that defined the target selection for BOSS spectroscopy was performed years prior to the development of the algorithm used in S14.  
Therefore, it would not be unexpected if some fraction of the BOSS targets do not correspond to the currently-identified host galaxy, resulting in an incorrect assumed redshift for some SNe.  
We find that the existing redshifts (either from the SN spectrum or from a non-BOSS host spectrum) of three SNe\,Ia disagree with those 
of their respective BOSS targets.    
For each of these cases, the BOSS spectrum is of a galaxy that is offset from the currently identified host by more than $8\arcsec$, indicating that 
the BOSS target is not the correct host. 
To avoid possible ambiguity, we remove these three SNe from our sample. For further discussion 
of BOSS targeting and host-galaxy mismatches see S14.

As all of Stripe 82 lies within the area observed by the SDSS-I/II spectroscopic survey, many of our transients have pre-existing host spectra.  The BOSS ancillary program targeted the location of the SN within the galaxy where spectroscopy of the host galaxy already exists in the SDSS database.  This paper derives global spectroscopic properties of the host galaxies, and thus preferentially uses SDSS spectra where they exist, as these spectra typically have higher $S/N$ than BOSS spectra due to their larger fiber width (3\arcsec\ diameter) and being centered on the host galaxy.

We will return to this point briefly in Section~\ref{sec:derivprop}, where we discuss the breakdown of spectra passing various cuts for data quality.


\section{SUPERNOVA SELECTION AND PROPERTIES}
\label{sec:SN}

We select our sample of photometrically-classified SNe\,Ia using the Photometric SN IDentification (PSNID) software (Sako et al. 2011) described in S14. PSNID uses the observed photometry of the SNe to first compute a Bayesian probability associated with each of the assumed three SN types (SN\,Ia, SN Ibc, and SN II), as well as parameters and errors assuming a SN\,Ia model, using a Markov Chain Monte Carlo.  The same procedures are then performed on a large simulated mixture of SN\,Ia and core-collapse SNe.  For each SN candidate in our sample, the measured SN\,Ia parameters (extinction, light-curve stretch, and redshift) are compared with those of the simulated set to calculate Cartesian distances to its neighbors which are used then to determine a nearest-neighbor probability.  The combination of the $\chi^2$-fit, Bayesian, and nearest-neighbor probabilities are used for the final classification.

In this work we use only those classifications from S14 where the host-galaxy redshift is included as a prior on the light-curve fit, which is important for precise placement of SNe\,Ia on the Hubble Diagram. We impose the PSNID selection criteria outlined in Section 4 of S14:  the PSNID fit probability is $\geq0.01$ for the SN\,Ia model; the Bayesian probability of being an SN\,Ia is $\geq0.9$; and the nearest-neighbor probability of being an SN\,Ia is greater than that of being a core-collapse SN.  We place an additional requirement on light-curve sampling, requiring the candidate to have at least one detection at $-5 \leq T_{\mathrm{rest}} \leq +5$ days and one at $+5 < T_{\mathrm{rest}} \leq +15$ days, where $T_{\mathrm{rest}}$ is the rest-frame time such that $T_{\mathrm{rest}}= 0$ corresponds to peak brightness in rest-frame B band. Imposing these criteria yields a sample of 824 photometrically-classified SNe\,Ia with a purity and efficiency of $\sim96$\% (determined from simulations; for more complete definitions of sample purity and efficiency see S14).

The photometrically-classified SNe selected by the above requirements, combined with the 499 Spec-Ia, define a maximally large sample of SNe\,Ia in SDSS-SNS.  As we are interested in host-galaxy correlations with the derived distance modulus to these SNe, we apply additional cuts to create a sample that can produce reliable distance estimates.  We fit these light curves using the implementation of {{\tt SALT2} }\citep{SALT2} in the SuperNova ANAlysis package \citep[{\tt SNANA};][]{Kessler09b}, keeping only SNe\,Ia that meet the following criteria: 
\begin{enumerate} 
\item{At least one detection before peak brightness ($T_\mathrm{rest}<0$).}
\item{At least one measurement with $T_\mathrm{rest} > +10$ days.}
\item{At least five detections between $-15 < T_\mathrm{rest} < +60$ days.}
\item{At least three filter-epoch detections with $S/N > 5$.}
\item{The measured color ($c$) and stretch ($x_{1}$) are within the elliptical cut outlined in \citet[Figure 6]{Campbell}.}
\item{$P_\mathrm{FIT} > 0.01$ where $P_\mathrm{FIT}$ is the {\tt SALT2} light-curve fit probability based on the $\chi^{2}$.}
\end{enumerate}

Distance moduli are then estimated using the code {\tt SALT2mu}  \citep{Marriner}, also a part of the {\tt SNANA} suite.  In the \salt\ model, the distance modulus is given by
\begin{equation}
\label{eqn:mu}
\mu_{\mathrm{SN}} = m_B - M_0 + \alpha x_1 - \beta c,
\end{equation}
where $m_B$ (peak apparent $B$-band magnitude), $x_1$, and $c$ are fit for each individual SN and $M_0$ (absolute magnitude), $\alpha$, and $\beta$ are global parameters 
of the SN sample. 
{\tt SALT2mu} computes $\alpha$ and $\beta$ (cosmology-independent corrections for the light-curve stretch and color) from a given SN\,Ia sample, allowing us to determine $\mu_{\mathrm{SN}}$ for each SN in the sample. 

\begin{figure}[tp]
\centering
\begin{tabular}{c}
\includegraphics[scale=0.32,angle=90]{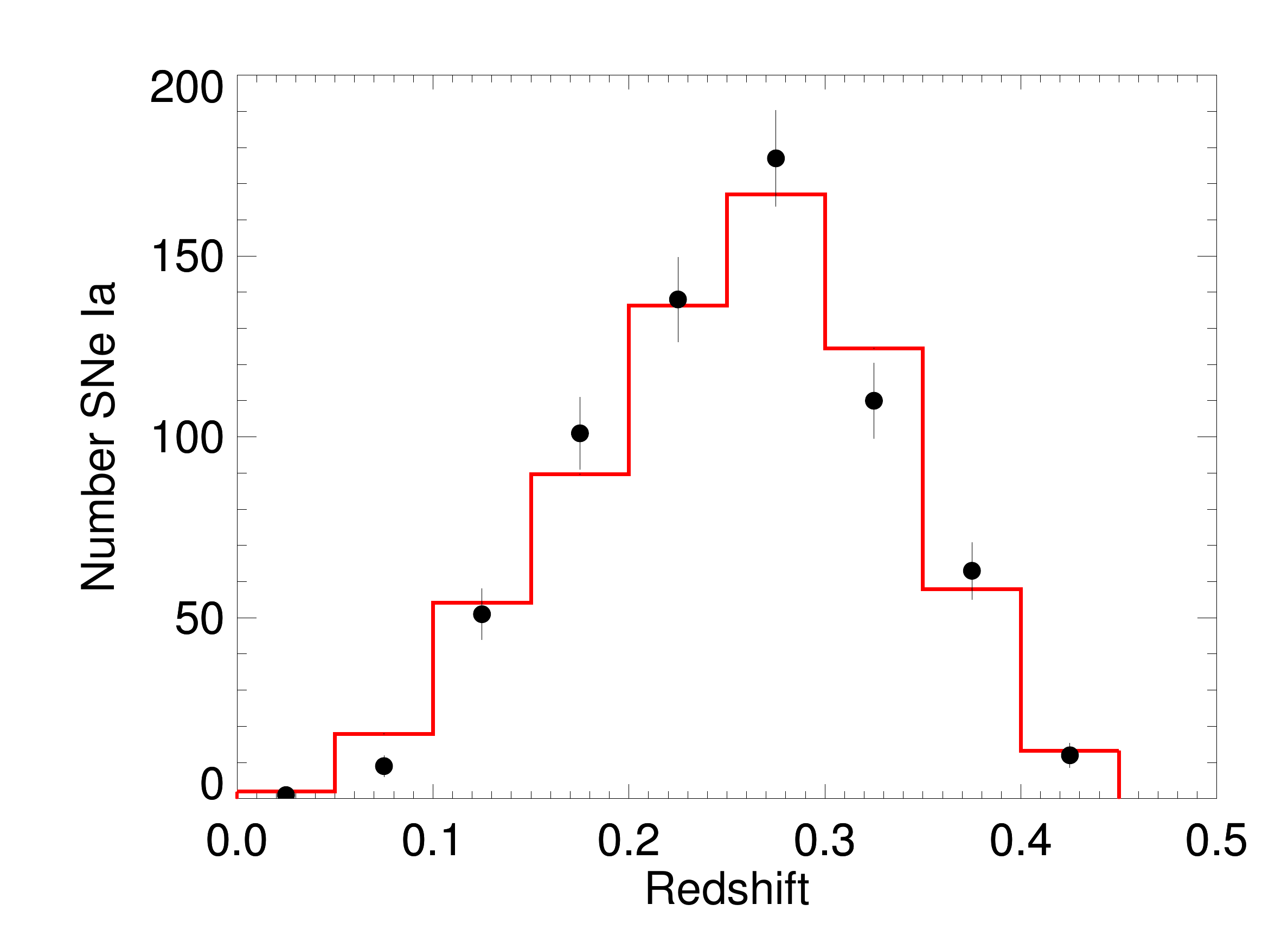} \\
\includegraphics[scale=0.32,angle=90]{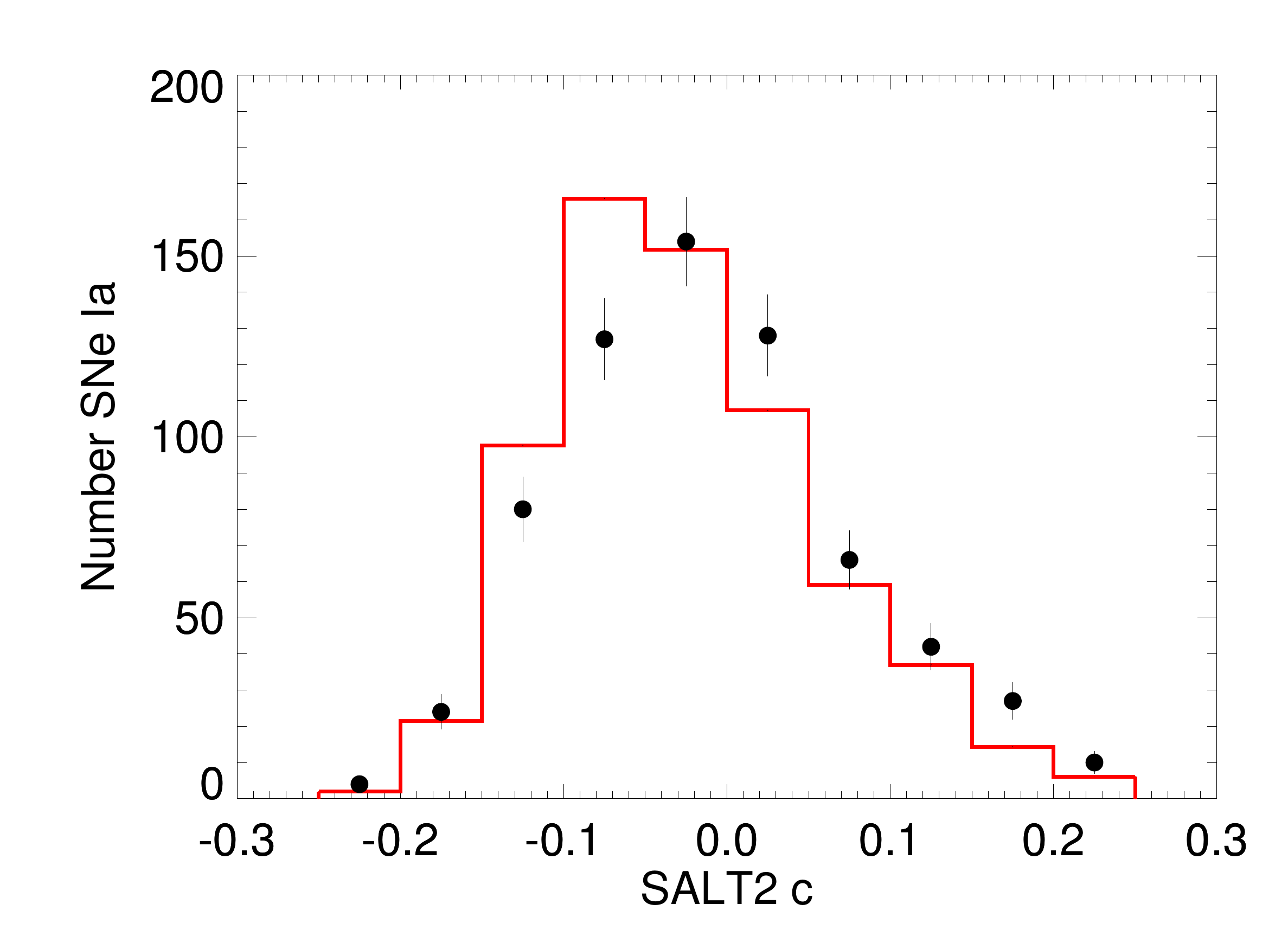} \\
\includegraphics[scale=0.32,angle=90]{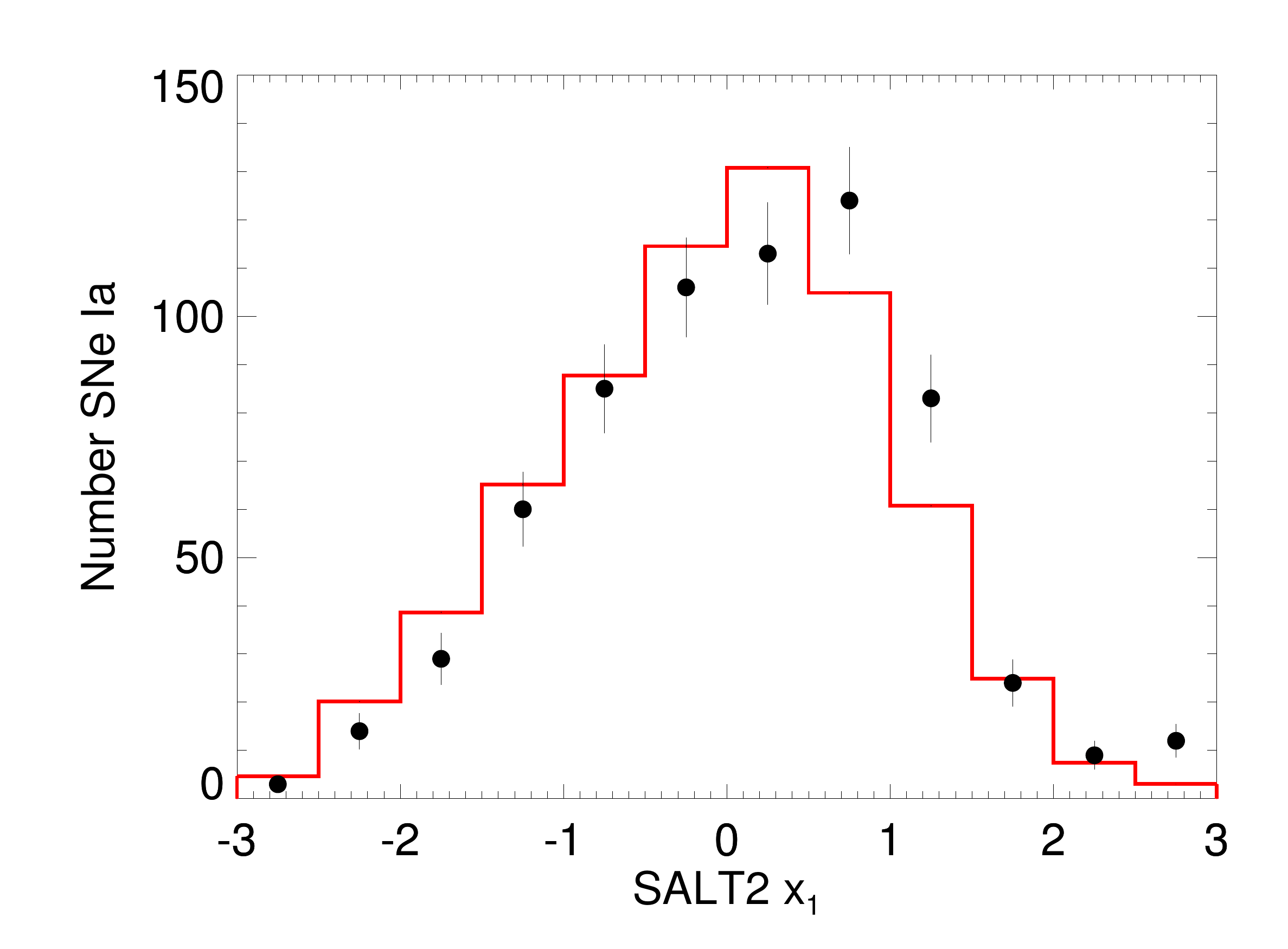} \\
\end{tabular}
\caption{Comparison of MC simulation (red histogram) and SDSS-SNS data (black points).  The MC distributions are normalized to the low-$z$ ($ z < 0.25$) data.  Error bars on the data points represent the square root of the number of SNe Ia in the respective bin.  Distributions are displayed for the redshift (top), {\tt SALT2} color (middle) and {\tt SALT2} stretch (bottom).}
\label{fig:datasim}
\end{figure}

This computation of the distance modulus, however, has not been corrected for selection effects (i.e. Malmquist Bias). The well-known Malmquist bias \citep{Malmquist} stipulates that for a magnitude limited survey, a given SN\,Ia may appear brighter due to random statistical fluctuations. These fluctuations can be seen to a greater distance and thus a larger portion will be detected in a magnitude-limited sample. To determine the correction for this effect, as well as others corrections stemming from {\tt SALT2} fitting (e.g., poor fits to low \emph{S/N} data), we run SDSS-like simulations (with approximately ten times the data statistics) and compare the expected ($\mu_{\mathrm{TRUE}}$) and observed ($\mu_{\mathrm{FIT}}$) distance moduli.  Realistic light curves are simulated using the {\tt SNANA}  code, where the MC is used to make detailed comparisons with the data using different models of intrinsic SN\,Ia brightness variations \citep{Kessler13}.  The simulations assume the best-fit flat $\Lambda$CDM cosmology of B14 ($\Omega_{\mathrm{M}} = 0.295$) and SN\,Ia are generated using the SALT-II model \citep{SALT2}.  As in \citet{Kessler13}, we simulate asymmetric Gaussian distributions for our input color and stretch.  The following parameters best match our data: $\bar{c}=-0.09$, $\sigma_{+,c}=0.13$, $\sigma_{-,c}=0.02$, $\bar{x_{1}}= 0.5$, $\sigma_{+,x_{1}}=0.5$, and $\sigma_{-,x_{1}}=1.5$.  Comparisons between the data and simulations are presented in Figure~\ref{fig:datasim}.  

\begin{figure}[tp]
\centering
\includegraphics[scale=0.37,angle=90]{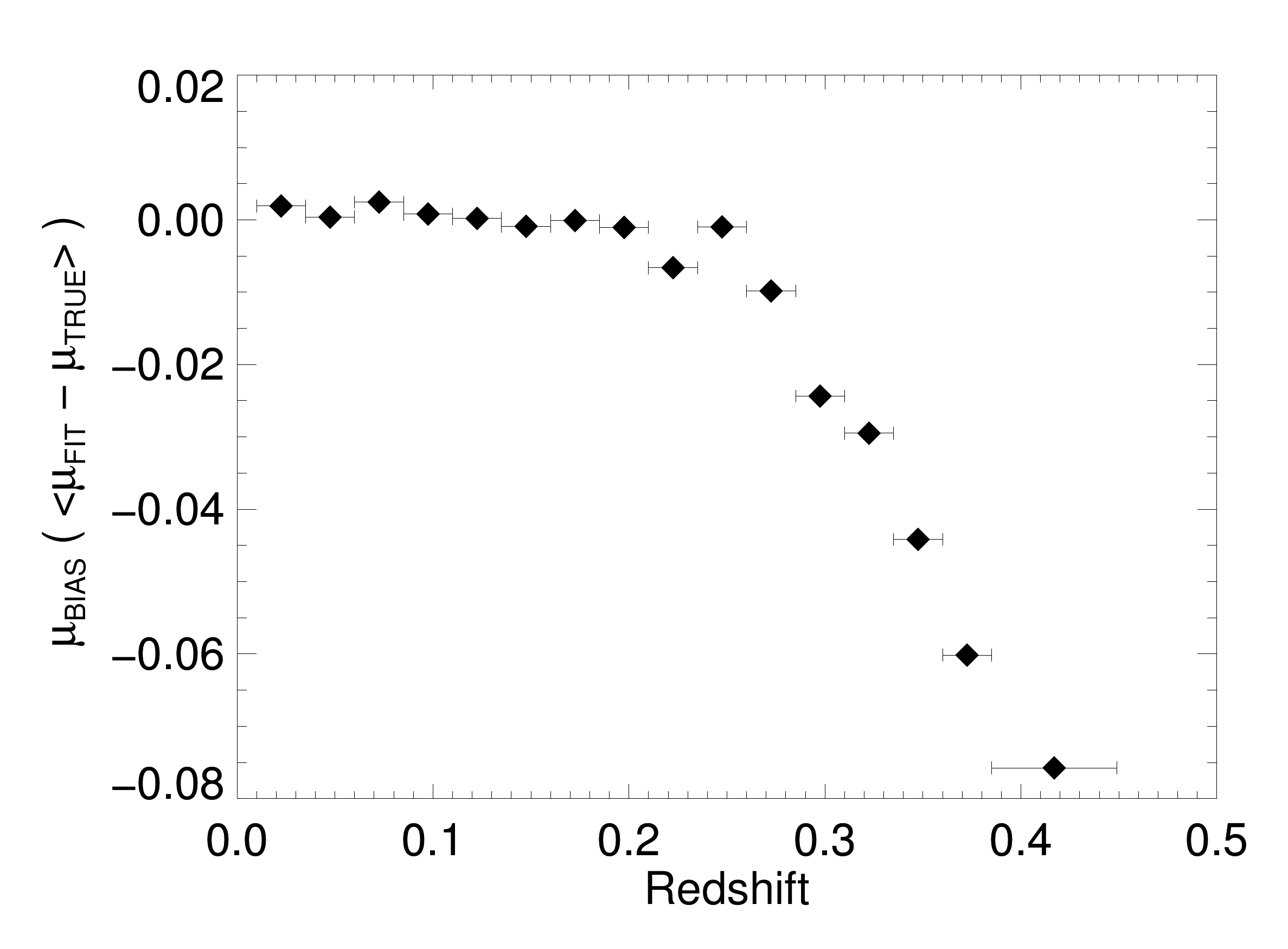}
\caption{Difference between the measured and true distance modulus (defined as $\mu_{\mathrm{BIAS}}$) from our simulations, as a function of redshift.  Data points are inverse-variance weighted averages in redshift bins of width 0.025 with error bars representing the width of each bin.  Each bin contains at least 500 SNe\,Ia.} 
\label{fig:mubias}
\end{figure}

The average difference in distance modulus as a function of redshift, which we define as $\mu_{\mathrm{BIAS}}$, is presented in Figure~\ref{fig:mubias}.  In the lower-redshift range ($z\lesssim0.3$) the bias is very small; however, as the redshift exceeds $z = 0.3$, the offset noticeably grows with redshift. In the higher redshift regime, the magnitude of the bias approaches that of our host-galaxy effects; therefore, correcting for this bias may potentially misconstrue any observed host-galaxy correlations. To ensure that our sample is not contaminated by this bias, we choose to limit the redshift of our SNe Ia to $z<0.3$. If we recompute the bias for this lower redshift sample only, we find $-0.006 < \mu_{\mathrm{BIAS}} < 0.008$ and conclude that this effect is negligible and does not require additional corrections.

As presented in Table~\ref{pmsamplecuts}, 473 SNe\,Ia meet the light-curve sampling, $c$ and $x_{1}$, $P_\mathrm{FIT}$ and redshift requirements.

\begin{deluxetable}{lrrrr}
\tablecaption{Cumulative PM Sample Definition   \label{pmsamplecuts} }
\tablewidth{0pt}
\tablehead{\colhead{Selection Requirements} & \colhead{Removed} & \colhead{Total} & \colhead{Phot-Ia} & \colhead{Spec-Ia}\\ 
\colhead{}& \colhead{SNe~Ia}&\colhead{}&\colhead{} }
\startdata
Total SDSS-SNS Transients & -- & 10,258 & -- & -- \\
S14 SNe\,Ia\tablenotemark{a} &8935  & 1323 &  824 & 499 \\
Non-peculiar SNe\,Ia & 8 & 1315 & 824 & 491 \\
Light-curve sampling & 534  & 770 & 434 & 336 \\
Elliptical $c$, $x_{1}$ cuts & 67 & 703 & 382 & 321 \\
$P_\mathrm{FIT} > 0.01$ &41 & 662 & 361 & 301 \\ 
$z < 0.3$ &189 & 473 & 215 & 258 \\
HR outlier rejection &7 & 466 & 208 & 258 \\
Host spectrum identified & 116 &  350 & 177 & 173 \\
Host, SN redshift agreement & 3  &  347 & 176 & 171   \\
Well-defined host mass & 2 & 345 & 176 & 169 \\
\enddata
\tiny
\tablenotetext{a}{This removes transients, such as core-collapse SNe, that were not identified as SNe\,Ia in S14.}

\end{deluxetable}

The elliptical cut in the $c$--$x_{1}$ plane removes much of the contamination from core-collapse (CC) SNe in the photometric sample.  We apply this cut on light-curve fit parameters to both the Phot and Spec-Ia samples, as we wish to maintain homogeneity across our combined sample and as these light-curve fit parameters are used to estimate the SN distance moduli.  Given our data, we find best-fit values of $\alpha= 0.14 \pm 0.012$ and $\beta = 3.11 \pm 0.140$.  In order to obtain $\chi^{2}_{red} \approx 1$, an intrinsic scatter of $0.167$ magnitudes must be added when performing the fit.  HRs for our SNe are then calculated from $\mu_{\mathrm{SN}}$ and $\mu_z$ computed with the assumed B14 cosmology.  However, we note that we do not incorporate this intrinsic scatter into the uncertainty on the distance moduli $\mu_{SN}$ used in this analysis. Rather, we independently fit for the intrinsic scatter when analyzing correlations between HR and host-galaxy properties. This is further explained in Section~\ref{sec:results}.

When examining the HR for our data, we notice a strong correlation between HR and $c$, particularly for $c < 0$; we do not observe such a correlation between HR and $x_{1}$.  Both trends are also apparent in our simulations and this trend with $c$ has been seen previously in SN surveys at both low- and high-redshift \citep{Sullivan11, Ganeshalingam13}.  We elect not to correct for this effect in our analysis as this is not done in previous works and we wish to compare our results in the most consistent manner possible.  A discussion of HR-$c$ corrections and the effect on our results can be found in Appendix B.

Figure~\ref{fig:HRdist} displays the distribution of HRs of those SNe Ia passing our selection requirements.  The mean of the distribution is $0.014$ magnitudes and the standard deviation is $0.228$.  We remove from our sample 7 SNe with HRs $>3\sigma$ from the mean (corresponding to HR $ <-0.668$ and HR $ > 0.697$) as it is highly unlikely these are normal SNe\,Ia. All SNe removed in this way are Phot-Ia; this outlier rejection method does not affect the number of spectroscopically-confirmed SNe\,Ia in our sample.  After removing these outliers, the mean and standard deviation of the HR distribution reduce to $0.002$ and $0.187$, respectively.  Imposing this requirement leaves 208 Phot-Ia and 258 Spec-Ia in our sample. As a check, we have examined the Hubble diagram of this sample and found that imposing these criteria removes the majority of potential contaminants and shows no noticeable redshift-dependent pollution. Overall, this Hubble diagram is much cleaner than what is presented in S14, due to the fact that we impose stricter $S/N$ requirements and temporal coverage of our SNe\,Ia light curves. 

\begin{figure}[tp]
\centering
\includegraphics[scale=0.37,angle=90]{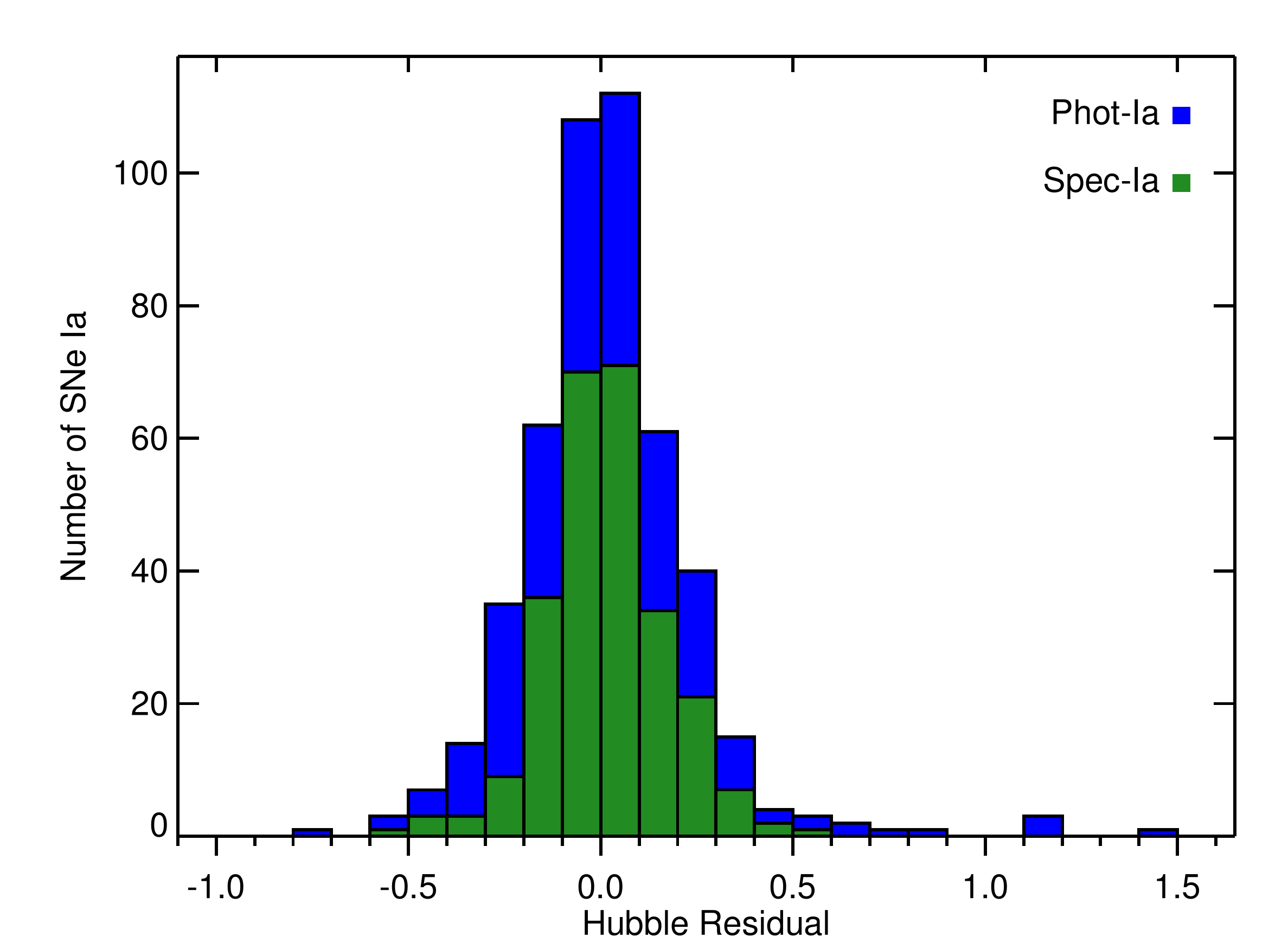}
\caption{Distribution of HRs calculated using the derived {\tt SALT2mu} distance moduli.  Histograms are stacked such that the Phot-Ia (blue) and Spec-Ia (green) add to the total number in a given bin. The mean of the distribution is $0.014$ magnitudes and the standard deviation is $0.228$.  We remove from our sample 7 SNe with HRs $>3\sigma$ from the mean (corresponding to HR $ <  -0.668$ and HR $ > 0.697$) as it is highly unlikely these outliers are normal SNe\,Ia.  All outliers removed in this way are Phot-Ia.  This reduces the mean and standard deviation to $0.002$ and $0.187$, respectively.}
\label{fig:HRdist}
\end{figure} 

Finally, we require that the SNe\,Ia have an observed host-galaxy spectrum and photometrically-derived host-galaxy mass with well-defined uncertainties (as described in Section~\ref{sub_sec:Mass}).  The requirement that each host has a BOSS or SDSS spectrum is necessary to ensure that we are correctly matching the SN\,Ia with its host. This requirement removes both Phot- and Spec-Ia with host-spectra followed up by programs other than BOSS or SDSS as well as hostless Spec-Ia. Although each host in our sample has an observed spectrum, we do not use spectral absorption features to obtain host masses (discussed in Section~\ref{sec:GalSpec_Method}) and instead rely on photometric mass measurements.

We remove those SNe\,Ia which do not meet these criteria and are left with a sample of \numpm, which we define as the PM (Photometric Mass) sample.  These cuts, in addition to all those previously described in this section, are outlined in Table~\ref{pmsamplecuts}.  The PM sample is one of two samples of SNe\,Ia we analyze in Section~\ref{sec:results}; further spectroscopic requirements imposed to cull the second sample are detailed in Section~\ref{sub_sec:spec}


\section{HOST GALAXY SPECTRAL ANALYSIS}
\label{sec:GalSpec}

We describe here our analysis of BOSS and SDSS-I/II spectra of the host galaxies of SNe\,Ia from the SDSS-SNS.  Section~\ref{sec:GalSpec_Method} outlines the procedure used to measure fluxes, equivalent widths, and amplitude-to-noise (the ratio of the peak flux of the emission line to the continuum; hereafter $A/N$) from the spectra, which we optimize and use instead of existing catalog data.  Section~\ref{sub_sec:spec} details the requirements, both physical (e.g., AGN contamination) and observational (e.g., $S/N$), we impose on the spectra to be included in our subsequent analysis of host-galaxy emission-line properties.   

\subsection{Methods}
\label{sec:GalSpec_Method}

\begin{figure*}[tp]
\centering
\includegraphics[scale=0.7,angle=90]{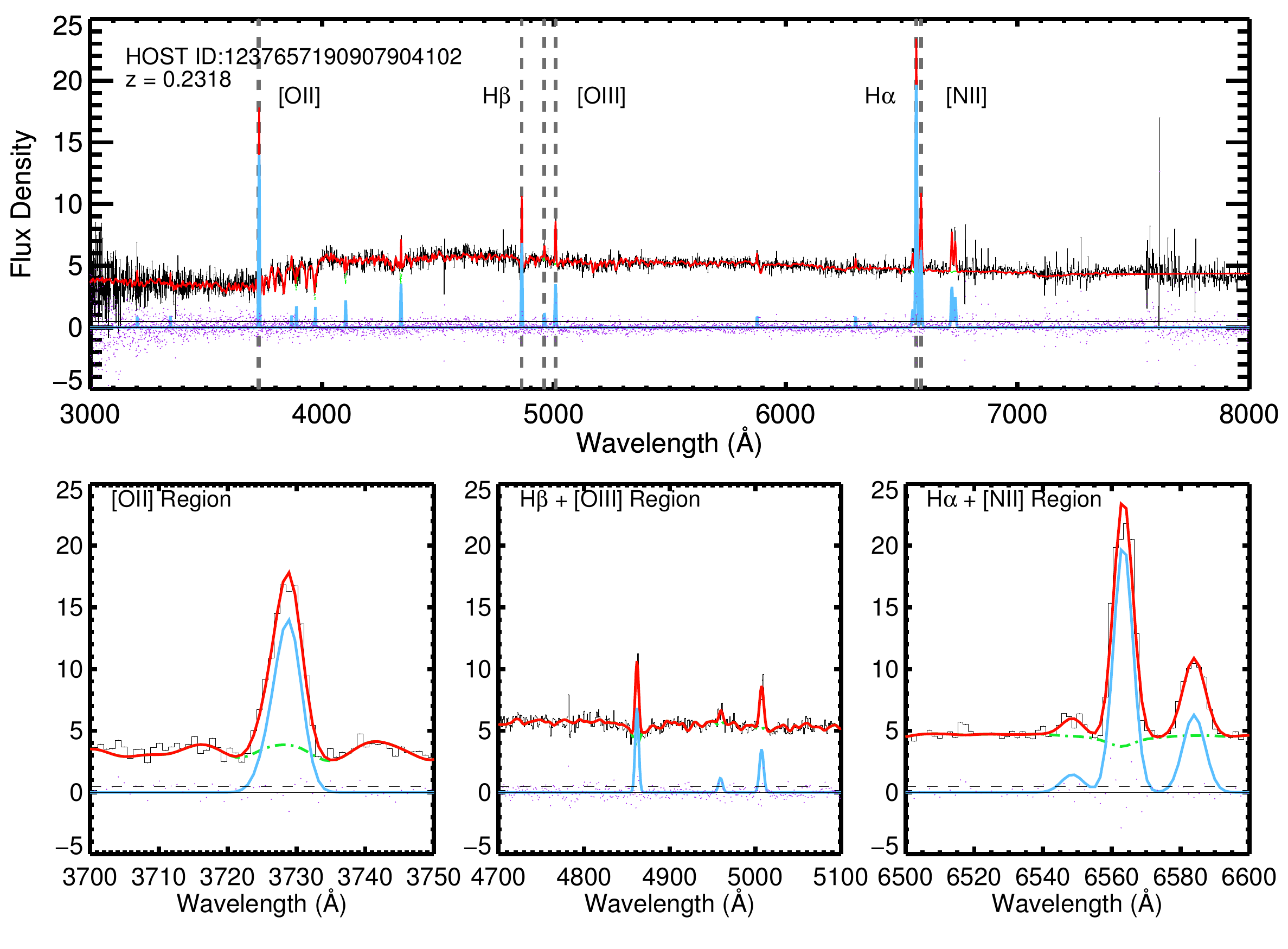}
\caption{A sample {\tt GANDALF} fit of the BOSS spectrum for the host of CID 13897.  Wavelengths in this spectrum are given in the rest frame.  Flux density is in units $10^{-17}\unt{erg}\unt{s}^{-1}\unt{cm}^{-2}\unt{\AA}^{-1}$.  The data are shown in black with the best-fit model overplotted in red.  The green dot-dashed line represents the continuum fit and the blue line shows the emission spectrum, which is obtained by subtracting the continuum model from the best-fit model.  Residual points between the data and the best-fit spectrum are also shown in purple.  Vertical dashed lines indicate the emission lines predominantly used in our analysis.  The three lower panels display the specific regions that contain these lines.}
\label{fig:gandalffit}
\end{figure*}

Emission-line properties of galaxy spectra obtained as part of the BOSS and SDSS-I/II programs are calculated using Version 1.8 (v1.8) of the code {\tt GANDALF} \citep[Gas AND Absorption Line Fitter;][]{Sarzi06}.  {\tt GANDALF} simultaneously fits for the stellar population and the emission-line spectrum, which prevents the presence of absorption lines from biasing the measurement of ionized gas emission.  {\tt GANDALF} uses pPXF \citep[penalized Pixel-Fitting;][]{Cappellari04} to measure the stellar kinematics of the galaxy while masking the emission-line regions.  The code then fits the gas kinematics (velocity and velocity dispersion) and measures emission-line fluxes for a user-determined set of (Gaussian) emission lines.  The effects of dust in the observed galaxy are corrected for by simultaneously fitting for extinction under the assumption of a \citet{Calzetti} reddening law.  A sample {\tt GANDALF} spectral fit is shown in Figure~\ref{fig:gandalffit}.

Our work with {\tt GANDALF} closely follows that of \citet[hereafter T13]{Thomas13}, which details the method used for measuring emission-line properties in SDSS DR9 \citep{Ahn}.  As in T13, our galaxy templates are simple stellar population (SSP) models from \citet[hereafter M11]{MS11}.  The particular set of models we use is built on the MILES stellar library, which is extended into the UV based on a theoretical library (necessary to constrain the blue end of our observed spectra).   Our template library is derived using a Salpeter initial mass function (IMF) \citep{Salpeter}, as an extended UV library for M11 is not available with Chabrier \citep{Chabrier} or Kroupa \citep{Kroupa} IMFs.  We resample the M11 galaxy templates to have a wavelength-independent resolution of $R=2000$.  This is an approximation to the true instrumental resolutions of both SDSS I/II and BOSS, which are wavelength-dependent.  Before conducting our analysis, we convert the observed spectra from the SDSS-standard vacuum wavelengths into air wavelengths. We additionally assume only a single metallicity (solar), and a subset of 19 of the 47 available galaxy ages in the model.  These choices are motivated by the fact that the primary goal is to remove the continuum; small variations in the underlying spectrum only matter to the extent that they affect the emission-line measurements.  It also results in a significant reduction in computation time.  We ran {\tt GANDALF} on a subset of our spectra using both the full and reduced sets of temporal templates and found that our results were in no way affected by this choice.  

We have made a few changes from the analysis of T13 that are optimized to our data set.  The most significant of these is how we tie spectral lines in the fitting procedure, fixing the velocity and width of the Balmer and forbidden lines to values derived for \Ha\ and \NII, respectively.  T13 does not adopt this procedure as \Ha\ and \NII\ are redshifted beyond the BOSS wavelength range at $z>0.45$, and their goal is a homogeneous derivation of emission-line fluxes across the entire BOSS sample.  Thus, they allow the velocity, width, and amplitude of each emission line to be fit freely.  However, all of the SNe\,Ia included in this analysis are below this redshift.  Therefore, we explicitly restrict our analysis to galaxies where we observe \Ha\ and \NII\ and take advantage of the constraining power added by tying the line velocities and widths together.

Unlike in T13, we first correct the observed spectra for the effects of dust absorption in the Milky Way before running {\tt GANDALF}.  We use the extinction values from \citet{Schlegel} and assume the \citet[CCM]{CCM} extinction law (with $R_{V}=3.1$). In addition, we use Case-B recombination \citep{Osterbrock}, which assumes the ratio of intrinsic \Ha\ to \Hb\ flux (the ``Balmer Decrement") of 2.86, to correct for host-galaxy extinction, while T13 utilizes the extinction output by {\tt GANDALF}, derived from a fit to the underlying galaxy continuum.  We find that in three cases, the observed \Hb\ flux output by {\tt GANDALF} is so large ($>10^{-13}\unt{erg}\unt{s}^{-1}\unt{cm}^{-2}\unt{\AA}^{-1}$) that the computed extinction value is unphysical. These large \Hb\ flux values are also unphysical, and so we remove these spectra from our sample.

The emission line file used in our {\tt GANDALF} fits is given in Table~\ref{ELF}.  This file allows the user to specify how to tie spectral lines together or fit them freely, and whether certain lines should be masked in the fit.  We note as an example that, unlike T13, we mask the Na I absorption feature when fitting the continuum.  For more details on how to create a user-specific emission line file, see \citet{Sarzi06}.  

We also make some adjustments to the {\tt GANDALF} code.  We have modified {\tt GANDALF} to return flux uncertainties for lines where the velocity and width of the species is tied to that of a stronger line.  {\tt GANDALF} v1.8 treats the uncertainty of the velocity and line width in these cases as zero, and thus computes no uncertainty.  We treat the uncertainties of the fitted parameters for these weaker lines in the same way as those to which they are tied.  In addition, {\tt GANDALF} v1.8 incorrectly measures the EW of spectral lines; the flux density of the continuum needs to be scaled up by a factor of $(1+z)$.  We include this correction, which is also discussed in T13, in our analysis.  Finally, we note that the stellar kinematics from pPXF are derived over the region $4000-6500\unt{\AA}$ in the rest frame of the galaxy.  This is the same band as in T13, although it is incorrectly stated in that work.  Comparisons between our {\tt GANDALF} results and those in the SDSS DR10, which include modifications on the published SDSS DR9 results as stipulated in T13, are presented in Appendix A.

\begin{deluxetable*}{c c c c c c c c c}
\tablecaption{{\tt GANDALF} Emission-Line Setup File \label{ELF}}
\tabletypesize{\scriptsize}
\tablewidth{0pt}
         \tablehead{ Line Index&Line Name  & Rest Wavelength & Action\tablenotemark{1} &  L-kind\tablenotemark{2} & A\_i\tablenotemark{3}  &   
          V\_g/i\tablenotemark{4}  & sig\_g/i\tablenotemark{5} & Fit-Kind\tablenotemark{6} \\ & & (\AA)  }
\startdata 
 0	&\spec{He}{II}	 & 3203.15 &m &l &1.000& 0&10&f \\
 1	&[\spec{Ne}{V}]  & 3345.81& m   &   l    &   1.000  &    0   & 10 &     f\\
 2	&[\spec{Ne}{V}] & 3425.81& m&l &1.000 & 0 &10  & f\\
 3	 &\OII	& 3726.03& m	&l 	&1.000	   &0	&10	&t25\\
  4	 &\OII	& 3728.73& m	&l 	&1.000	   &0	&10	&t25\\
  5	 &[\spec{Ne}{III}]& 3868.69& m	&l 	&1.000	   &0	&10	&f\\
  6	 &[\spec{Ne}{III}] &3967.40& m	&l 	&1.000	   &0	&10	&f\\
  7      &H5   &	 3889.05 &m     & l       &1.000      &0	&10    &	f\\
  8      &H$\epsilon$     & 3970.07& m     & l       &1.000      &0    &10     & f\\
  9      &H$\delta$    &4101.73& m      &l       &1.000      &0    &10      &t24\\
 10     & H$\gamma$    &4340.46& m      &l       &1.000      &0    &10     & t24\\
 11      &\OIII  &4363.15 &m      &l       &1.000      &0    &10     & f\\
 12	 &\spec{He}{II}	& 4685.74& m	&l	&1.000	   &0	&10	&f\\
 13      &[\spec{Ar}{IV}]  &4711.30 &m      &l       &1.000      &0	&10    	&f\\
 14      &[\spec{Ar}{IV}]  &4740.10 &m     & l       &1.000      &0	&10   	&f\\
15	 &H$\beta$ 	&4861.32 &m	&l	&1.000	   &0	&10	&t24\\
16	 &\OIII	& 4958.83 &m	&l	&1.000	   &0	&10	&t25\\
 17	 &\OIII	&5006.77 &m	&l	&1.000	   &0	&10	&t25\\
 18      &[\spec{N}{I}]    &5197.90& m	&l	&1.000	   &0	&10	&f\\
 19      &[\spec{N}{I}]    &5200.39& m      &l	&1.000	   &0	&10	&f\\
 20	 &\spec{He}{I}	&5875.60 &m	&l	&1.000	   &0	&10	&f\\
 21      &[\spec{O}{I}]	& 6300.20 &m	&l	&1.000	   &0	&10	&f\\
 22      &[\spec{O}{I}]	 &6363.67 &m	&l 	&1.000	   &0	&10	&f\\
 23	 &\NII	 &6547.96 &m	&l	&1.000	   &0	&10	&t25\\
 24	 &H$\alpha$	 &6562.80 &m	&l 	&1.000	   &0	&10	&f\\
25	 &\NII	 &6583.34 &m	&l	&1.000	   &0	&10	&f\\
26	 &\SII	 &6716.31 &m	&l	&1.000	   &0	&10	&t25\\
27 	 &\SII	 &6730.68 &m&	l	&1.000	  & 0	&10	&t25\\
 90  &    sky &    5577.00& m&      l    &   1.000&      0   & 10&      f\\
 91  &    sky   &  6300.00& m  &    l  &     1.000  &    0  &  10  &    f\\
 92    &  sky    & 6363.00& m    &  l&       1.000    &  0&    10    &  f\\
 100  &    \spec{Na}{I}   &  5890.00& m &     l   &   -1.000&      0   & 10&      t101\\
 101    &  \spec{Na}{I} & 5896.00& m    &  l&      -1.000   &   0&    10   &   f \\
	  \enddata
	
	\tiny
	\tablenotetext{1}{The ``action" sets whether each of the listed lines should be fit (f), ignored (i), or \\
	whether the surrounding spectral region should be masked (m).  As {\tt GANDALF} runs, the ``action" is changed by the code; e.g., if the ``action" is set to `m', the line will be masked when fitting for the continuum, then changed to `f' when fitting for the emission lines.  The subsequent fields in the setup file are only used when the ``action" is set to `f'.}
	\tablenotetext{2}{The line-kind ``l-kind" allows {\tt GANDALF} to identify whether or not a line should be treated as belonging to a doublet or multiplet.  All lines can be treated individually (l) or can be tied to the strongest element of their multiplet (dXX), where XX is the line index.  If a line is identified as part of a doublet or multiplet, its amplitude is fixed to that of the strongest element via A\_i.}
	\tablenotetext{3}{Used to set the relative emission (A\_i $> 0$) or absorption (A\_i $<0$) strength of lines in a multiplet.  If a line is to be treated individually, A\_i is set to unity.}
	\tablenotetext{4}{Initial estimate for line velocity, km s$^{-1}$.}
	\tablenotetext{5}{Initial estimate for line velocity dispersion, km s$^{-1}$.}
	\tablenotetext{6}{Indicates if the position and width of the line is found freely (f) or tied (tXX) to another line, where XX is the line index.}
	
	\end{deluxetable*}

Recent analyses of SN\,Ia host-galaxy spectra by J13 and P14 used {\tt GANDALF} to extract absorption spectra 
as well as emission lines.  Absorption spectra can be used to estimate galaxy age and stellar metallicity but require 
that the spectra be of sufficient $S/N$ to measure absorption-line indices.  
J13 used host-galaxy spectra from SDSS-II ($z \lesssim 0.2$) while P14 obtained most of their host spectra from Gemini 
observations ($z<0.09$).   The redshift limit for these samples is much lower than for our sample presented here (and in the case of 
P14, the host observations were taken using telescopes with larger apertures), and thus their host 
spectra are higher $S/N$.  Like J13, we make use of SDSS-II 
spectra; however, the majority of our spectra are from BOSS and are generally lower $S/N$ (see discussion in Section~\ref{sub_sec:apeff}).  Therefore, for this work we analyze only emission-line spectra and do not attempt to extract properties from absorption spectra.  
As noted in T13, one could attempt to do so by stacking spectra to increase the S/N, but we leave this exercise for future study.

\subsection{Selection Criteria}
\label{sub_sec:spec}

Here we describe the requirements placed on our host-galaxy spectroscopy which allow us to take the emission-line fluxes, measured as described in the previous section, and derive reliable host-galaxy properties in Section~\ref{sec:derivprop}.

To ensure accurate spectral line fits and emission-line fluxes, T13 requires $A/N > 2$  for the \Ha, \Hb, \OIII, and \NII\ lines.  However, we have many cases where these four emission lines are detected and yet not all their $A/N > 2$.  Requiring $A/N > 2$ for only the \Ha\ and \Hb\ lines removes the bulk of our low $S/N$ spectra, as well as the majority of our passive-galaxy sample, without sacrificing the large sample size.  Therefore, we impose this $A/N$ criterion on the Balmer lines only.

\begin{figure}[tp]
\centering
\includegraphics[scale=0.37,angle=90]{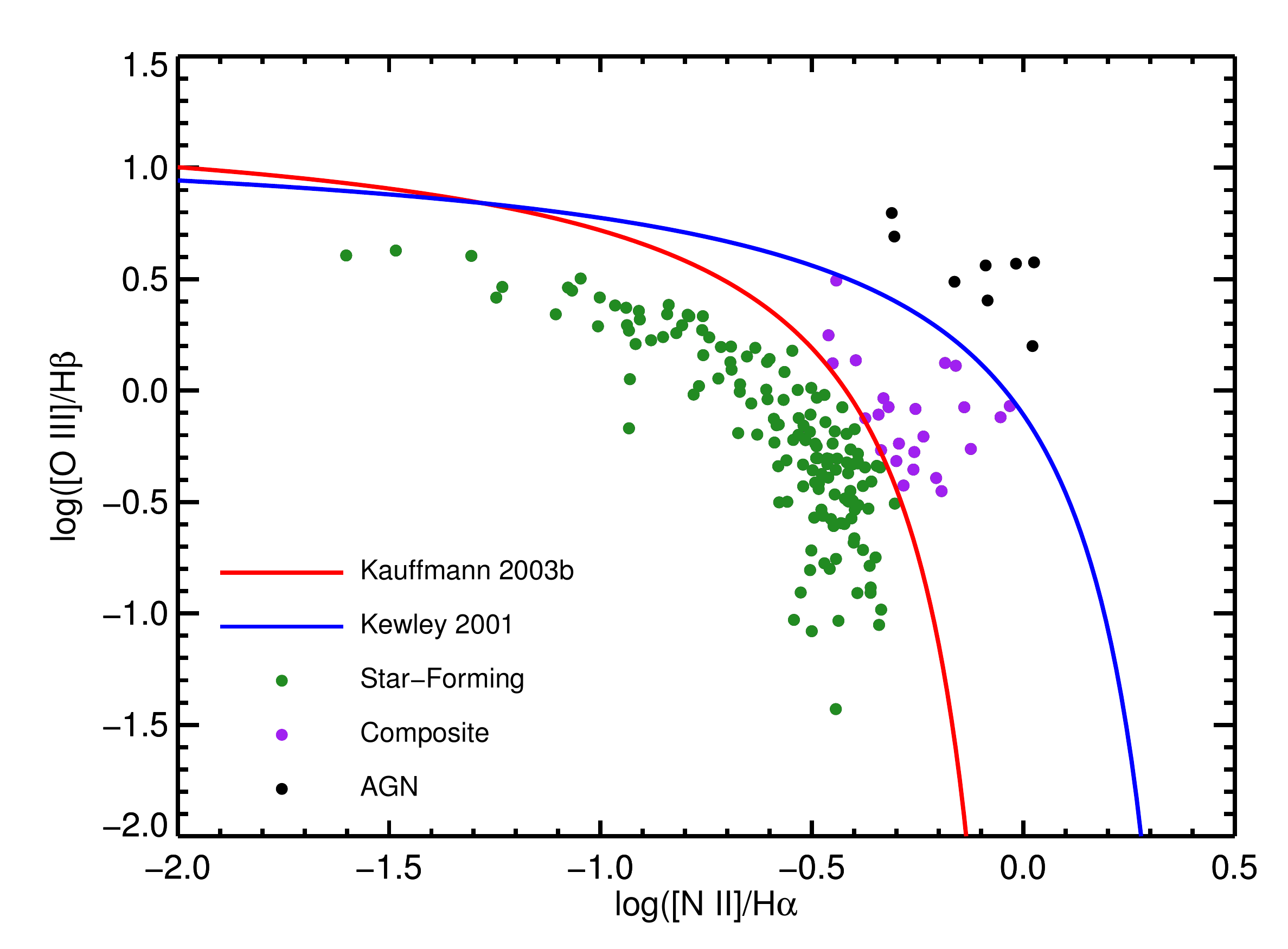}
\caption{BPT diagram for host galaxies of our SNe\,Ia.  The galaxies displayed here have passed selection criteria through $A/N$ cuts, as outlined in Tables~\ref{pmsamplecuts} and \ref{mzssamplecuts}.  We have trimmed the axes to better focus on the bulk of our sample; therefore, some star-forming hosts and AGN may not be shown.  Galaxies to the right of the blue curve (Kewley 2001) are deemed AGN (black points), while those to the left of the red curve (Kaufmann 2003b) are regarded as star-forming (green points).  Those galaxies that lie between the two curves (purple points) are labeled ``Composite".  We continue our analysis using galaxies to the left of the blue curve, although not all will be included in the final sample for analysis.} 
\label{fig:BPT}
\end{figure}

We then use BPT diagnostics \citep{BPT} to separate the star-forming galaxies from those dominated by active galactic nuclei (AGN).  This classification requires an analysis of the optical diagnostic plane spanning log(\OIII/\Hb) versus log(\NII/\Ha).  We first utilize the hyperbolic division of the plane in \citet{Kewley01}, then adopt the stricter division presented in \citet{Kauffmann} to select star-forming galaxies more carefully.  Hosts for which \citet{Kewley01} and \citet{Kauffmann} disagree are deemed ``Composite", as in \citet{Brinchmann}.  It is crucial to separate the AGN-dominated spectra as their emission lines are produced by different physical processes, and thus will produce inaccurate metallicity estimates.  The BPT diagram for our sample after imposing $A/N$ cuts is presented in Figure~\ref{fig:BPT}. 

In Table~\ref{mzssamplecuts} we list the cuts applied in this section which reduce the PM sample, given in Table~\ref{pmsamplecuts}, to a sample of \nummzs\ SN\,Ia host galaxies for which we produce (see Section~\ref{sec:derivprop}) reliable measurements of mass (M), metallicity (Z), and specific star-formation rate (S); we refer to this as the MZS sample.  The $A/N$ cut is the most significant, reducing our sample by $\approx50\%$, demonstrating the difficulty in measuring emission-line properties from low $S/N$ data.  We note the final cut in Table~\ref{mzssamplecuts} (not described in this section) is a requirement on the fraction of galaxy light obtained within the BOSS/SDSS fiber.  This is necessary to ensure the properties derived from our spectra are \emph{global} host-galaxy properties.  As this cut is not based on the spectroscopy itself, but rather on host-galaxy photometry, it is detailed in Section~\ref{sub_sec:apeff}.  

\begin{deluxetable*}{l r r r r}
\tablecaption{Cumulative MZS Sample Definition} 
\tablewidth{0pt}
\tablehead{Selection Requirements & Removed & Total & Phot-Ia & Spec-Ia \label{mzssamplecuts} \\ 
\colhead{}& \colhead{SNe~Ia}&\colhead{}&\colhead{} }
\startdata
PM Sample & -- &  345 & 176 & 169 \\
\tablenotemark{a}Observed \Hb\ flux $< 10^{4}$& 3 & 342 &176 & 166 \\
\Ha\ and \Hb\ $A/N > 2 $ &  149 &  184 &  88 &  96 \\
Star-forming or ``Composite" host &  9 &  175 &  80 &  95 \\
$0.2 \leq$ $g$-band fiber fraction $ < 1$ &  31 &  144 &  78 &  66 \\
\enddata
\tiny
\tablenotetext{a}{Flux density in units of $10^{-17}\unt{erg}\unt{s}^{-1}\unt{cm}^{-2}\unt{\AA}^{-1}$} \\
\end{deluxetable*}


\section{DERIVED HOST GALAXY PROPERTIES}
\label{sec:derivprop}
In this section we describe the methods used to derive the host-galaxy properties, both spectroscopic and photometric, used in this analysis.  Sections~\ref{sub_sec:met} and \ref{sub_sec:sfrmeth} detail the processes for computing, respectively, gas-phase metallicities and star-formation rates from the measurements obtained in Section~\ref{sec:GalSpec}.  In Section~\ref{sub_sec:Mass} we describe the source for our host-galaxy masses.  We discuss fiber aperture effects -- what biases may be present, how we correct for them, and their impact on sample selection -- in Section \ref{sub_sec:apeff}.  
\subsection{Metallicity}
\label{sub_sec:met}

There are several methods for estimating gas-phase metallicity ($Z \equiv \log(\mathrm{O/H})+12$) from emission-line fluxes.  Although the metallicities from each method do not have the same absolute values, relative values tend to remain consistent (i.e., a galaxy with low metallicity in one method will have low metallicity in another).  \citet[hereafter KE08]{KE08} summarize these techniques and derive conversions from one metallicity calibration into another.  In this analysis we adopt the calibration of \citet[hereafter KD02]{KD02}, as recommended by (and updated in) KE08.  

The KD02 algorithm is split into upper (high $Z$) and lower (low $Z$) branches based on the ratio of the \NII\ and \OII\ line fluxes obtained from the galaxy spectrum (\OII\ $= [$\spec{O}{II} $\lambda 3727]+[$\spec{O}{II} $\lambda 3729]$; \NII\ $= [$\spec{N}{II} $\lambda 6584]$).  For galaxies with log(\NII/\OII) $>-1.2$, the metallicity is found via the real roots of the polynomial 
\begin{align}
& \log(\text{\NII/\OII}) = 1106.8660 - 532.15451Z \nonumber \\ & + 96.373260Z^2 - 7.8106123Z^3 + 0.2392847Z^4
\end{align}
The systematic accuracy of this method on the high-$Z$ branch, as stated in KE08, is $\sim 0.1\unt{dex}$.

For galaxies with log(\NII/\OII) $< -1.2$, the KD02 method derives metallicities using an average of two distinct $R_{23}$ calibrations (for a more complete discussion of $R_{23}$ see KE08) with a systematic uncertainty of $\sim0.15\unt{dex}$.  The first method utilizes the iterative procedure of \citet[hereafter KK04]{KK04} in the lower $R_{23}$ branch, while the second \citep{M91} is based on the photoionization code {\tt CLOUDY} \citep{Ferland98} with associated analytic solutions from \cite{Kobulnicky99}.  We require that a solution is found using both techniques to determine an accurate metallicity. 

\subsection{Star-Formation Rate}
\label{sub_sec:sfrmeth}

The \Ha\ line flux is used to determine the star-formation rate (SFR) of our host galaxies, as it traces luminosity from young ($\sim10^{6}$ yrs), massive (M $>10\Msun$) stars \citep{Kennicutt}.  It also allows for a direct coupling of nebular emission to instantaneous SFR, independent of any previous star-formation history.  As outlined in \citet{Kennicutt}, the star-formation rate for a galaxy with a Salpeter IMF can be found by
\begin{equation}
\label{eqn:sfr}
\mathrm{SFR}\ (\Msun \unt{yr}^{-1})=7.9\E{-42}L(\mathrm{H}\alpha)\ (\mbox{erg s}^{-1})\\
 \end{equation}
 where the H$\alpha\ $luminosity is determined using the line flux and the assumed B14 cosmology.  \citet{Brinchmann} have shown that the conversion factor between $L(\mathrm{H}\alpha)$ and SFR is dependent on the mass and metallicity of the galaxy.  To account for this variation, as in D11, we assume a systematic uncertainty in log(SFR) of 0.2.  

We note that we correct our SFR values for aperture effects (see Section~\ref{sub_sec:apeff}).
In addition, we compute the  specific star-formation rate (sSFR) by dividing the SFR by the photometrically-derived 
galaxy stellar mass, which is described in the following subsection.

To test the validity of our methods, we compare our metallicity and sSFR measurements to those reported in D11, as they also extract emission-line fluxes from BOSS and SDSS host-galaxy spectra and also compute metallicity using the KD02 algorithm.  We find that for the 39 hosts which overlap in the two samples, we recover the gas-phase metallicity and SFR measurements reported in D11. The distribution of the difference between our measurements and those of D11 shows no bias and has an approximately Gaussian distribution; 95\% of the sample agrees to within $2\sigma$.

\subsection{Host Mass}
\label{sub_sec:Mass}
Stellar masses for our host galaxies are taken from S14 and were computed using the method of 
\citet{Gupta11}.  This method employs model spectral energy distributions (SEDs) generated on a fixed grid 
using the Flexible Stellar Population Synthesis code \citep[FSPS;][]{Conroy09, Conroy10}.  
Synthetic photometry computed from these model SEDs in the SDSS $ugriz$ bands were compared to SDSS 
photometry of our host 
galaxies\footnote{Obtained from the DR8 Catalog Archive Server (CAS) at 
http://skyservice.pha.jhu.edu/casjobs/} while fixing the redshift to the spectroscopic value.  
For more details on the FSPS model parameters used and on the exact method of estimating stellar mass, 
see \citet{Gupta11}.   Systematic uncertainties in stellar mass estimates for normal galaxies are generally $<0.2$ dex 
\citep{Conroy13}.  At best it is 0.1 dex (25\%), and so we incorporate this 0.1 dex into our systematic 
uncertainty.

\subsection{Aperture Effects}
\label{sub_sec:apeff}

As we are deriving some galaxy properties from fixed-aperture spectra, we require a parameter that indicates the degree to which each spectrum is representative of a global average.  To do this we compute in $ugriz$ for each spectrum the ratio of flux observed within the fiber (the fiberMag) to the total flux of the target galaxy based on a profile fit (the modelMag).  The fiber and model magnitudes are taken from the SDSS Catalog Archive Server.  We refer to the derived ratio in each band as the fiber fraction.  Because our sample consists of spectra from both 2\arcsec\ and 3\arcsec\ diameter fibers, we compute fiber fractions for both cases.  

Based on the $g$-band fiber fraction, we remove the star-forming and ``Composite" spectra whose properties are are not indicative of the global average of the target galaxy.  First, we find that some hosts have a $g$-band fiber fraction greater than one. Although objects are deblended before the modelMag is computed, this is not the case for the fiberMag; thus, we obtain fiber fractions $>1$.  After visual inspection of these cases, we conclude that these hosts have bright, nearby neighbors that contribute to the observed fiber magnitude.    Since these spectra include contamination from a galaxy other than the target, the derived properties cannot be assumed to be representative of the SN\,Ia host.  Second, all hosts with a $g$-band fiber fraction $< 0.2$ are removed from our sample. At these low fiber fractions too little of the galaxy is being measured to compute a global, rather than core, metallicity \citep{Kewley05}.  These two aperture cuts, as mentioned in Section~\ref{sub_sec:spec}, finalize our MZS sample at \nummzs\ galaxies (Table~\ref{mzssamplecuts}).

\begin{figure}[tp]
\centering
\includegraphics[scale=0.37,angle=90]{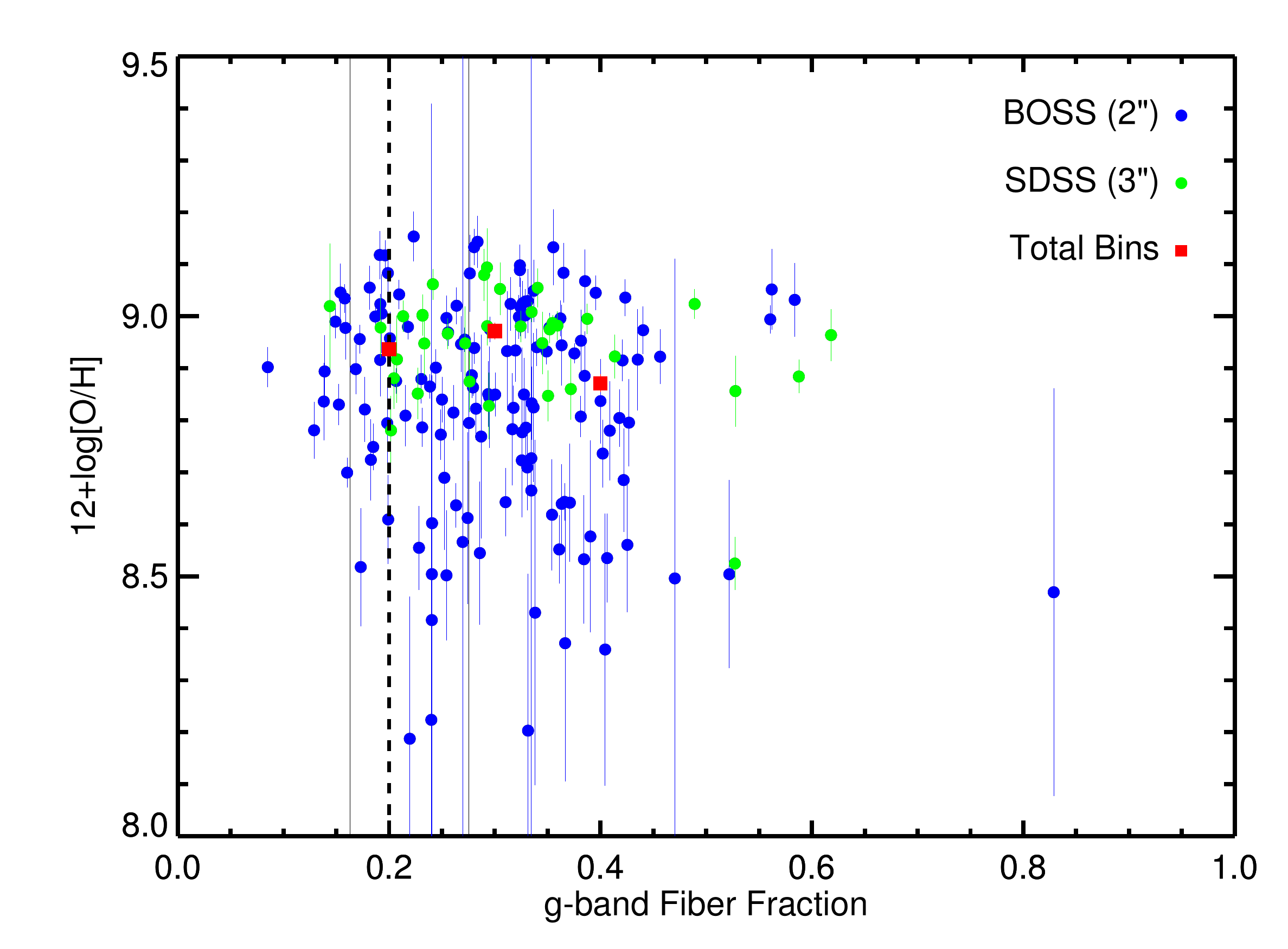}
\caption{Host metallicity as a function of $g$-band fiber fraction for hosts that satisfy BPT cuts.  The dashed line at $g$-band fiber fraction = 0.2 represents the threshold fiber fraction above which the derived gas-phase metallicity is considered indicative of the global average \citep{Kewley05}.  Inverse-variance weighted binned averages, of approximately equal-sized bins, are plotted in red.  There is a slight ($0.07 \unt{dex}$) decrease in metallicity with increasing fiber fraction.}
\label{fig:fiberfrac}
\end{figure}

Figure~\ref{fig:fiberfrac} shows the derived host gas-phase metallicities as a function of $g$-band fiber fraction, with the dashed line indicating the lower-limit for inclusion in the MZS sample.  We compute inverse-variance weighted averages over three bins of $g$-band fiber fraction (such that the bins are approximately equally sized), and find little correlation between $g$-band fiber fraction and gas-phase metallicity.  This indicates that our use of different physical scales does not have a significant effect on our metallicity, and thus we make no aperture-based corrections.  

We also use the $u$-band fiber fraction to adjust our estimate of the SFR based on the measured \Ha\ line flux \citep{Gilbank}.  Because our emission-line flux measurements are affected by the fixed aperture size, the \Ha\ flux we measure is not a global representation of the entire galaxy.  Therefore, to obtain a more reasonable estimate of the total SFR for the host, the \Ha\ flux measurement is corrected by dividing by the $u$-band fiber fraction as in \citet{Gilbank} (see Appendix A).  

Another important aperture effect to consider is that our analysis uses both SDSS and BOSS spectra, with 3\arcsec\ and 2\arcsec\ fiber diameters respectively.  For 19 of our SNe\,Ia, the hosts were targeted by both SDSS and BOSS; we use spectra from these observations to compare the derived metallicities.  We find the difference between the metallicity measurements to be within $0.1\unt{dex}$ (equivalent to systematic uncertainties) for 83\% of hosts, approximately Gaussian, and centered at zero.  This indicates that our sample suffers no metallicity bias due to aperture effects.

\begin{figure}[tp]
\centering
\includegraphics[scale=0.39,angle=90]{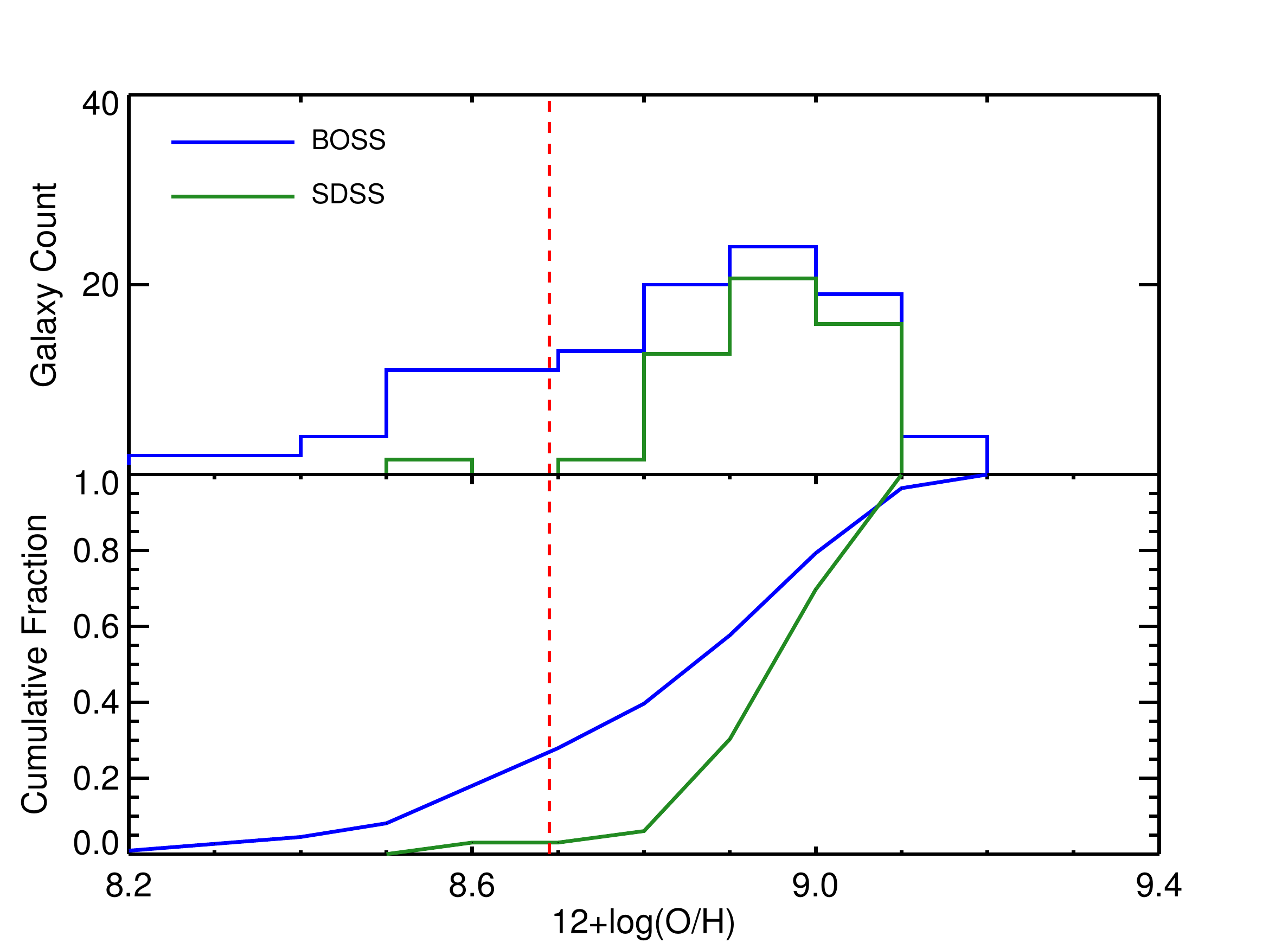}
\caption{The distribution of host gas-phase metallicities for SDSS (green) and BOSS (blue) galaxies in our MZS sample, with total number counts shown in the top panel and the corresponding cumulative distribution function in the bottom panel.  To focus on the bulk of our sample, we leave out one host with $Z<8.2$ from this figure.  The vertical dashed line at 12+log(O/H) = 8.69 represents the solar metallicity value, shown for comparison.  The SDSS spectra are systematically higher metallicity than the BOSS spectra due to how targets were selected for the two samples.  }
\label{fig:boss_sdss_met_dist}
\end{figure}

The majority of the host-galaxy spectra we use were obtained from BOSS rather than from SDSS-I/II.  Priority for BOSS targets was given to galaxies with a 3\arcsec\ $r$-band fiber magnitude $<21.25$, though some galaxies 
fainter than this limit were observed \citep{Olmstead14}.  By contrast, SDSS-I/II spectra were obtained from the SDSS Legacy 
Survey and other targeted surveys within SDSS, many of which had much brighter limiting magnitudes.  As a result, 
the SDSS spectra tend to have higher signal-to-noise ratio and their corresponding galaxies are at lower redshift.  In addition, since they are the brightest galaxies at a given redshift, they are generally more massive and more metal-rich.  
This effect is displayed Figure~\ref{fig:boss_sdss_met_dist}.  The BOSS spectra peak at slightly lower metallicity 
compared to the SDSS spectra while also extending much farther into the low-metallicity regime.  The median metallicity for the BOSS spectra is $Z=8.85$, while the median metallicity for the SDSS spectra is $Z=8.97$.  It is important to remember that this offset is an effect of target selection, not a bias due to the fiber aperture size, as we have demonstrated from hosts present in both spectroscopic samples.

Where spectra exist for both BOSS and SDSS galaxies, we choose to use the SDSS spectrum for our analyses in Section~\ref{sec:results}.  In addition to being higher $S/N$ spectra on average, all SDSS spectra targeted the core of the galaxy, while some spectra from the BOSS ancillary program targeted the location of the SN itself \citep{Olmstead14}.  In all cases where only BOSS spectra exist for a galaxy, the fiber was centered on the galaxy core.  Together with the cuts in this section and examination of potential sources for aperture bias, this selection creates a consistent, high-quality set of data for our analyses.
 
\section{RESULTS}
\label{sec:results}

In Table~\ref{alldata}, we present our derived SN\,Ia and host-galaxy properties for all data used in this analysis.  All \numpm\ of these SNe\,Ia have passed SN light-curve quality cuts, have an identified host-galaxy spectrum, and have a photometrically-derived host mass (the PM sample; Table~\ref{pmsamplecuts}).  For a subset of \nummzs\ of these SNe\,Ia, the MZS sample, we have spectroscopically measured global host-galaxy metallicities and star-formation rates. Table~\ref{mzssamplecuts} summarizes the requirements placed on this sample.  The full version of Table~\ref{alldata} is available in the electronic version of this work.  

The derived host-property uncertainties quoted in Table~\ref{alldata} do not include any systematic uncertainties previously discussed ($0.1\unt{dex}, 0.2 \unt{dex},$ and $ 0.1\unt{dex}$ for metallicity, SFR, and stellar mass, respectively).  Similarly, error bars in subsequent plots (e.g. Figures~\ref{fig:snprophostpropz} and \ref{fig:hrmasspmmzs}) reflect only statistical uncertainties  for clarity.  However, when fitting for linear trends, systematic uncertainties are added in quadrature to the quoted statistical uncertainties.  As S14 reports asymmetric mass uncertainties, we choose the larger value as the single, conservative estimate.  

In the following analysis, we discuss our derived host properties and SN\,Ia properties, as well as explore correlations between them. We use the IDL LINMIX routine, which employs the linear regression model presented in \citet{Kelly07}, to assess the strength of observed correlations:
\begin{equation}
\label{eqn:linmix}
\hat{y} = m\hat{x}+ b + \epsilon
\end{equation}
Here, $m$ is the fit slope, $b$ is the fit intercept, and $\epsilon$ is the scatter about the best-fit regression line. As described in \citet{Kelly07}, we assume $\epsilon$ is drawn from a normal distribution with mean zero and variance $\sigma^2$. Throughout this work we report the intrinsic dispersion ($\sigma$) and its uncertainty, computed by taking the square root of the posterior distribution of the best-fit variance. We define the significance of a non-zero slope as $m/\sigma_m$, where $m$ is best-fit slope and $\sigma_{m}$ is the error on the slope. LINMIX allows for uncertainties in the dependent and independent variables (assuming Gaussianity) and employs a Bayesian approach using Markov Chain Monte Carlo (MCMC).  Posterior distributions for at least 10,000 iterations of the MCMC are used to determine the regression coefficients and their errors. For completeness, we report the median and standard deviation of the posterior distributions of the best-fit slope, intercept, and dispersion in our results tables. This method of linear fitting was chosen over other linear regression techniques (such as least-squares) as we find that the LINMIX fits provide more realistic estimates for our fit parameter errors. 

We also use the Spearman rank correlation coefficient and corresponding significance test to study the relationship between SN\,Ia and host-galaxy properties. This is a nonparametric measure of statistical dependence that requires that the relationship between the two variables of interest is monotonic, but not necessarily linear. The value of the coefficient, $\rho$, ranges from $-1$ to $+1$ with $|\rho|=1$ indicating a perfectly monotone relation. The null hypothesis for this test states that there is no correlation between the dependent and independent variable; the associated $p$-value describes the chance that random sampling of the data would have generated the observed correlation.  While this technique provides important insight into our SN Ia-host-galaxy correlations, we must be cautious as it does not account for large differences in the measurement errors of different data points when computing the correlation coefficient.

A general outline is as follows: Section~\ref{sub_sec:hostpropsum} describes our derived host-galaxy properties.  Section~\ref{sub_sec:lightcurvesum} discusses the stretch and color of our SNe\,Ia and correlations between these parameters and host-galaxy properties.  Section~\ref{sub_sec:hrhostprop} examines the individual relations between HR and host-galaxy mass, gas-phase metallicity, and specific star-formation rate, separately.  In Section~\ref{sub_sec:hrhostpropbins} we explore the interplay between these host properties and how they affect trends with HR when fit simultaneously. 

\subsection{Host-Galaxy Properties}
\label{sub_sec:hostpropsum}

\begin{figure}[tp]
\centering
\includegraphics[scale=0.38,angle=90]{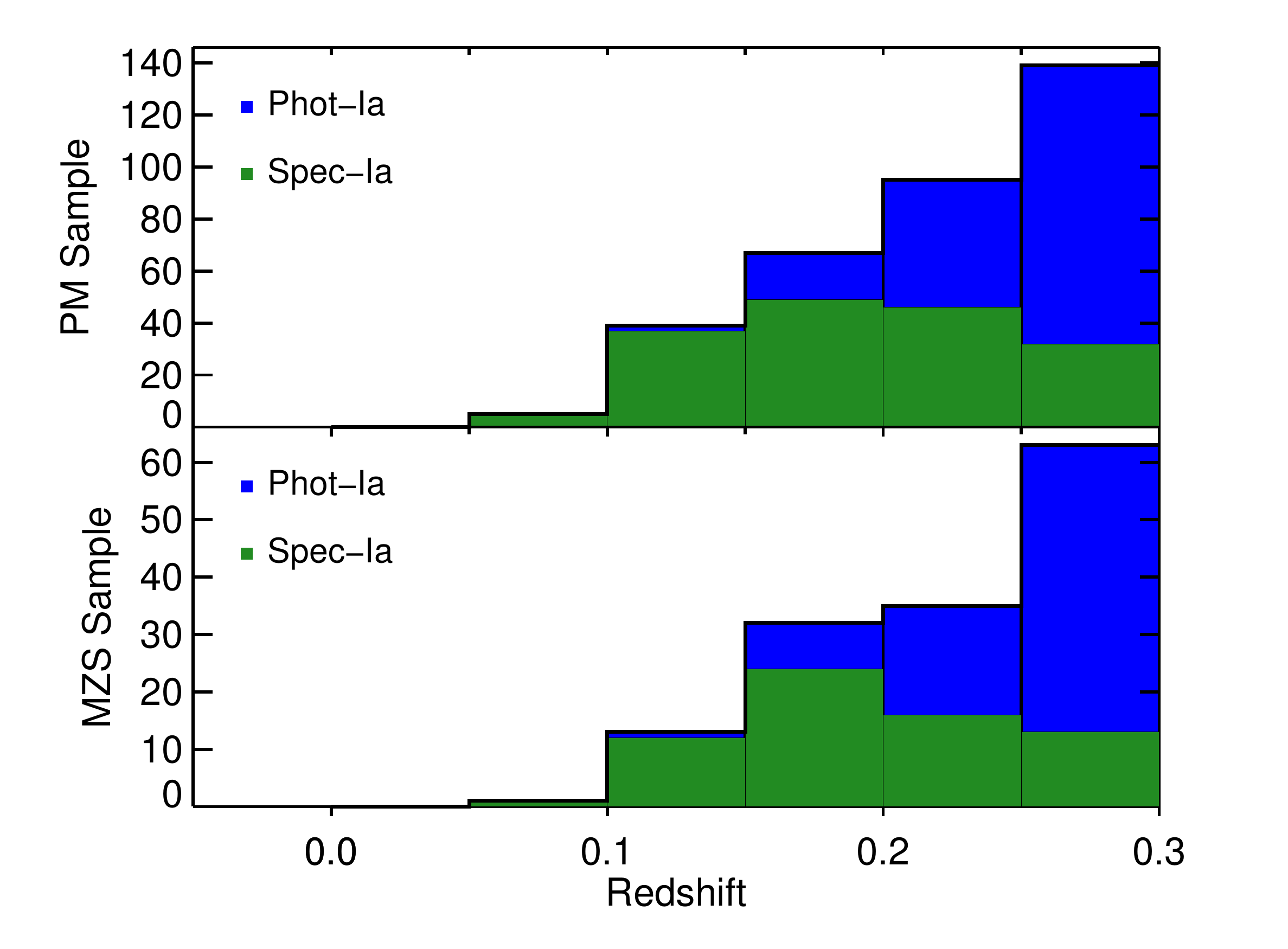}
\caption{Redshift distributions of the PM and MZS samples.  Histograms are stacked such that the number of Spec-Ia (green) and Phot-Ia (blue) shown in each bin add to the total number of SNe\,Ia in that bin. The mean and median redshifts of the PM and MZS samples are each $z=0.24$. For both samples, the median redshift of the Spec-Ia is $0.19$ and the median redshift of the Phot-Ia is $0.26$.}
\label{fig:redshiftdist}
\end{figure}

The redshift distributions of the PM and MZS hosts are shown in Figure~\ref{fig:redshiftdist}. The mean and median redshifts for both the PM and MZS samples is $z=0.24$, and the shapes of the redshift distributions are consistent.  The median redshifts of the Spec-Ia and Phot-Ia in sub-samples are $0.19$ and $0.26$, respectively, in both the PM and MZS.  We thus conclude that the requirements we impose on our host-galaxy spectroscopic data when creating the MZS sample does not result in any redshift bias relative to the PM sample. 

\begin{figure*}[tp]
\centering
\includegraphics[scale=0.6,angle=90]{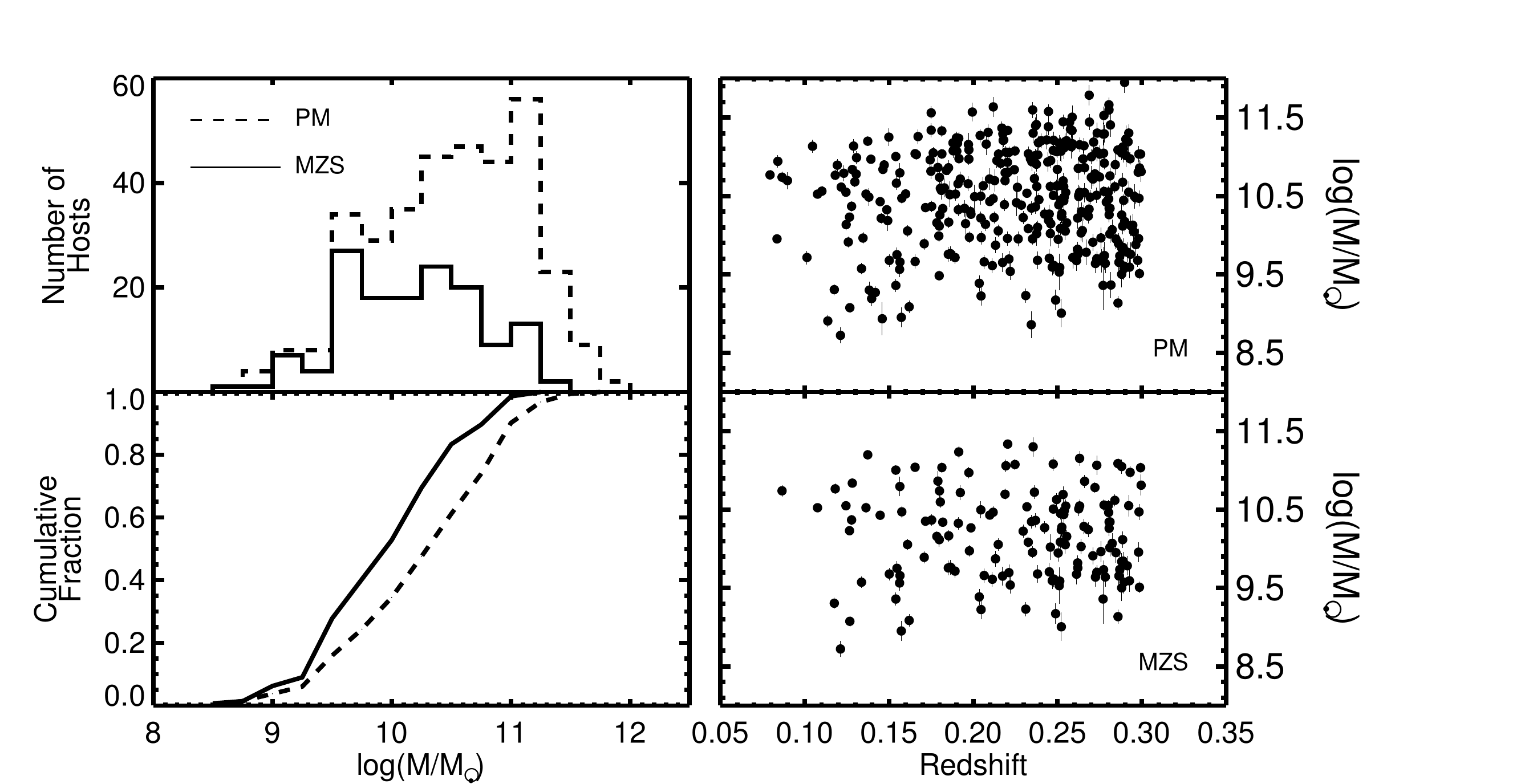}
\caption{Mass distributions of our PM (dashed) and MZS (solid) galaxies are displayed in the top left panel. The mean of the PM and MZS mass distributions (in log(M/\Msun) are $10.5$ and $10.2$, respectively. The bottom left panel presents the cumulative fraction of hosts as a function of mass.  The right panels show our galaxy masses as a function of redshift.}
\label{fig:massdist}
\end{figure*}

We present in Figure~\ref{fig:massdist} the host-galaxy stellar mass distribution for both our PM and MZS samples, both as a whole and as a function of redshift.  While the MZS host-galaxy sample only contains star-forming galaxies through the requirement of measurable emission lines, the PM sample consists of both star-forming and elliptical galaxies.  The inclusion of elliptical galaxies, which have a higher mass on average, results in the PM sample spanning a slightly larger range in masses with a higher mean mass (log(M/\Msun)$=10.5$) than the MZS sample (log(M/\Msun)$=10.2$). We also see in the right panels of Figure~\ref{fig:massdist} that there is no noticeable trend of host mass with redshift for our sample over this redshift range, indicating that our sample has no strong differential bias with redshift.

\begin{figure}[tp]
\centering
\begin{tabular}{c}
\includegraphics[scale=0.37,angle=90]{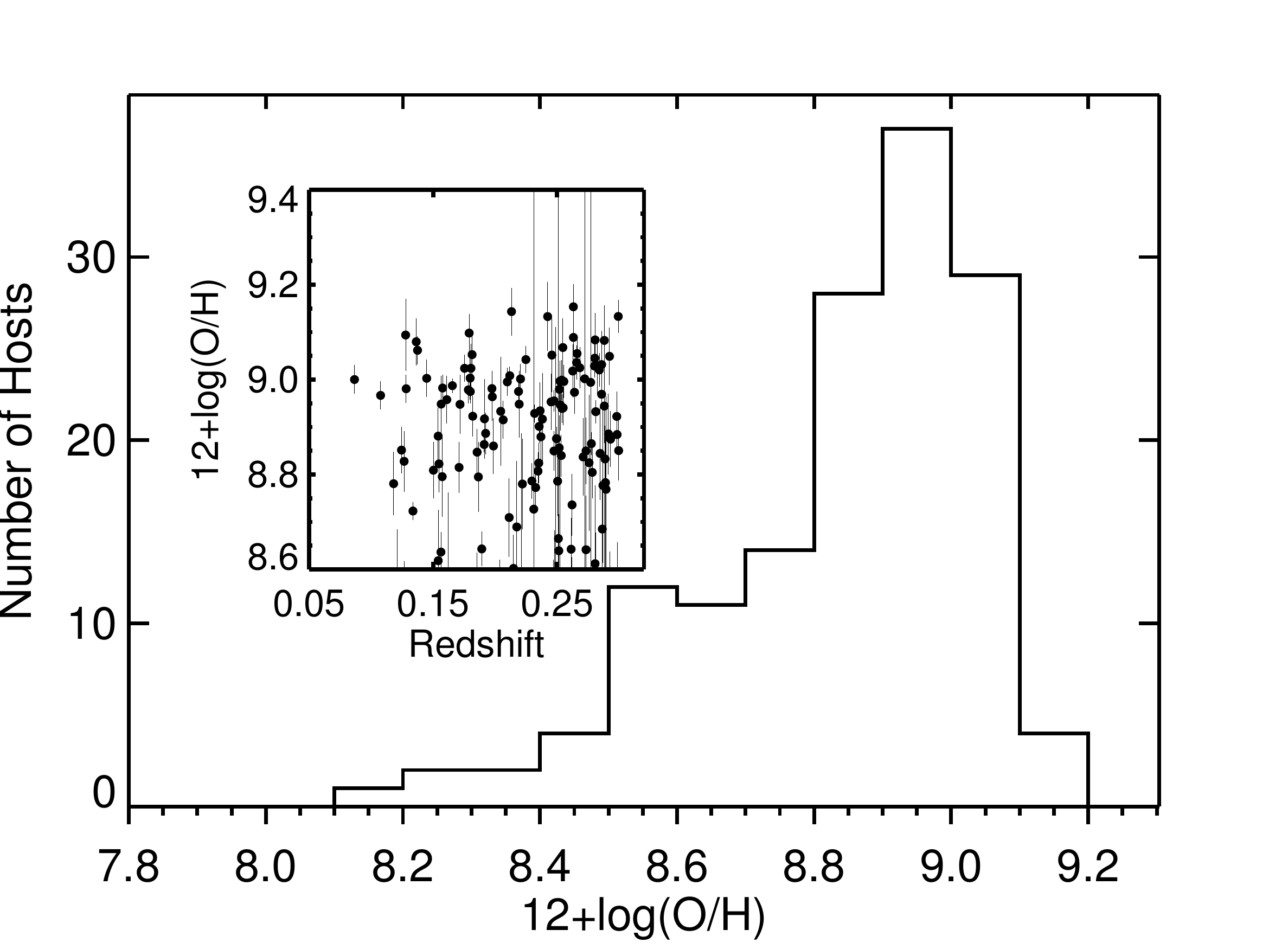} \\
\includegraphics[scale=0.37,angle=90]{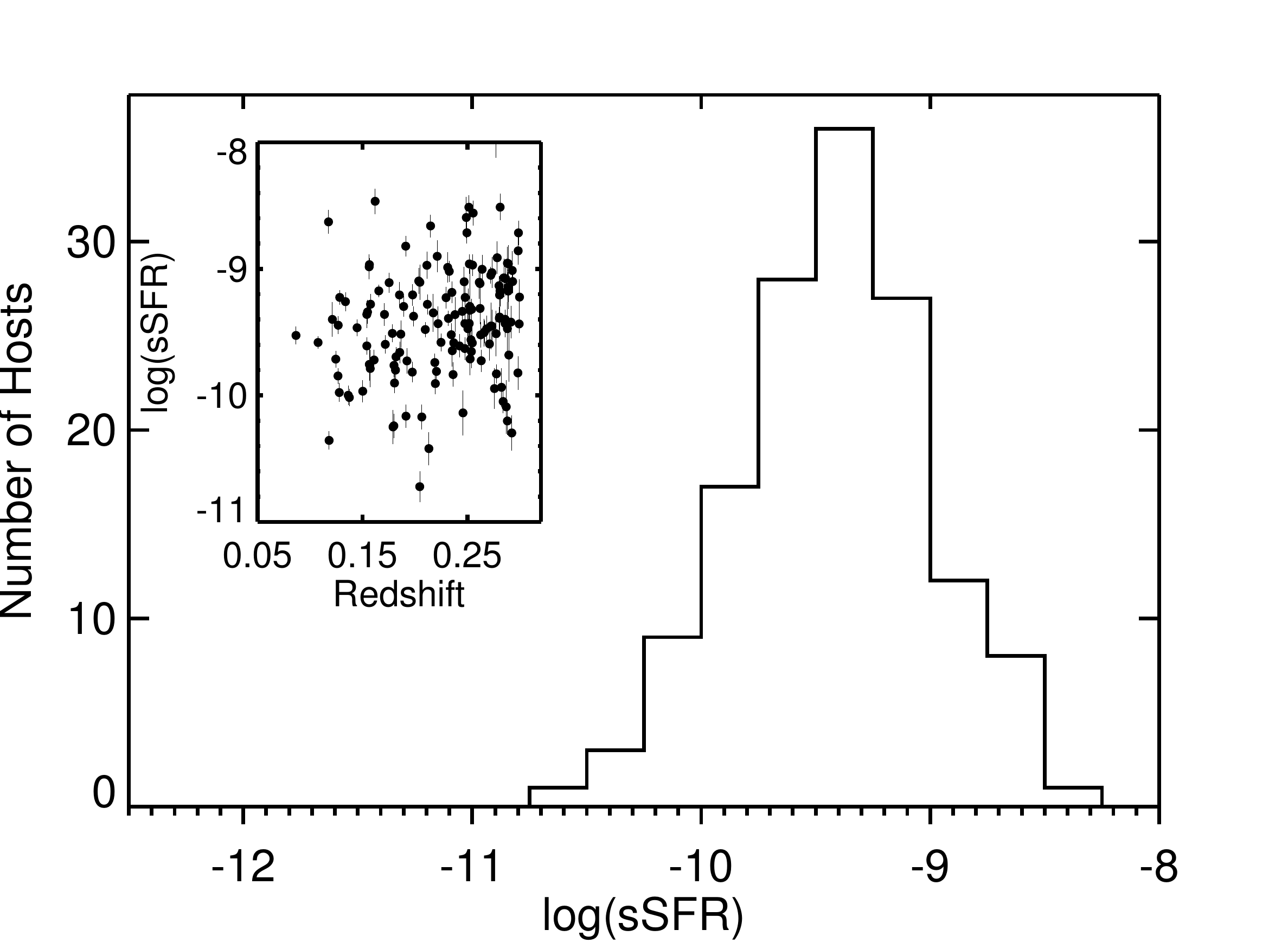}
\end{tabular}
\caption{The left panel shows the metallicity distribution of galaxies in our MZS sample.  The mean of the metallicity distribution is $Z=8.84$.  The right panel shows the sSFR distribution of galaxies in our MZS sample.  The mean of the sSFR distribution is log(sSFR)$=-9.43$.  The inset figures of both panels display the respective host properties as a function of redshift.  Axes of the inset figures have been adjusted to focus on the metallicity and sSFR redshift dependence; as such, some data points are excluded from the plots. }
\label{fig:metssfrdist}
\end{figure}

In Figure~\ref{fig:metssfrdist} we show the distributions of metallicity and specific star-formation rates from our MZS sample.  The mean gas-phase metallicity for our sample is $Z=8.84$, and the mean sSFR is log(sSFR)$=-9.43$.  While the sSFR distribution is roughly Gaussian, the  metallicity distribution is negatively skewed, although there are few galaxies with sub-solar metallicities even in the long low-metallicity tail.  As shown in the inset panels in Figure~\ref{fig:metssfrdist}, we see no evolution of metallicity or sSFR with redshift.

As we use different IMFs, methods, selection criteria, and calibration techniques, we cannot directly compare our results to previous studies. 
However, we can qualitatively assess how our host-property distributions compare to those of other surveys.  The peak host-galaxy mass in the PM sample is consistent with that in the PTF (P14), SNFactory (C13), SNLS \citep{Sullivan10}, and Pan-STARRS1 \citep[PS1;][]{Scolnic14}.  

We notice our host-galaxy mass distribution contains relatively fewer galaxies with $\log(M/M_\odot) \la 9.0$. We attribute this primarily to the BOSS targeting criteria and the use of the SDSS DR8 catalog for host identification. Given that our Phot-Ia sample depends on redshifts from BOSS, which only targeted hosts brighter than a certain magnitude, we expect this sample to be biased against SNe in low-luminosity (low-mass) hosts. We also lose low-mass hosts due to the $r$-band magnitude limit of 22.2 for SDSS DR8, which is the catalog used to select host galaxies in S14.{\footnote{Though a deep co-added image catalog exists for SDSS Stripe 82 \citep{SDSS-Coadd}, these images contain SN light for SNe occurring in 2005. Ideally, SN surveys in the future should create custom co-added images excluding images with SNe and use these for host identification and host-galaxy studies.} In addition, our choice of mass-fitting technique may also contribute to the dearth of low-mass hosts. We use FSPS masses in this work which are shown in Figure 23 of S14 to be $\approx 0.3$ dex higher than the masses derived from ZPEG (a code commonly used by other works). Therefore, we note that our reduced host-mass range may affect our derived trends with HR (Section~\ref{sub_sec:hrhostprop}).

In the MZS sample, the derived metallicities of P14 for PTF host galaxies are biased substantially lower than our metallicities, but as the typical offset between the calibration used by us and in that work is 0.2-0.3 \unt{dex}, the range of measured values are consistent. C13 uses a calibration that typically returns a wider range of metallicities, and this is seen in their results compared to this work.  However, although C13 also finds the peak of their distribution at 12+log(O/H)$\approx9.0$, they have a greater fraction of their host-galaxies at sub-solar than can be explained through calibration techniques alone. In addition, we find that the sSFR distribution of the MZS sample also exhibits a lack of low-sSFR hosts when compared to other studies. One reason for this difference is that some studies \citep{Sullivan10, Childress13} with hosts with lower star-formation rates rely on host photometry, rather than spectroscopy, to obtain SFR measurements and are thus not limited by spectral quality requirements.

The differences in these property distributions likely stem from our spectral quality requirements. We impose a cut on the $A/N$ of the \Ha\ and \Hb\ lines to ensure good spectral quality, but by doing so reject those spectra with lower emission-line flux measurements. If we remove this $A/N$ criterion, an additional 41 hosts would be included in the MZS sample. Of these 41, 26.8\% have sub-solar metallicity.  Additionally, we find that 58.5\% of the 41 additional hosts have low sSFR ($\mathrm{log(sSFR)} < -10$). Adding these hosts into our sample would not significantly impact the fraction of low-metallicity hosts, but would raise the fraction of low-sSFR hosts from 9.7\% to 20.5\%. However, we believe the quality of these spectra is not sufficient to produce reliable host-property estimates, and so we do not include these in our sample.

\subsection{SN\,Ia Light-Curve Properties}
\label{sub_sec:lightcurvesum}
 
SN\,Ia light-curve parameters such as color ($c$) and stretch ($x_{1}$) -- the essential calibration tools for using SNe\,Ia as distance indicators -- have long been known to correlate with host environment \citep{Hamuy96,Gallagher05}.  Figure~\ref{fig:snprophostpropz} shows the SN\,Ia stretch and color as a function of our derived host-galaxy properties.  We observe the correlations seen by \citet{Howell09} and \citet{Sullivan10}: more massive galaxies host fainter, redder SNe\,Ia.  We also find that SNe\,Ia with higher $c$ occur in galaxies with lower specific star-formation rates.  Since the SN\,Ia color parameter contains information not just on the intrinsic color of the SN but also effects of host-galaxy dust extinction, it is expected that both massive galaxies and those with low specific-star formation should host redder SNe\,Ia.  It is interesting to note that we find low-metallicity galaxies tend to host only blue SNe\,Ia, to an extent not seen in low-mass or high-sSFR galaxies (properties that are correlated with low metallicity).  This metallicity-color relation is consistent with what is found in C13 and P14.

To quantify the strengths of these correlations, we perform a Spearman rank test on each combination of SN\,Ia and host property displayed in Figure~\ref{fig:snprophostpropz}. In each of the six cases, the correlation coefficient is non-zero; however, only the SN\,Ia stretch-host mass correlation exhibits enough evidence to reject the null hypothesis ($\rho = -0.308$, $p=5.305\times10^{-9}$). 

 \begin{figure*}[htb]
\centering
\includegraphics[scale=0.65,angle=90]{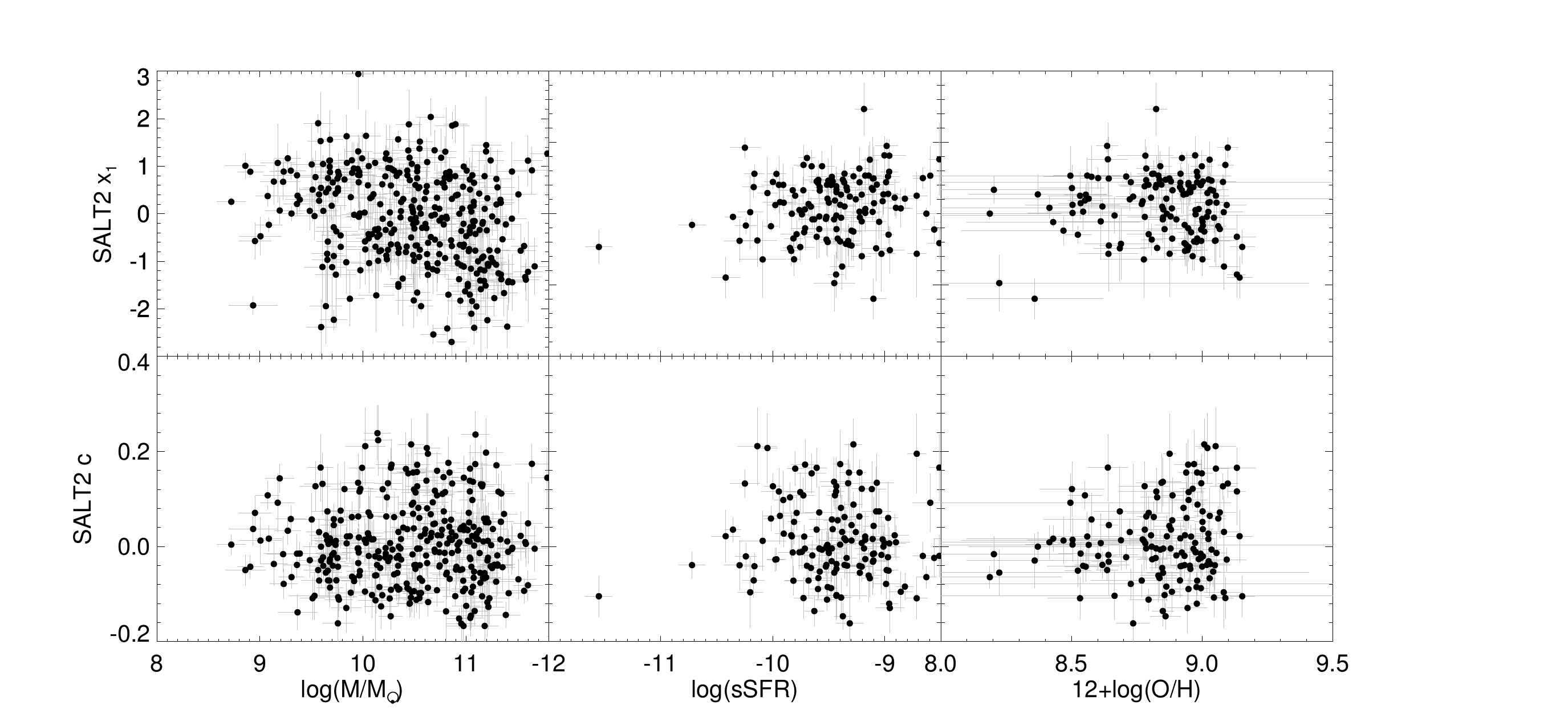}
\caption{SNe\,Ia color ($c$) and stretch ($x_{1}$) as a function of derived host properties. The left panel displays data from the PM sample; the middle and right panels show data from the MZS sample. Axes have been truncated to focus on the bulk of the data.}
\label{fig:snprophostpropz}
\end{figure*}

\subsection{HR as a Function of Host-Galaxy Properties}
\label{sub_sec:hrhostprop}

We now examine whether the stretch- and color-corrected luminosities of SNe\,Ia (and thus HRs) show correlations with properties of their host galaxies.  Linear fits to the data using the LINMIX routine are shown on the figures included in this section and the corresponding results are reported in Table~\ref{hostproptrends}.  Spearman rank correlation statistics for each linear fit are also presented in Table~\ref{hostproptrends}. We note that the posterior distributions of the model parameters of these fits are roughly Gaussian.  To determine the model parameters of these fits, we choose the point estimator to be the median of the posterior distribution, limiting the effects of outliers in the distribution.  Errors on the fit parameters are obtained using the standard deviation of the respective posterior distribution.  Host-galaxy properties are also split to create low and high mass (metallicity, sSFR) bins which are then used to compute the difference between the HR in these bins (``HR step"). The split point of each property is chosen to be the median of its respective distribution, thus creating two bins of equal number. We define the ``HR step" as the difference between the high and low-binned inverse-variance weighted averages. When computing the significance of the step (the mean and uncertainties on the mean), we fit for the unknown intrinsic scatter that ensures $\chi^2/\mathrm{DoF} \approx  1$ after the step is removed.  These bins are also included in relevant figures in this section.  We note that when we refer to the over or under-luminosity of SNe\,Ia in this section, this refers to the luminosity after light-curve corrections have been applied.  

\begin{deluxetable*}{l l c r r r r r r r r r}
\tablecaption{LINMIX Linear Fit Results for HR as a Function of Derived Host-Galaxy Properties \label{hostproptrends}}
\tabletypesize{\tiny}
\tablewidth{0pt}
\tablehead{\colhead{Host} & \colhead{Sample} & \colhead{$N$\tablenotemark{a}} & \colhead{Split\tablenotemark{b}} & \colhead{HR Step} & \colhead{Slope} & \colhead{Intercept} & \colhead{$\sigma$} & \colhead{Sig\tablenotemark{c}} & \colhead{$\rho$} & \colhead{$p$-value} \\ \colhead{Property} & & & \colhead{Value} & \colhead{[mag]} & & &\colhead{[mag]} \\}
\startdata
Mass & PM & \numpm & 10.5& $0.048\pm0.019$& $-0.055\pm0.015$ &$0.570\pm0.160$& $0.121\pm0.009$&$3.62\sigma$ & $-0.1708$ & $0.0015$ \\
Mass & MZS & \nummzs & 10.2 & $0.082\pm0.030$ & $-0.071\pm0.029$&$0.728\pm0.293$ & $0.136\pm0.014$&$2.46\sigma$ & $-0.2094$& $0.0118$ \\
12+log(O/H) & MZS & \nummzs &  8.9 & $0.057\pm0.031$&$-0.579\pm0.409$&$5.162\pm3.641$& $0.125\pm0.021$ &$1.42\sigma$ & $-0.1811$& $0.0299$ \\
sSFR & MZS & \nummzs & $-9.4$ & $0.013\pm0.031$ & $0.019\pm0.046$&$0.190\pm0.437$&$0.140\pm0.014$ & $0.42\sigma$& $0.0965$ & $0.2500$ \\ 
\enddata
\tiny
\tablenotetext{a}{Sample size.}
\tablenotetext{b}{Value used to create high and low-mass (metallicity, sSFR) bins of equal number. The median of the respective host-property distribution. }
\tablenotetext{c}{Significance of a non-zero slope.}

\end{deluxetable*}

Figure~\ref{fig:hrmasspmmzs} shows HR as a function of mass for the PM sample.  Using LINMIX, a non-zero slope of the linear fit is detected at $3.6\sigma$. We also take the difference between the inverse-variance weighted averages of the high and low-mass bins and measure the ``HR step" to be $-0.048 \pm 0.019$ magnitudes.  A similar trend is present in the MZS sample; the best-fit slope and the HR step are both shown to be consistent within $1\sigma$ that of the full PM sample. Our results show that more massive galaxies host over-luminous SNe\,Ia, supporting previous findings \citep{Lampeitl10,Sullivan10,Gupta11,Childress13,B14}.

The results of the Spearman rank correlation test for both the PM and MZS samples further support the significance of the HR-mass relation. In both cases, we find $\rho\approx -0.2$, which indicates that more massive galaxies host overluminous SN Ia. For both samples, there is a less than $2\%$ chance that this correlation is due to chance, and thus we again conclude that this correlation is significant.
 
\begin{figure}[tp]
\centering
\includegraphics[scale=0.37,angle=90]{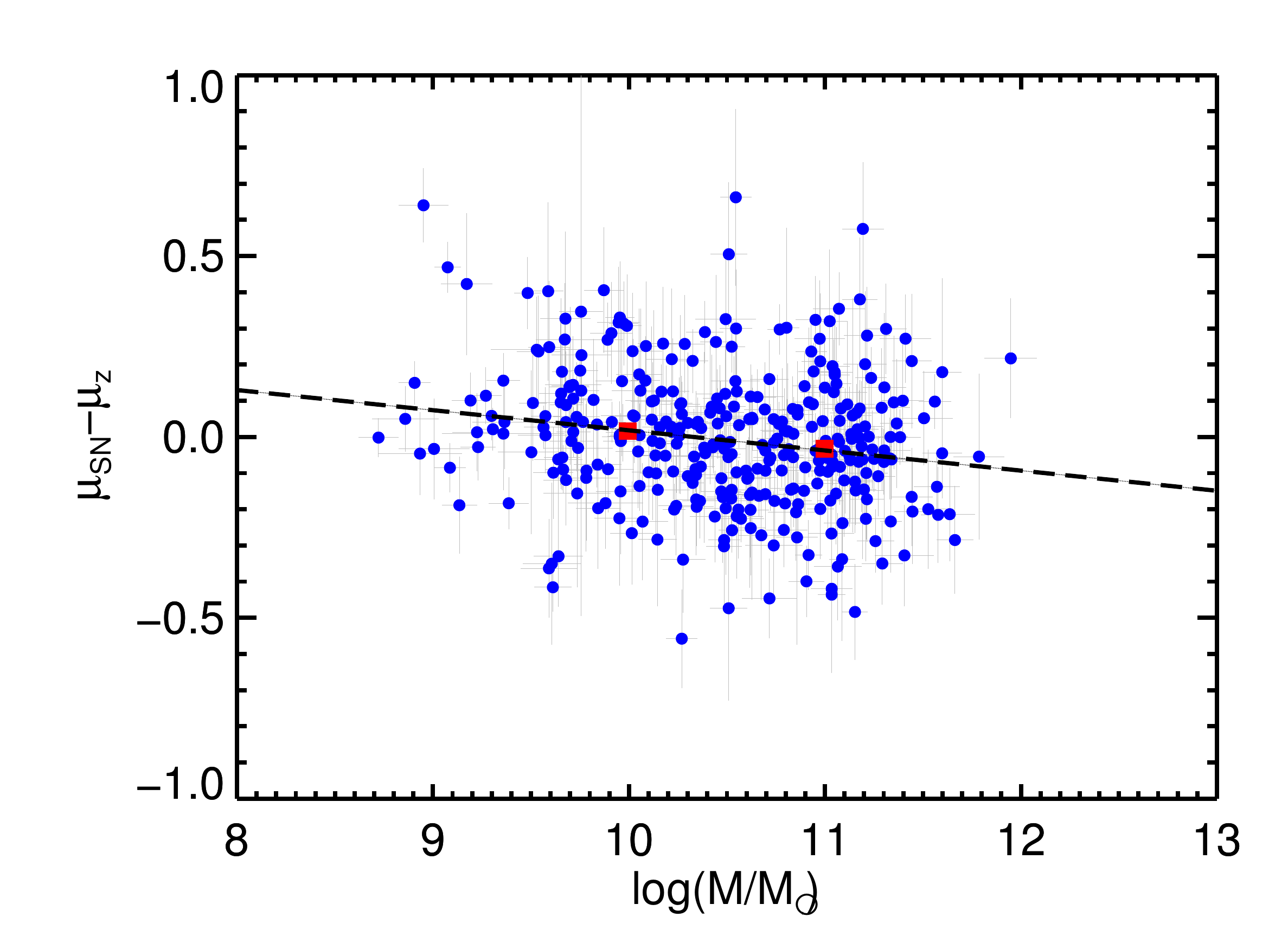}
\caption{HR as a function of host-galaxy mass for the PM sample.  The LINMIX linear fits to the data are shown in dashed black; red squares represent inverse-variance weighted binned averages, with bins split at log($M$/\Msun)$=10.5$.  The significance of a non-zero slope is $3.6\sigma$ and the difference in HR between the high and low-mass bins is 0.048 magnitudes. This result indicates that more massive galaxies host over-luminous SNe\,Ia.} 
\label{fig:hrmasspmmzs}
\end{figure} 

Several recent studies suggest that HR as a function of host-galaxy mass resembles a smoothly-varying step function rather than a line.  To explore this idea of a ``mass step", we fit an empirical continuous step function to our data in the PM sample.  We choose a function of the form 
\begin{equation}
\label{eq:HR}
\mathrm{HR}=A\left(\frac{2}{1 + e^{-B(x-C)}} - 1\right)
\end{equation}
where the parameter $A$ controls the amplitude, $B$ controls the steepness of the step, and $C$ indicates the step position.  The independent variable, $x$, is the host mass, $\log(M/\Msun)$.  We use the IDL routine MPFITFUN \citep{MPFIT} to perform a least-squares fit, using input parameters motivated by results in previous works and find that the resulting best-fit to the data is highly sensitive to the choice of input parameters.  We also compute the best-fit to inverse-variance weighted average bins of varying bin width and minimum number of SNe\,Ia per bin and find that choice of bin width and number of SNe\,Ia per bin significantly affects the best-fit results. 

Therefore, we choose to explore the shape of the HR-mass relation using nonparametric regression. We employ the {\tt loess} routine in the R statistical software package, based on {\tt cloess} of \citet{Cleveland92}. This method of locally-weighted smoothing combines linear regression in a $k$-nearest-neighbor based model and relies on a user-input bandwidth, also known as the span ($\alpha$), to determine the proportion of the data to be used in each local regression (i.e., a fit at some point $x$ is computed using its neighbors and contributions from neighboring points are weighted based on their distance from $x$). While this method cannot produce an empirical model, it does illustrate the general shape of the data.

The results of the {\tt loess} regression are presented in Figure~\ref{fig:hrmassnpreg}, with the HR axis truncated to better focus on the fit. The best-fit to the data is shown in red with an approximate corresponding $1\sigma$ confidence interval. This method of local regression is sensitive to edge effects, but has no consequence on the resulting best-fit for the bulk of the data. Therefore, the behavior of the best-fit at the low and high-mass extremes must be interpreted with caution. After testing multiple spans, we determine a span that responds best to fluctuations in the data of $\alpha=0.6$. 

\begin{figure}[tp]
\centering
\includegraphics[scale=0.37,angle=90]{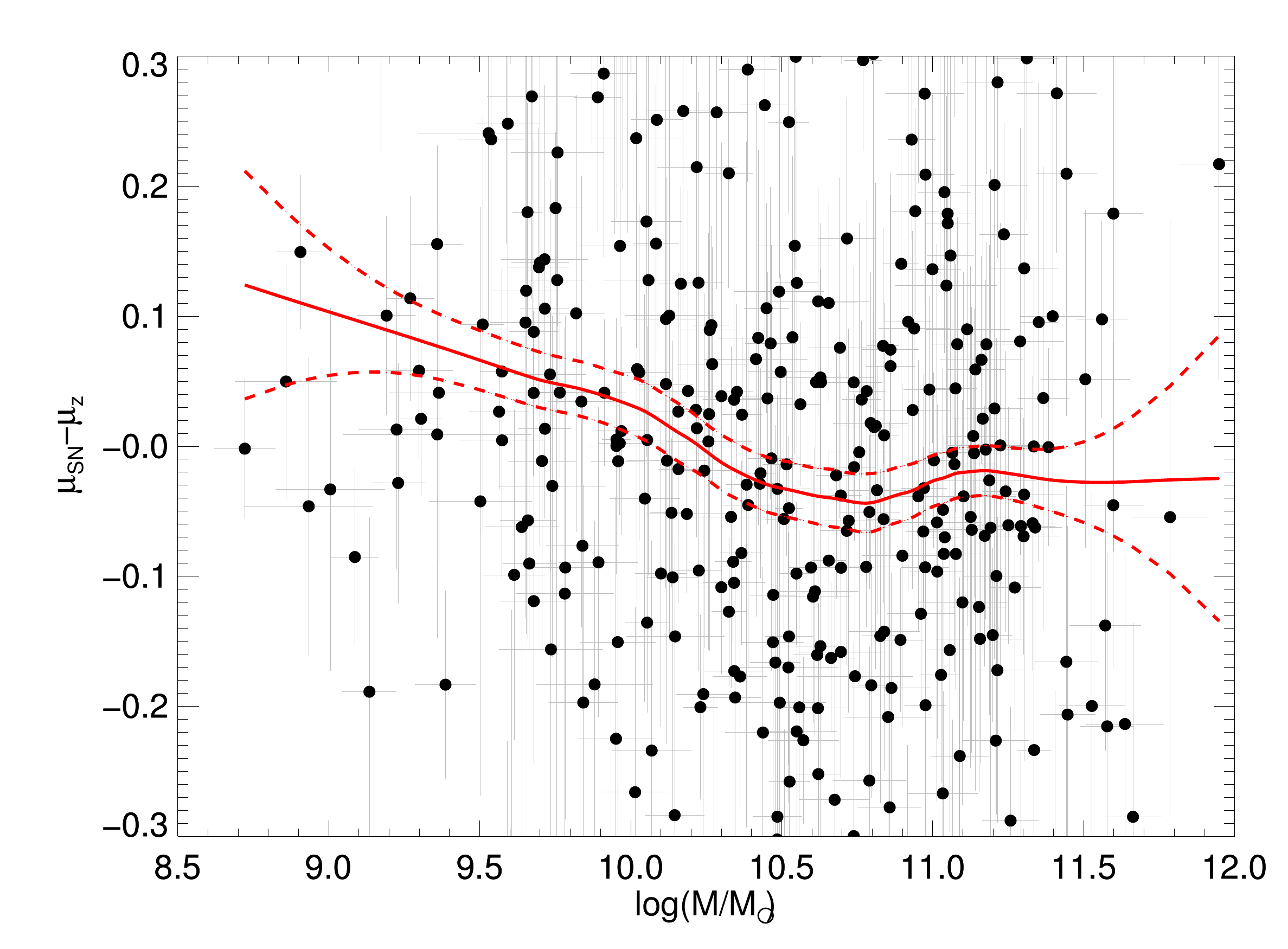}
\caption{Non-parametric regression fit of HR as a function of host-galaxy mass. The best-fit is presented in red with the approximate corresponding $1\sigma$ confidence interval.  A span of $\alpha=0.6$ was used for the fit. }
\label{fig:hrmassnpreg}
\end{figure} 

As shown in Figure~\ref{fig:hrmassnpreg}, there appears to be a relatively smooth transition region in the HR-mass relation between $10.0 \lesssim $ log($M/\Msun$) $ \lesssim 10.4$. However, because of the sensitivity of the fit at the edges, the shape of the ``step regions" is not well represented.  In addition, the computation of the best-fit did not include measurement error, which may affect the observed behavior. The shape of the HR-mass relation is similar to the behavior reported in C13 and J13; the slope of the transition region in J13, C13, and this work is roughly $-0.2$. We note that our results should be correlated with what is presented in J13 and C13 as their analyses utilize a subset of the SDSS SNe\,Ia. Despite the shortcomings of our chosen fitting technique, the nonparametric fit is an interesting interpretation of the HR-mass relation and a more rigorous treatment should be considered for future studies.

We next examine the correlation between HR and host-galaxy gas-phase metallicity; the results are shown in Figure~\ref{fig:hrmet}.  The best-fit linear relation has a negative slope with $1.4\sigma$ significance, suggesting that more metal-rich galaxies host more over-luminous SNe\,Ia.  Examining the difference between our low and high-metallicity bins reveals an ``HR Step" of $0.057$ magnitudes with $1.86\sigma$ significance. When analyzing this relation using the Spearman coefficient, we find a statistically significant correlation ($\rho=-0.1811$, $p=0.0299$) between HR and gas-phase metallicity. Although the LINMIX results do not recover a significant correlation, the other statistical analysis tools indicate that there is a significant difference between the low and high-metallicity populations. This suggests that the behavior of HR-metallicity relation may not be adequately represented by the LINMIX linear fit. 

\begin{figure}[tp]
\centering
\includegraphics[scale=0.37,angle=90]{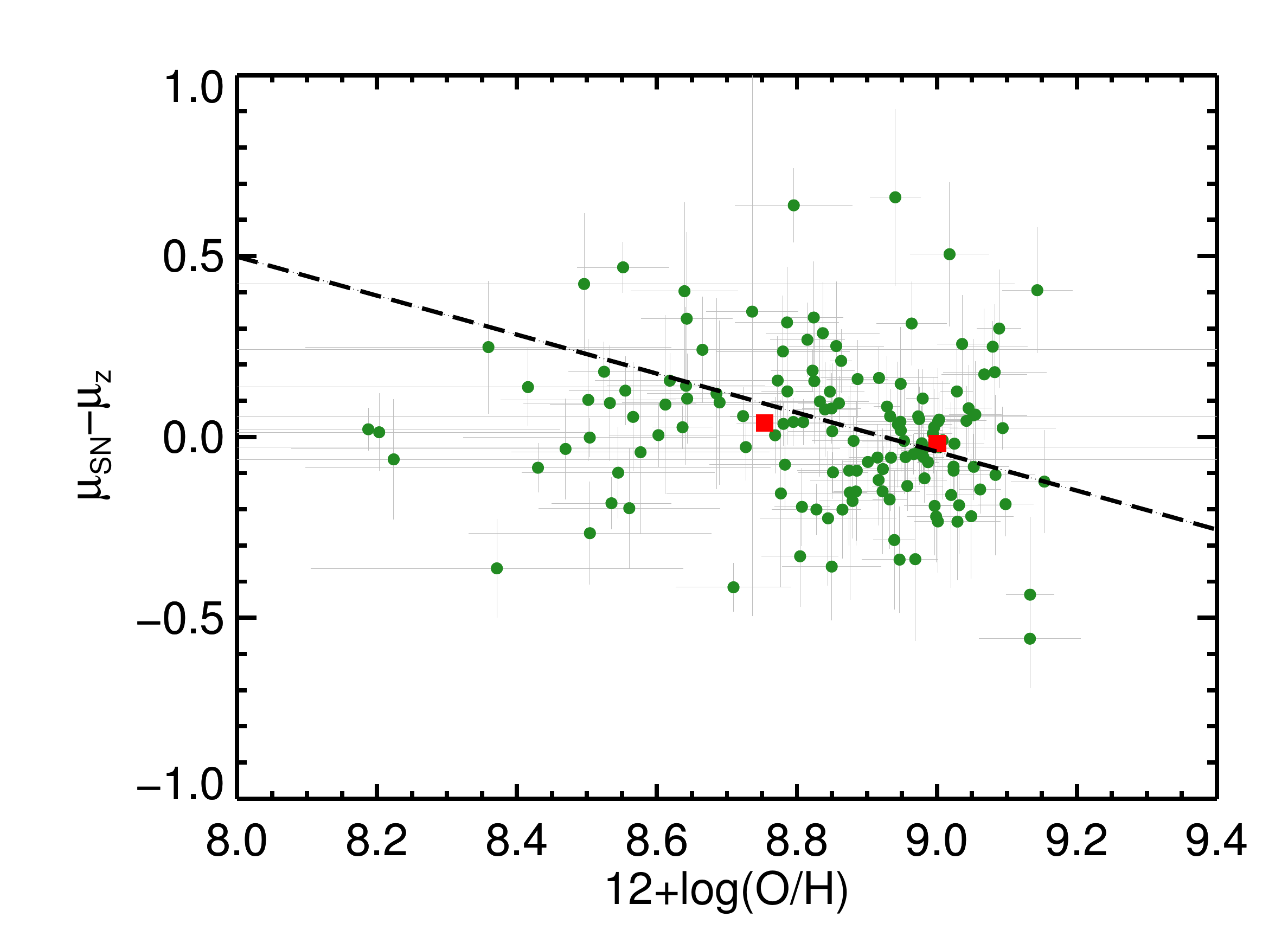}
\caption{HR as a function of gas-phase metallicity for the MZS sample.  These points are separated at 12+log(O/H) = 8.9 to create high and low-metallicity bins.  Red squares indicate the inverse-variance weighted average of these bins.  The difference between the binned averages is $0.057$ magnitudes.  The linear LINMIX fit to the data is shown in dashed-black; there is a $1.4\sigma$ significance of a non-zero slope which suggests that more metal-rich galaxies host over-luminous SNe\,Ia.}
\label{fig:hrmet}
\end{figure} 

Finally, we investigate HR as a function of sSFR.  These results are shown Figure~\ref{fig:hrssfr}.
The significance of this trend deviating from a non-zero slope as determined by LINMIX, however, is only 0.42$\sigma$.  In addition, the difference between the average HR in the high and low-sSFR bins is $0.013$ magnitudes with $0.42\sigma$ significance. The trend seen here is the weakest correlation observed between HR and host-galaxy properties. The results of the Spearman correlation test ($\rho=0.0965$, $p=0.25$) suggest that we do not have enough evidence to reject our null hypothesis; the HR-sSFR trend resembles a random sampling of uncorrelated variables.

\begin{figure}[tp]
\centering
\includegraphics[scale=0.37,angle=90]{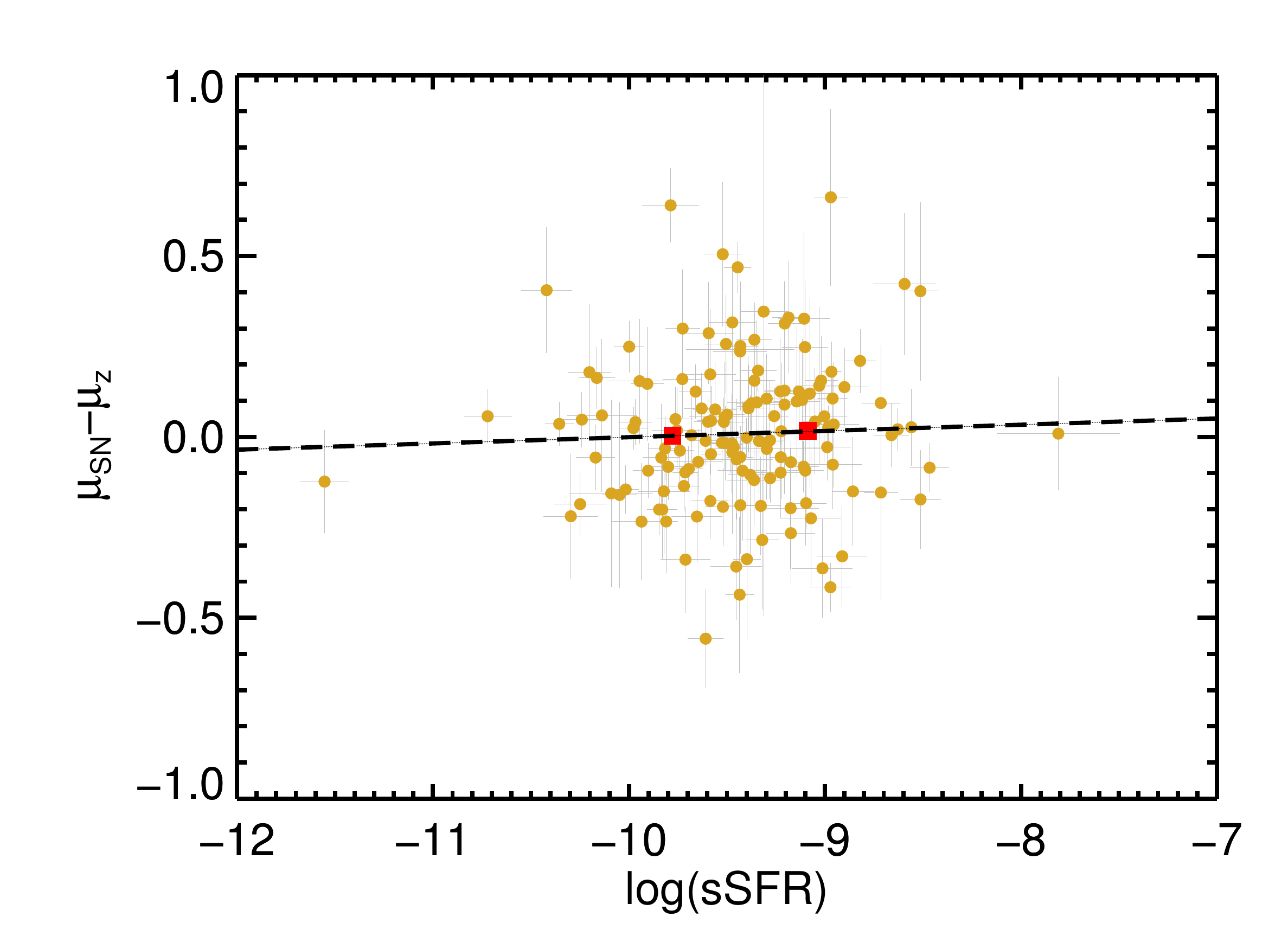}
\caption{HR as a function of specific star-formation rate for the MZS sample.  Points are separated at log(sSFR) = $-9.4$ to create high and low-sSFR bins.  Red squares indicate the inverse-variance weighted average of these bins.  The difference between the binned averages is 0.013 magnitudes.  The linear LINMIX fit to the data is shown in dashed-black; there is a $0.42\sigma$ significance of a non-zero slope.  This slight correlation suggests that galaxies with lower specific star formation rates host over-luminous SNe\,Ia. } 
\label{fig:hrssfr}
\end{figure} 

As discussed in Section~\ref{sub_sec:hostpropsum}, our cut on the \Ha\ and \Hb\ $A/N$, which is imposed to ensure spectral quality, removes 41 hosts from the MZS sample. If we add these hosts back into the MZS sample and recompute the slope of the HR-sSFR relation we find a slope of $0.021\pm 0.02$, which is within $0.1\sigma$ of the slope observed using the MZS sample. A Spearman rank test on this new sample also shows little evidence of an HR-sSFR correlation ($\rho = 0.1337$, $p=0.0695$). This indicates that the $A/N$ requirement and the lack of very low-sSFR hosts does not have a strong effect on our trend of HR with host sSFR.


\subsection{HR as a Function of Multiple Host-Galaxy Properties Simultaneously}
\label{sub_sec:hrhostpropbins} 
Our wealth of data allows not only an investigation of HR as a function of an individual host-galaxy property, but as a function of several host properties simultaneously. We perform linear fits of HR using combinations of two, and a combination of all three, derived host-galaxy parameters. For these linear fits, we include the relation with SFR as a opposed to sSFR as the sSFR and mass uncertainties are correlated. We then examine the HR-metallicity and HR-sSFR relation in several mass bins and also after correcting for the HR-mass relation.  Since mass appears to have the most dominant effect on HR, removing this dependence could provide important insight into the degeneracy of our host-galaxy properties. 

We first use the LINMIX package for multiple linear regression to determine the best-fit relation between HR and multiple host-galaxy parameters.  When using all three host properties, this function takes the form: 
\begin{align}\label{eq:3param}
& \mathrm{HR}=a\times\mathrm{log}(M/\Msun)+b\times\mathrm{(12+log(O/H))} \nonumber \\
& +c\times\mathrm{log(SFR)}+d + \sigma^2
\end{align}
where the coefficients $a$, $b$, $c$, $d$, and $\sigma^2$ are the parameters to be fit.  These same coefficients are fit using combinations of two host properties: i.e.,
\begin{equation} \label{eq:mz}
\mathrm{HR}=a\times\mathrm{log}(M/\Msun)+b\times\mathrm{(12+log(O/H))}+d +\sigma^2 
\end{equation}
\begin{equation} \label{eq:ms}
\mathrm{HR}=a\times\mathrm{log}(M/\Msun)+c\times\mathrm{log(SFR)}+d + \sigma^2
\end{equation}
\begin{equation} \label{eq:zs}
\mathrm{HR}=b\times\mathrm{(12+log(O/H))}+c\times\mathrm{log(SFR)}+d +\sigma^2
\end{equation}
We assume the errors on the host-parameters are uncorrelated. 

When fitting for Equation~\ref{eq:3param}, repeated trials (i.e, running LINMIX multiple times) do not yield the same fit results. For each fit parameter, results between trials are consistent within $1\sigma$ but can have dramatically different values (e.g., $a=-0.374\pm31.68$ compared to $a=0.53\pm2.67$).  We perform 20 trials of the same linear fit and find a substantial variance between fit-parameter outputs for each trial and strong skewness in the fit-parameter distributions.  Although repeated fit-parameter outputs are not identical, the results of each fit are consistent with no significant correlation between HR and all host-galaxy properties.  

In addition, we perform 20 trials of each of the fits using two host-galaxy properties (Equations~\ref{eq:mz}-\ref{eq:zs}).  Fit-parameter distributions with similar variance and skewness are observed using Equations~\ref{eq:mz} and~\ref {eq:zs}; these fits are also consistent with no correlation.  The output fit parameters using host-galaxy mass and SFR are nearly identical between the different trials, and the mass component is significant at $\approx 1\sigma$, again suggesting that the fit is consistent with no correlation.

We find that the large errors on our model parameters are due, in part, to an inappropriate choice of interval estimator.  Upon further analysis, we find that many of the LINMIX model parameter posterior distributions are highly non-Gaussian with strong skewness and high kurtosis.  While we continue to use the median of the distribution as our point estimator, we recompute a new interval estimator rather than use the standard deviation; we find the interval, about the median, that contains approximately 68\% of the distribution.  We take the average of the lower and upper bounds and use this as the uncertainty.  Using this method, we obtain more reasonable errors on our fit parameters (i.e., $a=-0.374\pm31.68$ becomes $a=-0.374\pm2.67$).  However, utilizing this new estimator does not generally affect the significances of correlations observed between HR and multiple host-galaxy properties simultaneously.

We also study the dependence of HR on metallicity as well as on sSFR while imposing different criteria on host mass to try to control for the apparently dominant effect of mass.  First, we remove the HR-mass dependence by adding the measured PM sample ``HR step" of 0.049 magnitudes to the HR of our higher-mass ($\log(M/\Msun) \ge 10.2$) MZS hosts.  We then re-fit HR as a function of metallicity and  also HR as a function of sSFR (this time including measurement errors again).  In both cases, the direction of the best-fit slope is the same as that fit with the entire MZS sample.  However, the significance of non-zero slopes in both cases is $<1\sigma$.  We next investigate HR as a function of metallicity and sSFR in mass bins.  Our first separation is into low and high-mass bins, split at $\log(M/\Msun)=10.2$, shown in Figure~\ref{fig:lowhimassbins}.  In each case, the significance of a non-zero slope for the best-fit to the data is $\lesssim 0.8\sigma$, which is consistent with flatness. 

Unfortunately, each of these tests are consistent with no correlation between HR and multiple host-galaxy properties. This is perhaps largely due to the variation in measurement errors between the properties; i.e. photometric stellar masses are much easier to estimate and have smaller uncertainties than spectroscopically-derived properties such as metallicity and SFR. We recommend that future surveys interested in studying these correlations obtain high S/N host-galaxy spectra for as many SNe Ia host galaxies. We also recommend further investigation of how to incorporate correlations, both physical and in measurement uncertainty, between various host-galaxy properties in future studies of this type. Hopefully, combining the results of these efforts will provide a better understanding of the physical mechanism driving these observed trends.

\begin{figure*}[tp]
\centering
\begin{tabular}{cc}
\includegraphics[scale=0.35,angle=90]{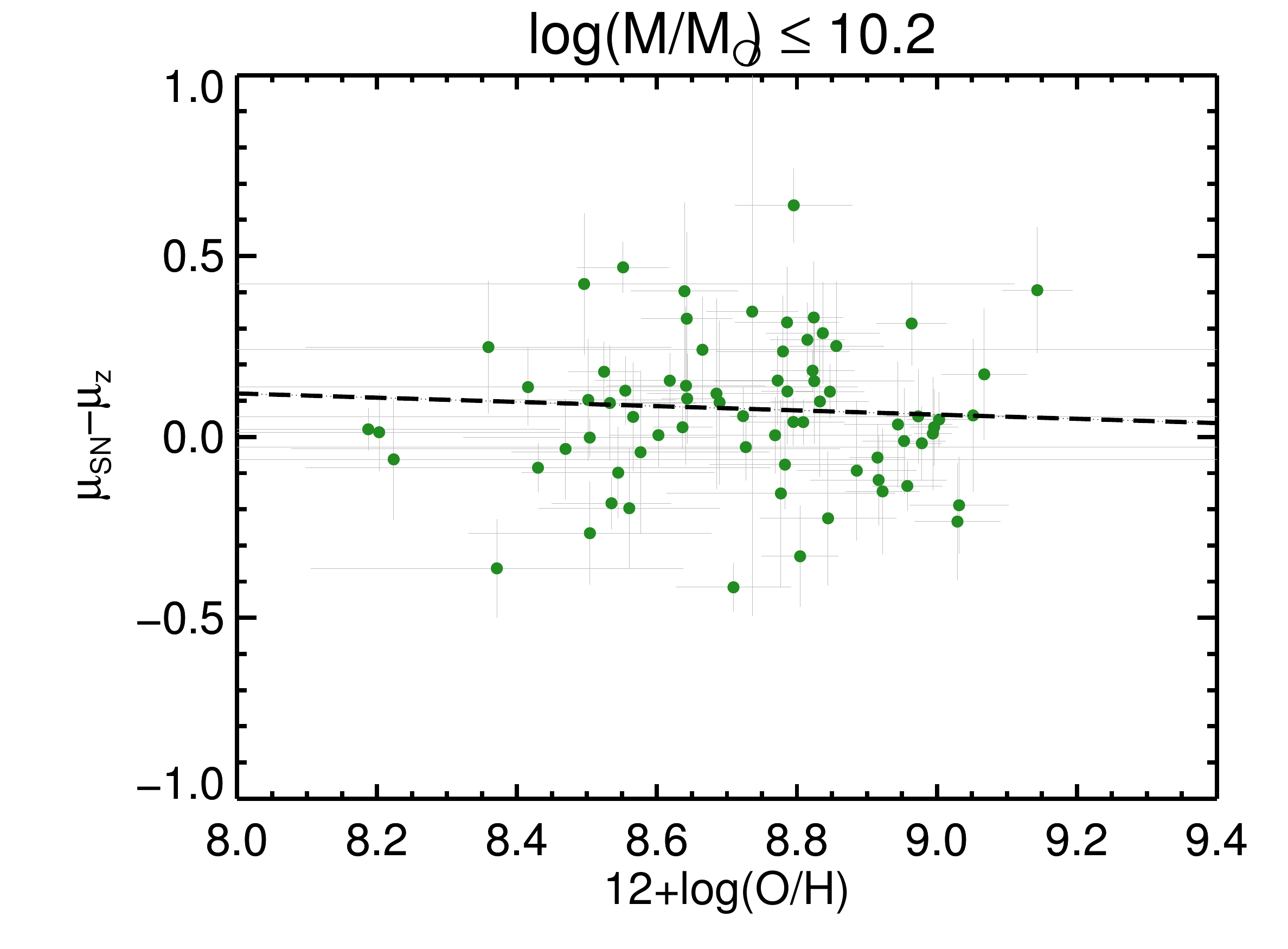}
\includegraphics[scale=0.35,angle=90]{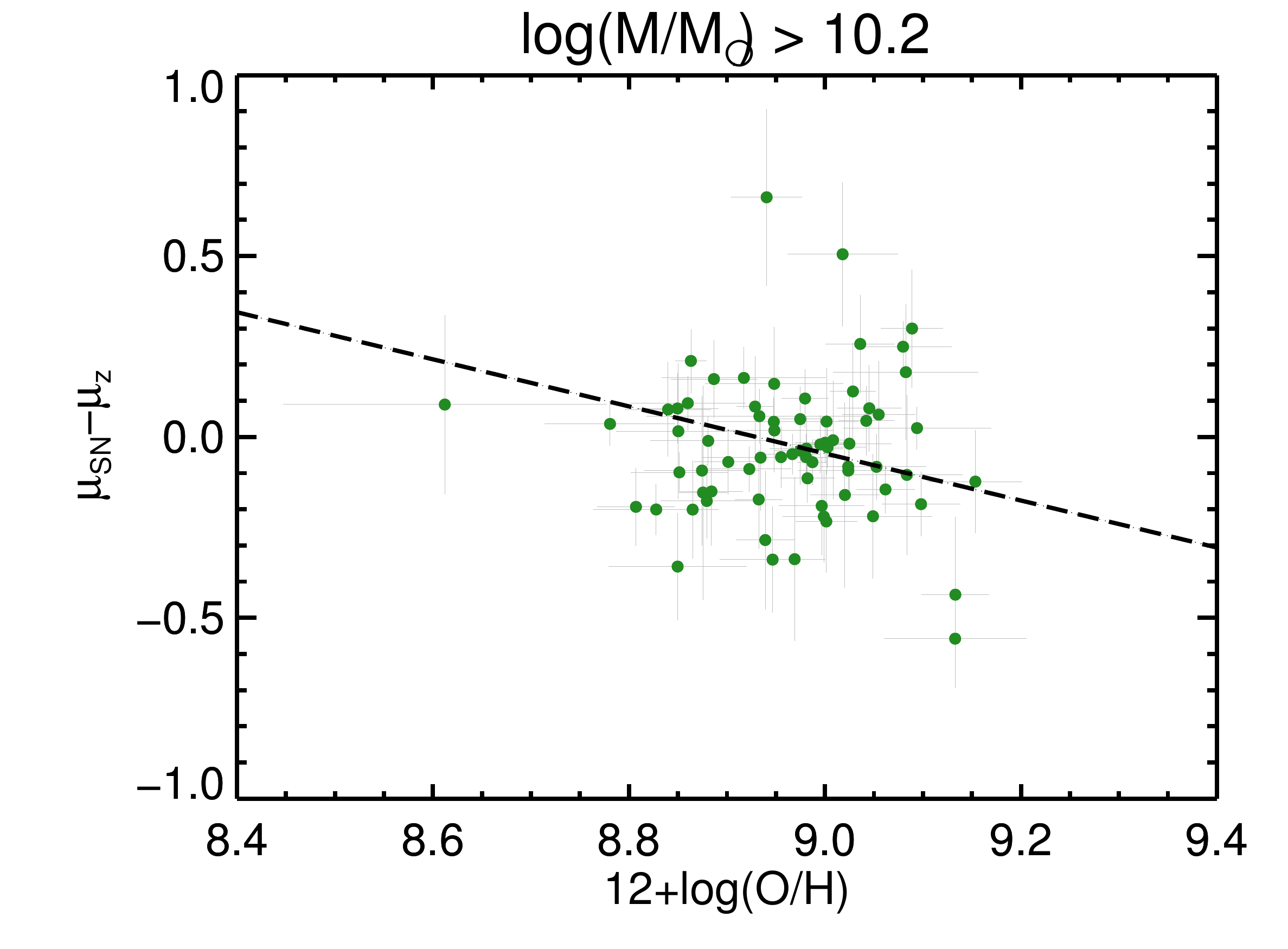}\\
\includegraphics[scale=0.35,angle=90]{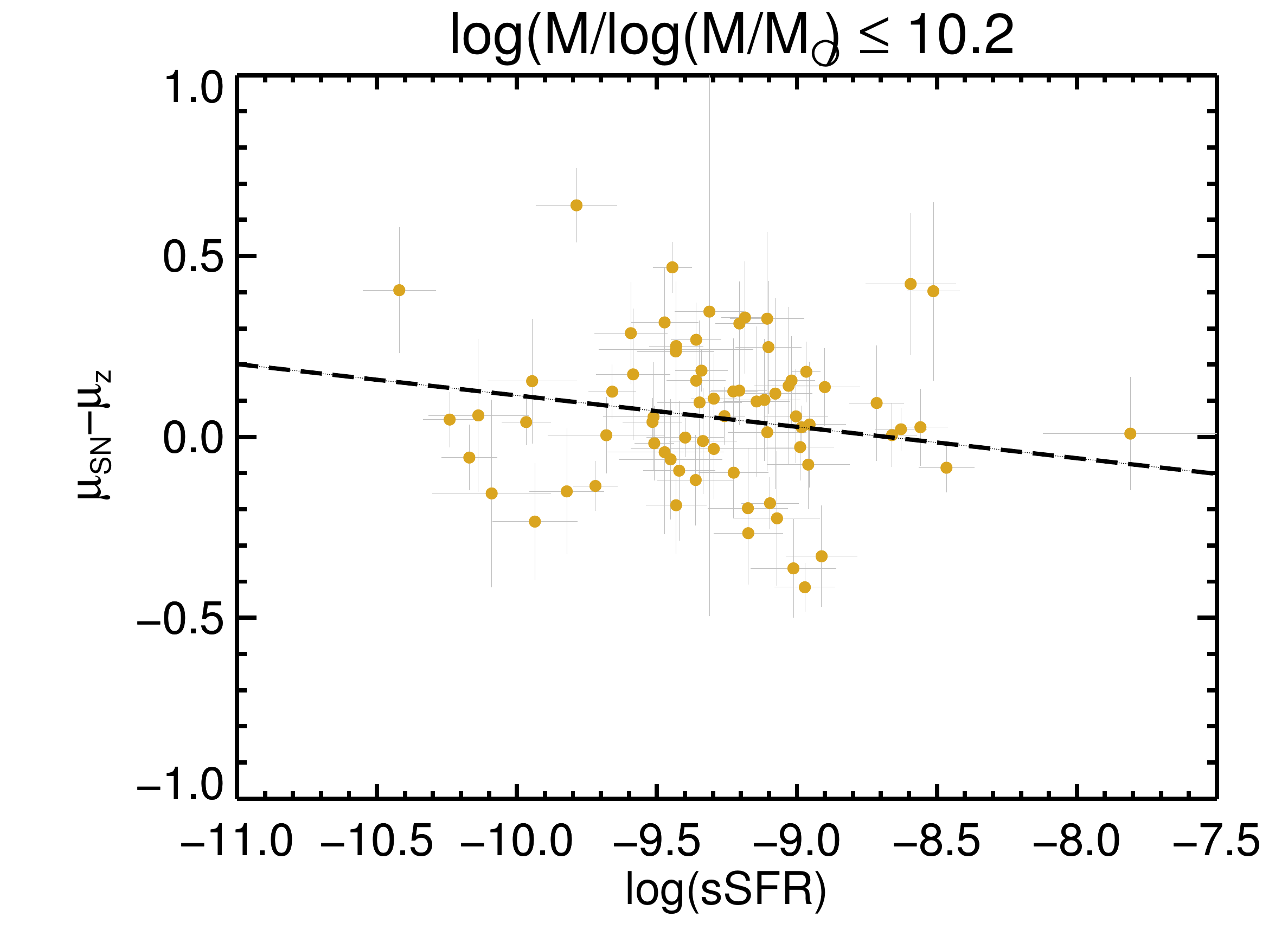}
\includegraphics[scale=0.35,angle=90]{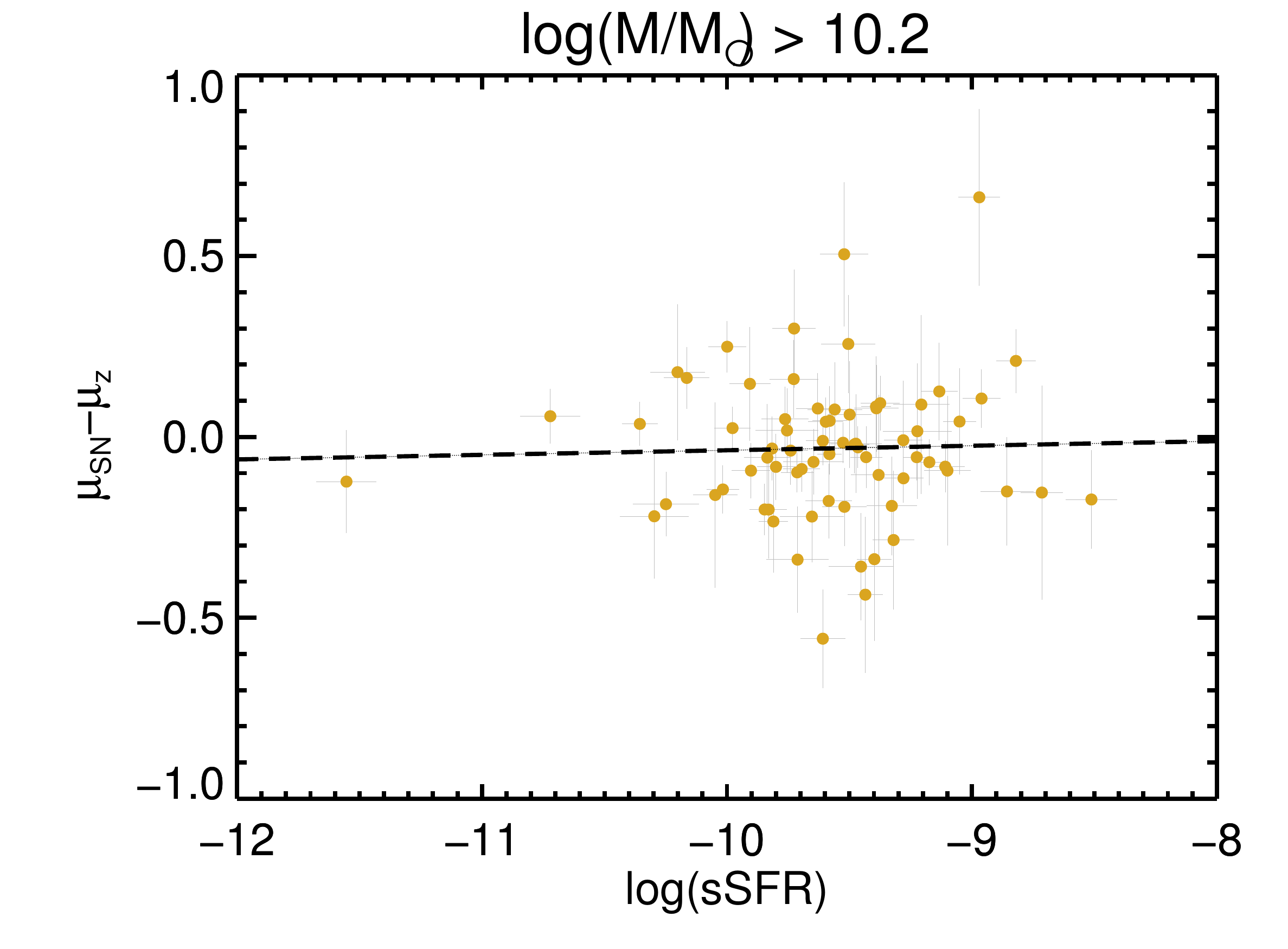}
\end{tabular}
\caption{HR as a function of metallicity and specific star-formation rate for the MZS sample in low and high-mass bins.  LINMIX linear fits to the data are shown in dashed-black.  In each case, the significance of a non-zero slope is $\lesssim 0.8\sigma$.}
\label{fig:lowhimassbins}
\end{figure*}

\section{DISCUSSION}
\label{sec:Disc}

In this section, we compare our linear fit results of HR as a function of host-galaxy properties to those reported in previous studies.  We also separate the PM and MZS samples into Spec-Ia and Phot-Ia subsets to assess the effect of including a sample of photometrically-classified SNe\,Ia on studies of HR and host properties.  Finally, we discuss the differences between the star-forming and passive galaxies in our PM sample.

 \subsection{Comparing with Previous Studies}
 \label{sub_sec:hrvhostpropdis}
 
We compare the correlations observed in this work between HR and host-galaxy properties with those reported in previous studies and present a sample of HR-host-galaxy correlations as well as fit significances as they are reported in the literature. In some cases, a linear best-fit to the data was not provided, and thus we provide the significance of the binned-average HR-step. Because of differences in metallicity calibrations and IMFs used for host-mass calculations, we encourage the reader to use caution when comparing linear fit results from all previous works directly.  However, we can consider the strengths and significances of the linear correlations between HR and host-galaxy properties to get a qualitative sense of how these studies compare.  We present this summary in Table~\ref{comp} and note that the list of works included only represents a subset of the literature.

\begin{deluxetable*}{ l l l r  r r r}
\tablecaption{Comparison of correlations found between HR and host galaxy mass (M), gas-phase metallicity (Z), and specific star formation rate (S) \label{comp} }
\tabletypesize{\scriptsize}
\tablewidth{0pt}
\tablehead{\colhead{Paper} & \colhead{SN Survey} & \colhead{Host} & \colhead{Sample}  & \colhead{HR Step} & \colhead{Slope}\tablenotemark{a} & \colhead{Slope\tablenotemark{b}} \\ &   & \colhead{Property} & \colhead{Size} &\colhead{(mag)} &  \colhead{(mag/dex)}& \colhead{Significance ($\sigma$)} }
\startdata
This work (PM) & SDSS-SNS  & M & \numpm& $0.048\pm0.019$ &$-0.055\pm0.015$ & $3.6\sigma$ (L) \\ 
This work (MZS) & SDSS-SNS & M & \nummzs &$0.082\pm0.030$&$-0.071\pm0.029$ & $2.5\sigma$ (L)  \\ 
Sullivan `10 & SNLS & M & 195  & $0.080\pm0.020$ & $-$$0.042\pm0.013$ &$3.3\sigma$ (L) \\ 
Lampeitl `10 & SDSS ($z \leq 0.21$) & M & 162 & $0.100\pm0.025$& $-$$0.072\pm0.018$ &$4.9\sigma$ (L) \\
Gupta `11& SDSS & M & 206 &$0.096\pm0.028$& $-0.057\pm0.019$ & $3\sigma$ (L)\\ 
Kelly `10 & CfA3 & M & 62 &$0.094\pm0.045$& $-0.150\pm0.060$ &$2\sigma$ (L)\tablenotemark{c}\\
C13 & SNf & M & 115 &$0.085\pm0.028$ & $-0.043\pm0.014$ &$3.1\sigma$ (L) \\ 
P14 & PTF & M & 50 &$0.085\pm0.047$& $-0.041\pm0.030$ &$1.4\sigma$ (L)\\
Scolnic `14 & Pan-Starrs1 & M & 112 &$0.040\pm0.032$& -- & $1.25\sigma$ (B) \\ 
\hline
This work (MZS) & SDSS-SNS & Z & \nummzs &$0.057\pm0.031$&$-0.579\pm0.409$ & $1.4\sigma$ (L) \\ 
Konishi `11 & SDSS & Z & 72 & $0.130\pm0.060$& -- &$1.8\sigma$ (B) \\
D11 & SDSS ($z \leq 0.15$)  & Z & 34 &$0.091\pm0.021$&--  & $1.3\sigma$ (L)\tablenotemark{d,e} \\ 
C13 & SNf & Z & 69 &$0.103\pm0.036$&$-0.106\pm0.043$ & $2.5\sigma$ (L)\\
P14 & PTF &Z& 36 &$0.115\pm0.046$&$-0.358\pm0.176$ & $2\sigma$ (L)\\
\hline
This work (MZS) & SDSS-SNS & S & \nummzs &$0.013\pm0.031$&$0.019\pm0.046$  & $0.4\sigma$ (L)\\ 
P14 & PTF & S & 48 &$0.070\pm0.041$&$-0.019\pm0.077$  & $0.25\sigma$ (L) \\
D11 & SDSS ($z \leq 0.15$) & S\tablenotemark{c} & 34 &-- &--& $1.2\sigma$ (L)\tablenotemark{d} \\ 
\enddata

\tiny
\tablenotetext{a}{Slopes presented use the sign convention where Hubble residuals are defined as $\mathrm{HR} = \mu_{\mathrm{SN}} - \mu_{z}$. This switches the sign of the values reported in \citet{Sullivan10} and \citet{Lampeitl10}.}
\tablenotetext{b}{We have included significances for linear fits (L) and well as differences in high and low-mass (metallicity, sSFR) bins (B) (for those papers which do not provide linear fit results).}
\tablenotetext{c}{Result quoted is from using \tt{SALT2}.}
\tablenotetext{d}{sSFR$_{\mathrm{phot}}$ \citep[see][]{D11}.}
\tablenotetext{e}{The uncertainty quoted on the HR Step is as reported as does not include intrinsic scatter.}

\end{deluxetable*}

As seen in Table~\ref{comp}, the results of this study confirm much of what is established in the literature.  In five studies using a sample of more than 100 SNe\,Ia, a significant linear correlation ($\gtrsim3\sigma$) was found suggesting more massive galaxies host over-luminous SNe\,Ia; it is possible that the three studies that did not detect such a correlation did not have large enough samples to detect as strong of an effect. Although the HR step with host-galaxy mass observed for the PM sample in this work is smaller than what is reported in several other studies, it is consistent at $\lesssim 1.7\sigma$. The trend observed between HR and host-galaxy gas-phase metallicity and sSFR is also consistent with existing results, particularly that the HR-sSFR correlation is the weakest observed.  

When comparing to D11, it is important to clarify that they computed two estimates of sSFR: ``sSFR$_{\mathrm{spec}}$" (using host-galaxy 
masses determined from the spectroscopic fit to the galaxy continuum) and ``sSFR$_{\mathrm{phot}}$" (using 
masses derived from host-galaxy photometry).  
In their study, they find a $ >3\sigma$ correlation between HR and sSFR$_{\mathrm{spec}}$.  Unfortunately, we are 
unable to compute spectroscopic masses (and thus sSFR$_{\mathrm{spec}}$) in our current emission-line analysis 
and suggest this for future study.  
However, D11 find that the correlation between HR and sSFR$_{\mathrm{phot}}$ is only significant at the
$1.2\sigma$ level.\footnote{As expounded in D11, the difference between sSFR$_{\mathrm{spec}}$ and sSFR 
$_{\mathrm{phot}}$ (and thus their trends with HR) might be due to corrections for aperture effects which are applied 
to sSFR$_{\mathrm{phot}}$ but not to sSFR$_{\mathrm{spec}}$.  See Section 4.2 of D11 for more details.}  
Given that the method we use to compute sSFR in this work is analogous to D11's sSFR$_{\mathrm{phot}}$, 
it is not unexpected that we see a significance of similar strength.

\subsection{Photometric vs. Spectroscopic SN\,Ia Subsets}
\label{sub_sec:spec_v_phot}
Here we consider the Phot and Spec-Ia subsets of the PM and MZS samples separately and re-compute correlations between HR and host-galaxy properties.  Figure~\ref{fig:hrps} displays the linear fits for the separate datasets and the fit results are presented in Table~\ref{psres}.  Generally, in each study of HR as a function of host property using just the Spec-Ia, the significance of a non-zero slope is $\lesssim2\sigma$.  The significance of a non-zero correlation between HR and host-galaxy mass using the Phot-Ia is $3.9\sigma$,while the significance of the relation using only the Spec-Ia is $1.5\sigma$. When using the Phot-Ia MZS subsample, the significances of a non-zero HR-metallicity correlation and non-zero HR-sSFR correlation are $1.6\sigma$ and $1.1\sigma$, respectively.  As evident in Figure~\ref{fig:hrps}, the correlation between HR and metallicity for the Phot-Ia may be best-fit by a non-linear function.  We find that in all cases of HR as a function of host property, the linear fits obtained for the Spec-Ia are in the same direction as those for the Phot-Ia.  The slopes of the linear fits for the Phot-Ia and Spec-Ia subsamples, for the HR-metallicity and HR-sSFR relations, are consistent within $1.3\sigma$. The slopes of the fits of the HR-mass relation between the Phot-Ia and Spec-Ia samples are consistent at $2.3\sigma$; however, both are consistent with the slope recovered using the full PM sample within $1.5\sigma$.

\begin{figure*}[tp]
\centering
\begin{tabular}{cc}
\includegraphics[scale=0.35,angle=90]{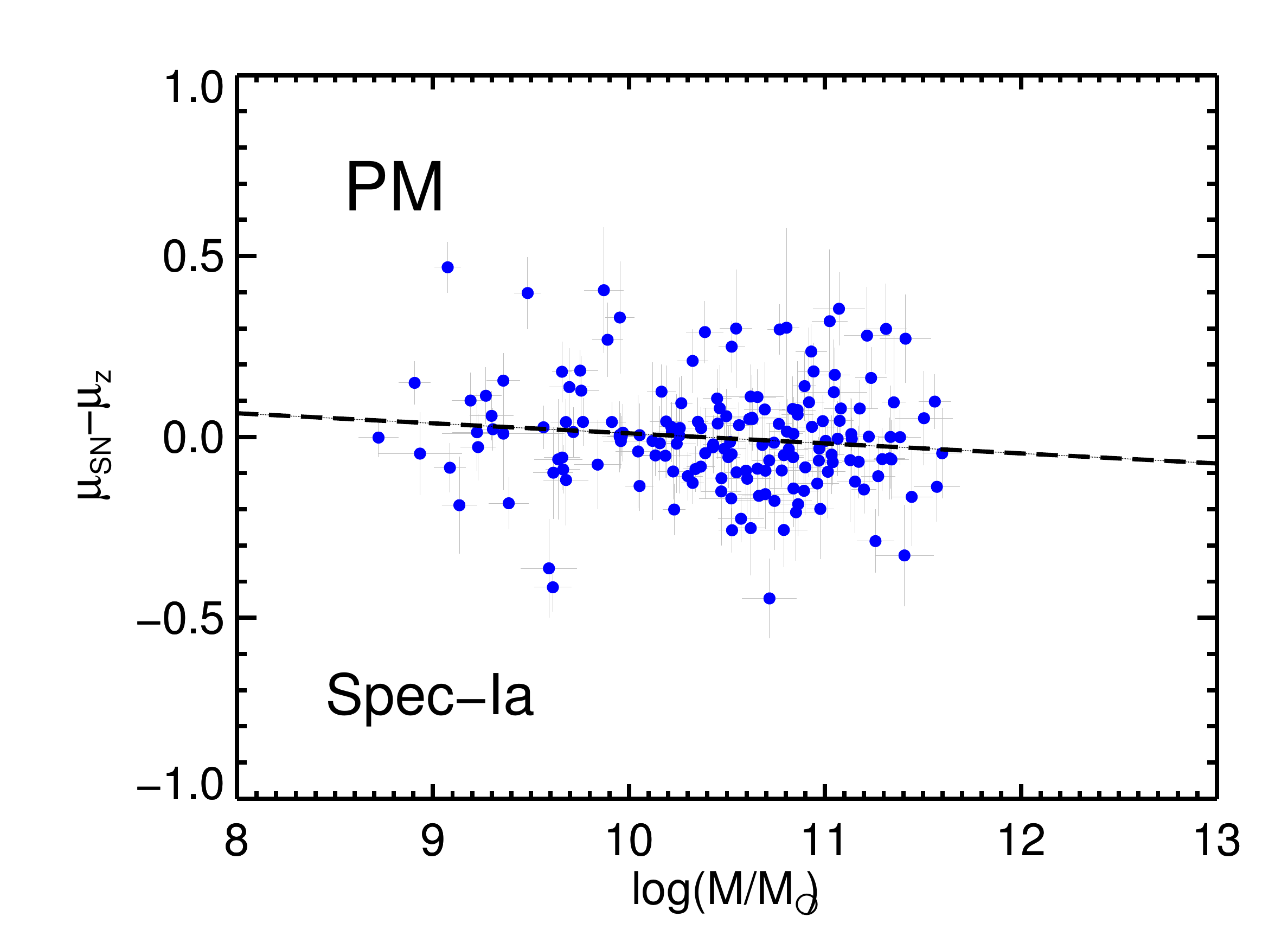}
\includegraphics[scale=0.35,angle=90]{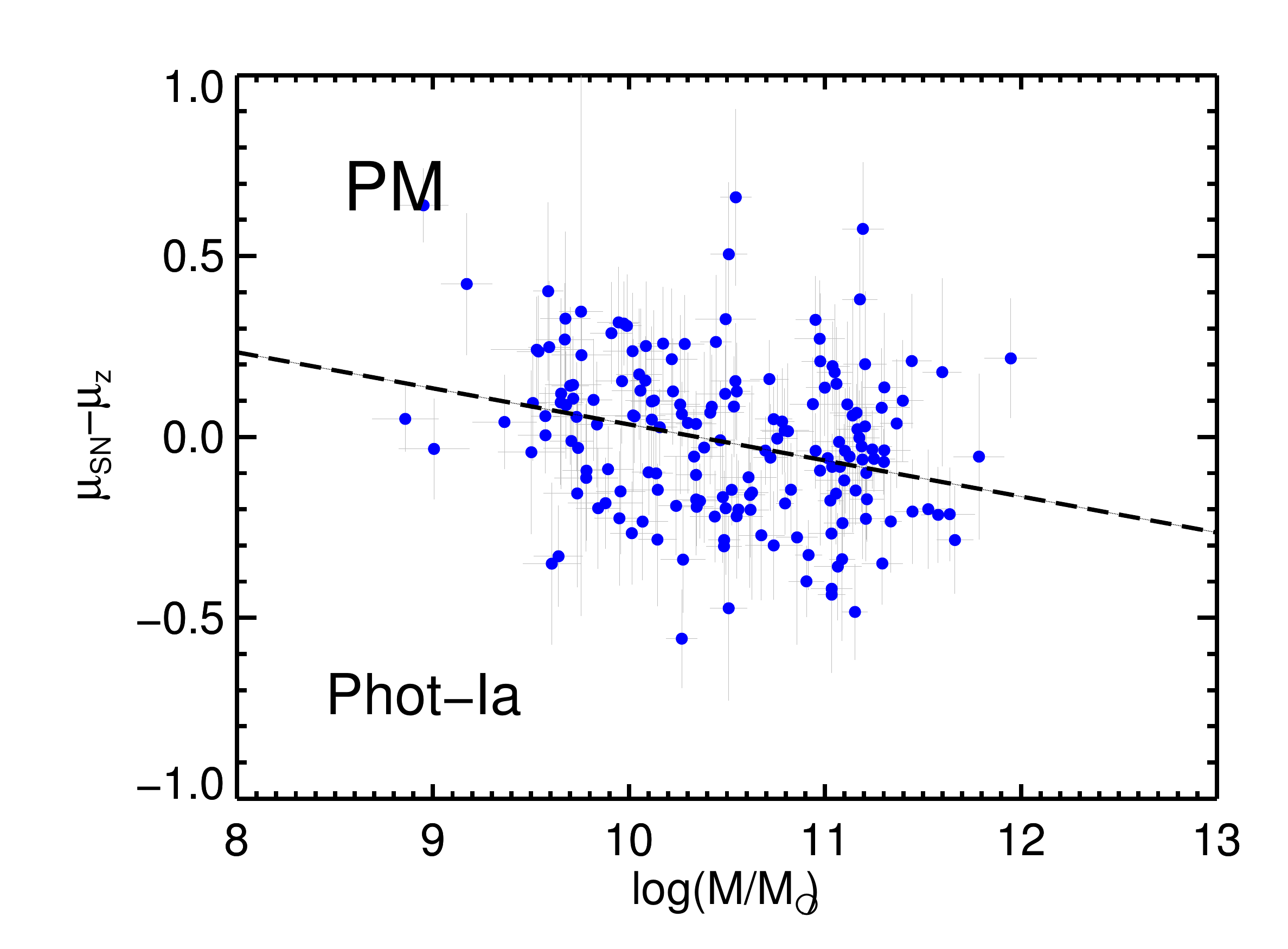}\\
\includegraphics[scale=0.35,angle=90]{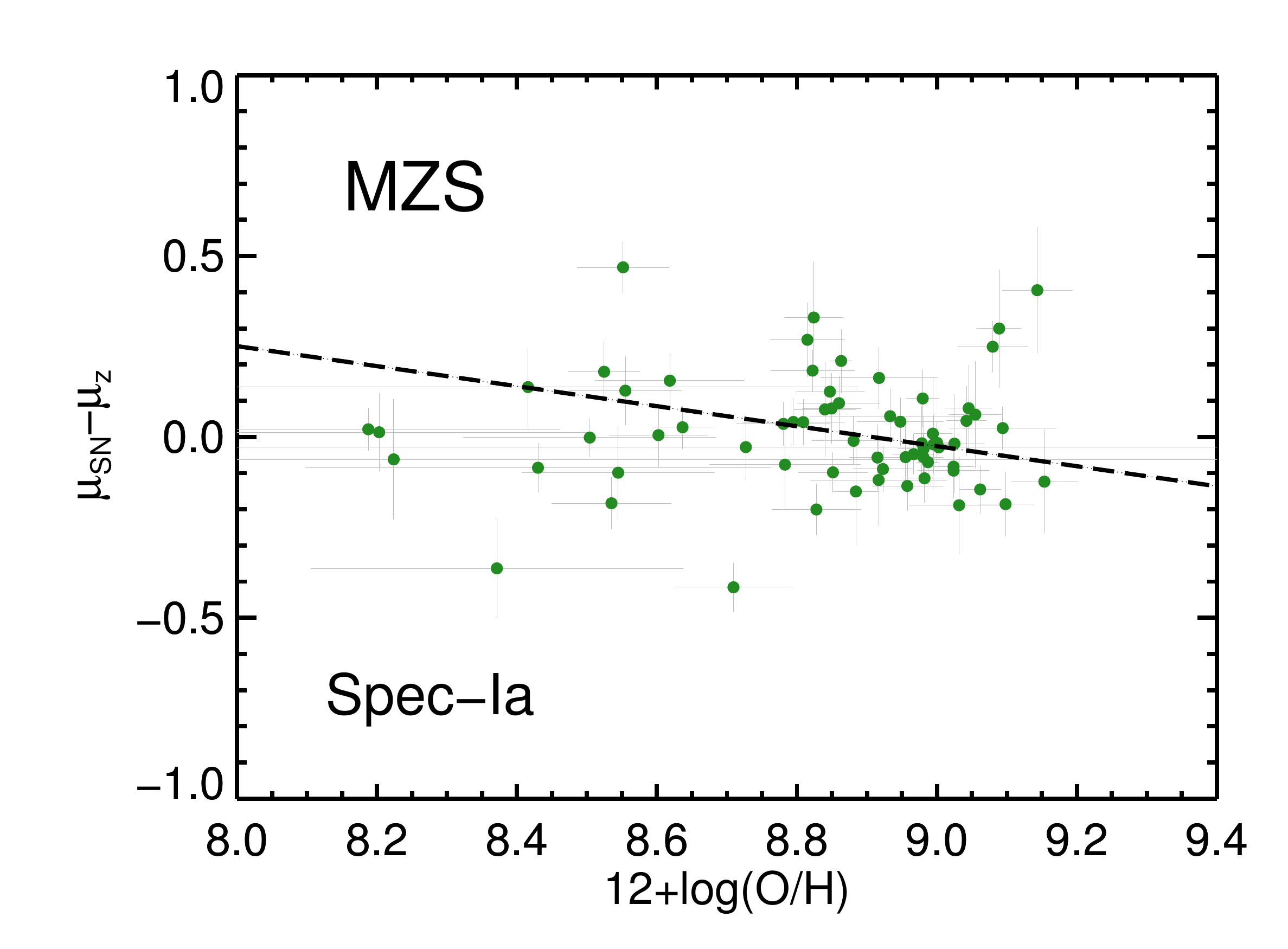}
\includegraphics[scale=0.35,angle=90]{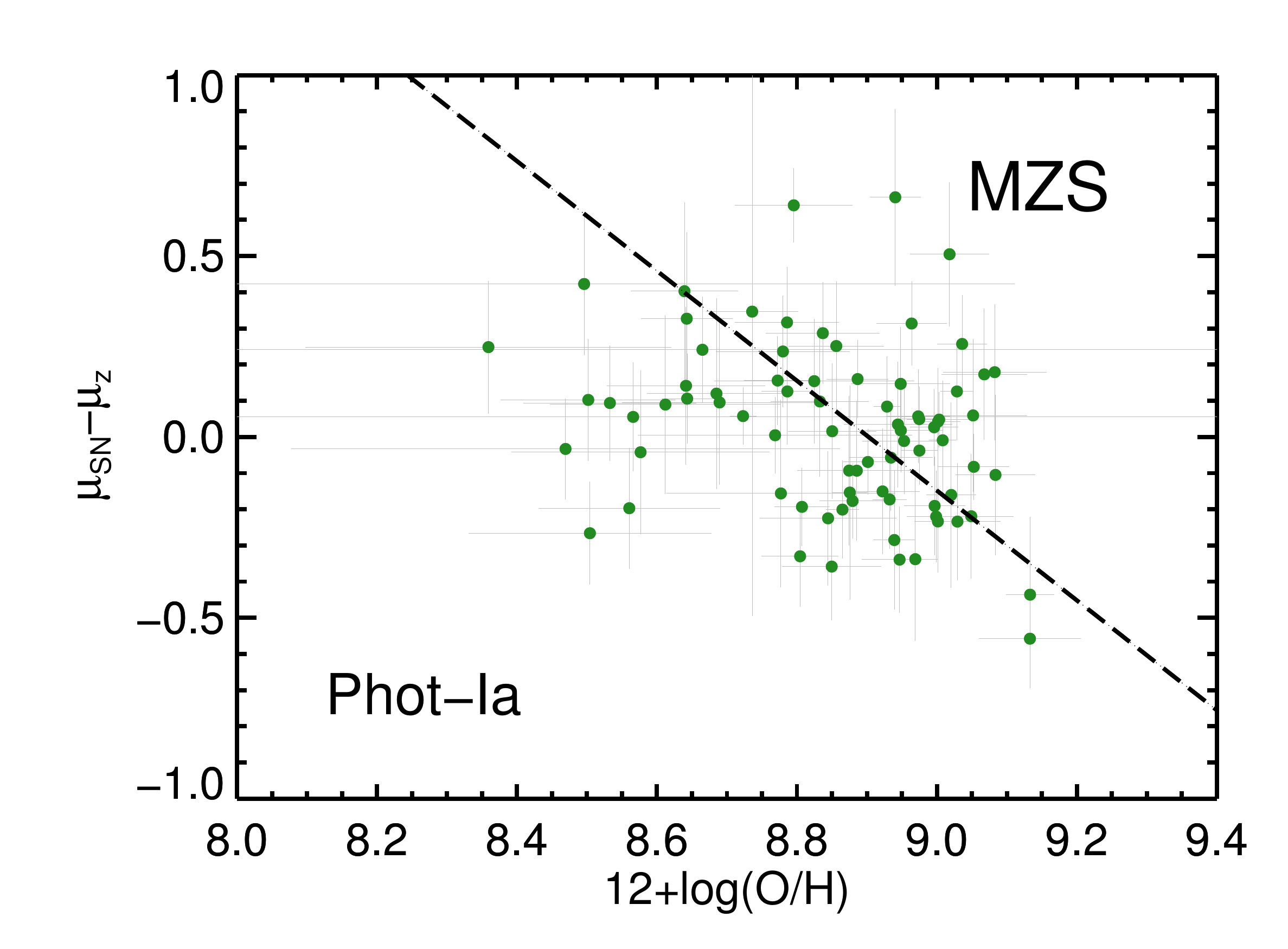}\\
\includegraphics[scale=0.35,angle=90]{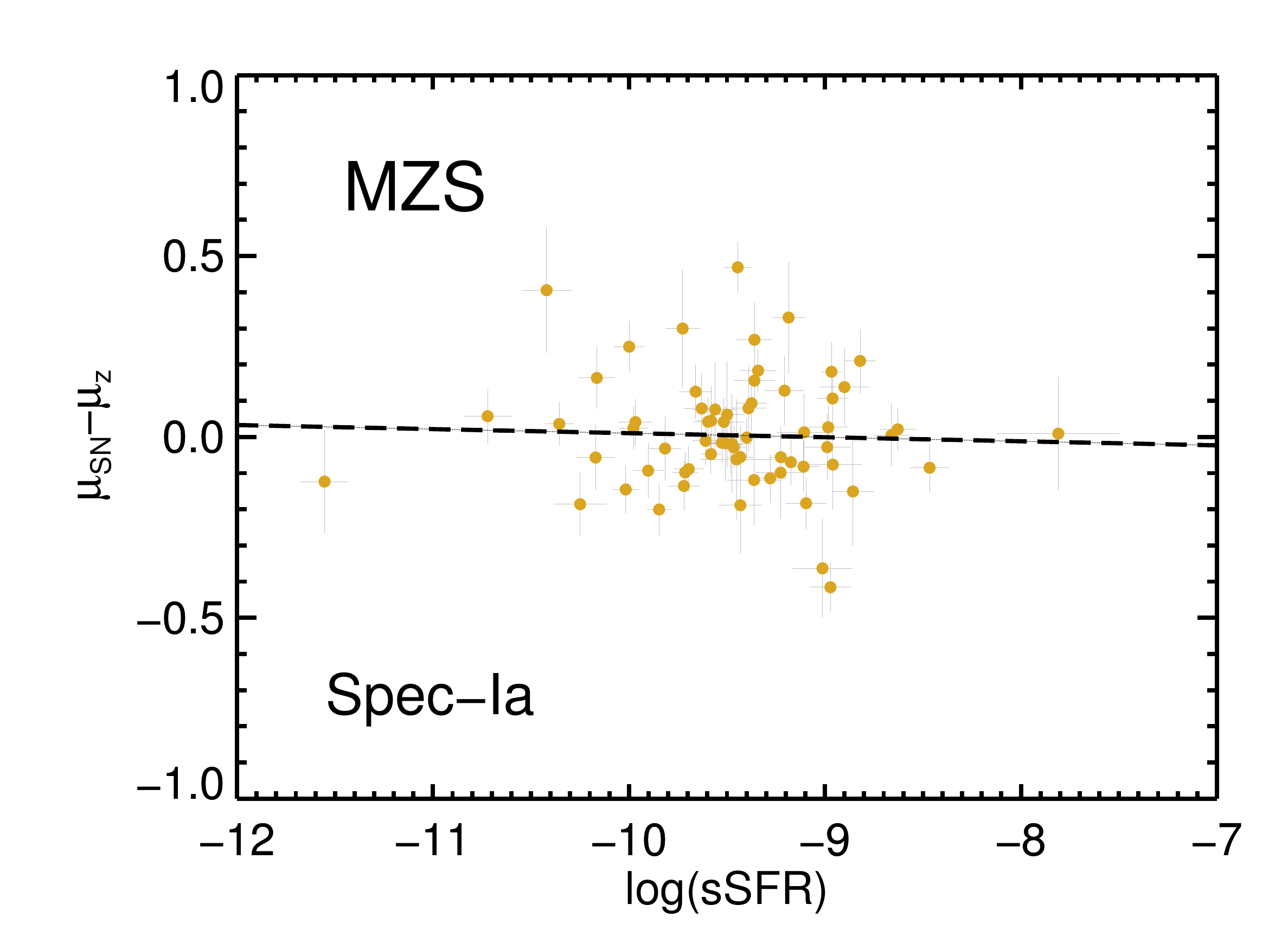}
\includegraphics[scale=0.35,angle=90]{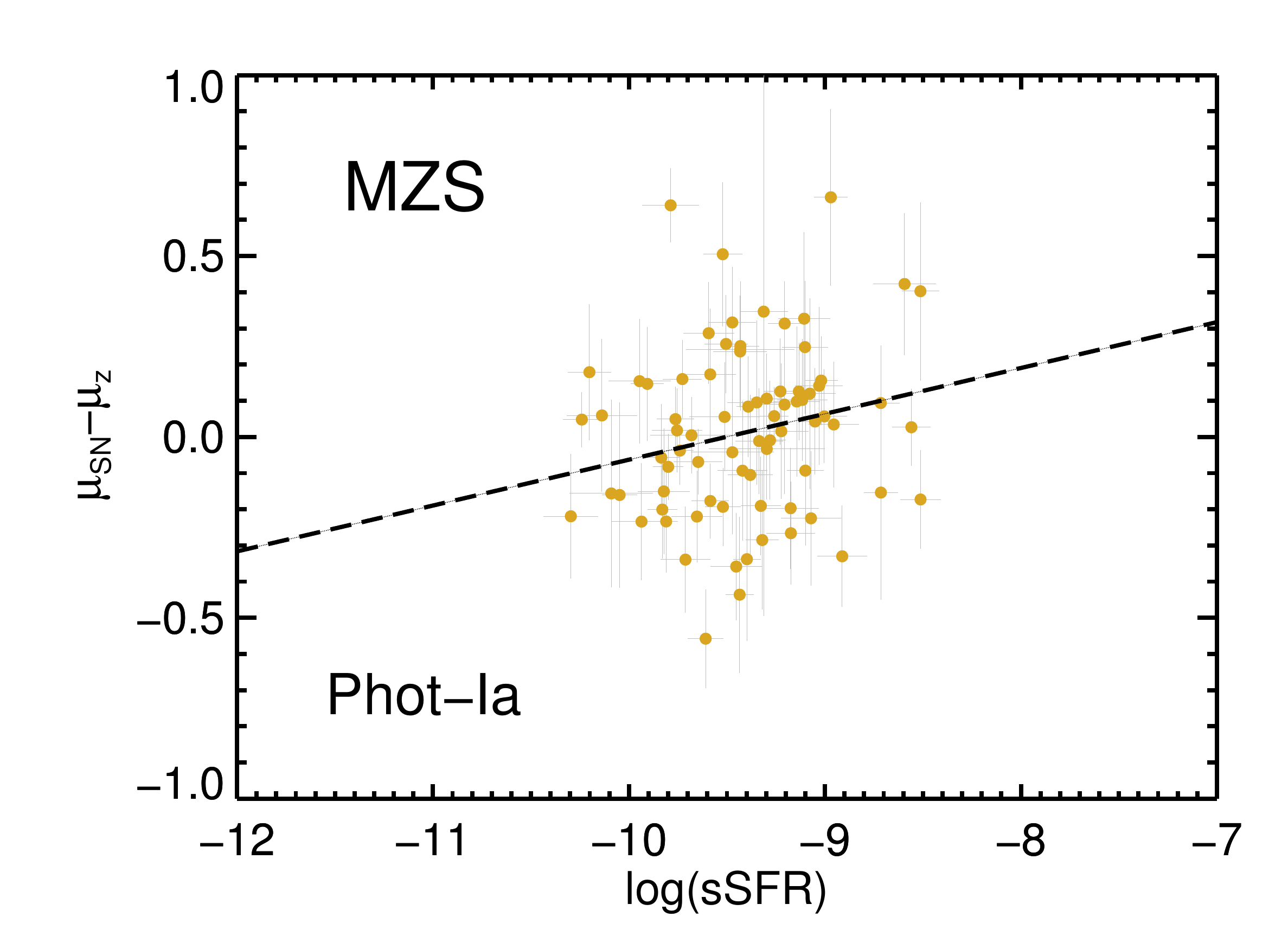}\\
\end{tabular}
\caption{HR as a function of derived host properties for the Spec and Phot SN\,Ia samples separately.  The first row displays HR as a function of mass for the PM sample and the lower two rows show HR as a function of derived host properties for the MZS sample. LINMIX fits to the data are shown in dashed-black.  Fit results are reported in Table~\ref{psres}. }
\label{fig:hrps}
\end{figure*}

\begin{deluxetable*}{l l c c r r r r r r}
\tabletypesize{\scriptsize}
\tablecaption{Fit Results for HR as a Function of Host-Properties: Spec-Ia and Phot-Ia \label{psres}}
\tablewidth{0pt}
\tablehead{ \colhead{Host Property}  &\colhead{Sample}& \colhead{SN Type\tablenotemark{a}} & \colhead{$N$\tablenotemark{b}} & \colhead{Slope} & \colhead{Intercept} & \colhead{$\sigma$ [mag]}&\colhead{Sig\tablenotemark{c}} & \colhead{$\rho$} & \colhead{$p$-value} } 
\startdata
Mass & PM & S & 169& $-0.028\pm0.018$&$0.287\pm0.188$&$0.113\pm0.010$&$1.54\sigma$ & $-0.0718$ &$0.3538$ \\
Mass & PM&  P & 176 & $-0.101\pm0.026$&$1.042\pm0.270$&$0.137\pm0.017$&$3.87\sigma$ &$-0.2496$ & $0.0008$ \\
12+log(O/H) & MZS& S & 66& $-0.277\pm0.250$&$2.464\pm2.240$&$0.119\pm0.019$&$1.11\sigma$ & $-0.0718$ & $0.5668$\\
12+log(O/H) & MZS& P&78&$-1.518\pm0.960$&$13.512\pm8.640$&$0.133\pm0.043$&$1.58\sigma$ &$-0.2797$ & $0.0132$ \\
sSFR & MZS& S & 66 & $-0.011\pm0.046$&$-0.102\pm0.440$&$0.126\pm0.017$&$0.24\sigma$ & $-0.0130$ & $0.9177$ \\
sSFR & MZS& P & 78 &$0.127\pm0.120$&$1.204\pm1.140$&$0.170\pm0.026$&$1.06\sigma$ &$0.1845$ & $0.1058$ \\
\enddata
\tiny
\tablenotetext{a}{Indicates Spec-Ia (S) or Phot-Ia (P).}
\tablenotetext{b}{Number of SNe\,Ia in the sample.}
\tablenotetext{c}{Significance of a non-zero linear slope.}
\end{deluxetable*}

The weaker HR-mass correlation in the Spec-Ia sample is a bit unexpected, especially when comparing to previous analyses using SDSS SNe\,Ia. In particular, we would expect a similar significance to that reported in \citet{Gupta11}, which uses a comparably-sized sample of spectroscopically-confirmed SNe\,Ia, also from the SDSS-SNS. However, we note that while many of the SNe\,Ia used in this analysis overlap with those in the \citet{Gupta11} sample, there are several key differences in our sample construction, namely: sample redshift cuts, SN\,Ia light-curve quality criteria, requirements on host-galaxy spectroscopy, and host-galaxy photometry used to compute stellar masses. We find that only 94 SNe\,Ia overlap between the \citet{Gupta11} sample and our PM sample. A comparison of the median of the best-fit LINMIX posterior slopes of each overlapping sample yields an agreement of $0.08\sigma$, indicating that sample construction, rather than methodology, plays a large role in the differing results between the two works.

Initially, we believed that the magnitude limit of the host spectroscopic follow-up may have biased our Spec-Ia host sample against low-mass hosts. To test this, we create a sample of SN\,Ia hosts using all criteria in Table~\ref{pmsamplecuts}, \textit{without imposing any requirements on the host spectra}, and compare this mass distribution to that of the Spec-Ia hosts. Using the two-sided Kolmogorov-Smirnov test, we find no significant difference between the Spec-Ia host-mass distribution and that of this new sample, even when only considering the low-mass hosts. This indicates that our spectral quality requirement does not change our results.

The disagreement between the Spec-Ia and Phot-Ia results when fitting for HR as a function of mass is also surprising, particularly if the Spec-Ia and Phot-Ia samples are indeed drawn from a homogenous sample of SNe\,Ia. To further explore the results, we plot the 68\% and 95\% confidence intervals of the slope and intercept LINMIX posterior distributions for both samples. As shown in Figure~\ref{fig:contour}, the two samples show poor agreement. We also see that the Phot-Ia slope is definitively negative and that both the slope and intercept distributions are wider than those of the Spec-Ia. 

\begin{figure}[tp]
\centering
\includegraphics[scale=0.47]{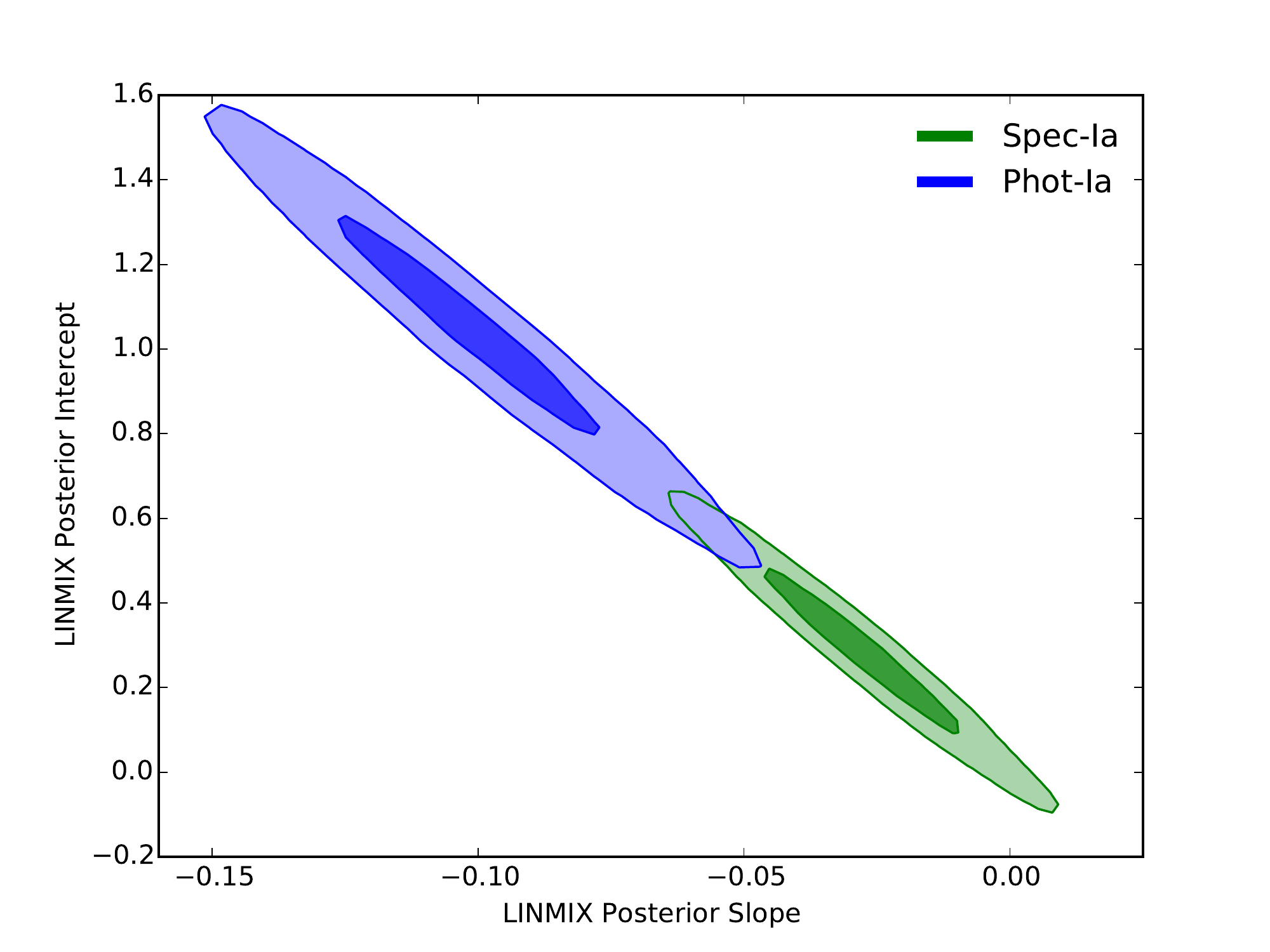}
\caption{Contour intervals showing the 68\% and 95\% confidence regions of the Spec-Ia (green) and Phot-Ia (blue) LINMIX posterior distributions for the HR-mass relation.}
\label{fig:contour}
\end{figure}
While the differences we observe between the Spec-Ia and Phot-Ia samples could be attributed to random statistical fluctuations, the contour plots strongly allude to a more fundamental discrepancy between the Phot-Ia and Spec-Ia subsamples. Issues with photometric typing, for example, may seriously affect the homogeneity of the two datasets, thus limiting the ability to perform comparable analyses with each independently.  While probing the differences between the Spec-Ia and Phot-Ia subsamples is beyond the scope of this work, we encourage future studies to explore this problem further.

\subsection{Star-forming and Passive Hosts in the PM Sample}
\label{actpas}

Although we require the host-galaxies in the MZS sample to have active star formation (as indicated by strong \Ha~emission), we do not require this of the PM hosts.  Therefore, the PM sample is comprised of both actively star-forming and passive galaxies.  Motivated by the fact that SN properties and rates are correlated with the amount of star formation in their hosts \citep[e.g.,][]{Hamuy00,Sullivan06}, we study correlations between HR and mass separately for star-forming and passive hosts using the FSPS photometric measurements of star-formation rate reported in S14 and the suggested division at log(sSFR$_{\mathrm{FSPS}}) = -12$.  We require a ``star-forming" galaxy to have log(sSFR$_{\mathrm{FSPS}}) > -12$ and a ``passive" galaxy to have log(sSFR$_{\mathrm{FSPS}}) \le -12$.  While this separation may not be absolute, it provides a reasonable estimate of star formation activity, yielding 259 star-forming hosts and 86 passive hosts.  We fit for linear trends of HR with host mass for these two groups separately; results are shown in Figure~\ref{fig:hrap}.  In star-forming galaxies, there is a $3.3\sigma$ significance of a non-zero slope; however, in passive galaxies, the significance of a non-zero slope is only $0.09\sigma$. This may be due, in part, to the fact that we lose the low-mass end of the mass distribution for the passive hosts, which significantly reduces the mass range for this subsample.  The inverse-variance weighted average HR of the star-forming and passive samples is calculated, including the best-fit intrinsic scatter, and we find that SNe\,Ia in the passive galaxies are $0.041$ magnitudes more luminous, with a confidence of $1.87\sigma$, than those in star-forming galaxies after light-curve correction.  This trend is consistent to $1.3\sigma$ with \citet{Lampeitl10} who also used SDSS SNe and reported a $\simeq 0.1$ mag difference between star-forming and passive hosts at the $2-3\sigma$ level. \footnote{We note that readers should approach the comparison to the \citet{Lampeitl10} results cautiously, as the sample construction (96 overlapping SNe\,Ia) and calculation of HRs differs significantly between the two works.}

\begin{figure*}[tp]
\centering
\includegraphics[scale=0.55,angle=90]{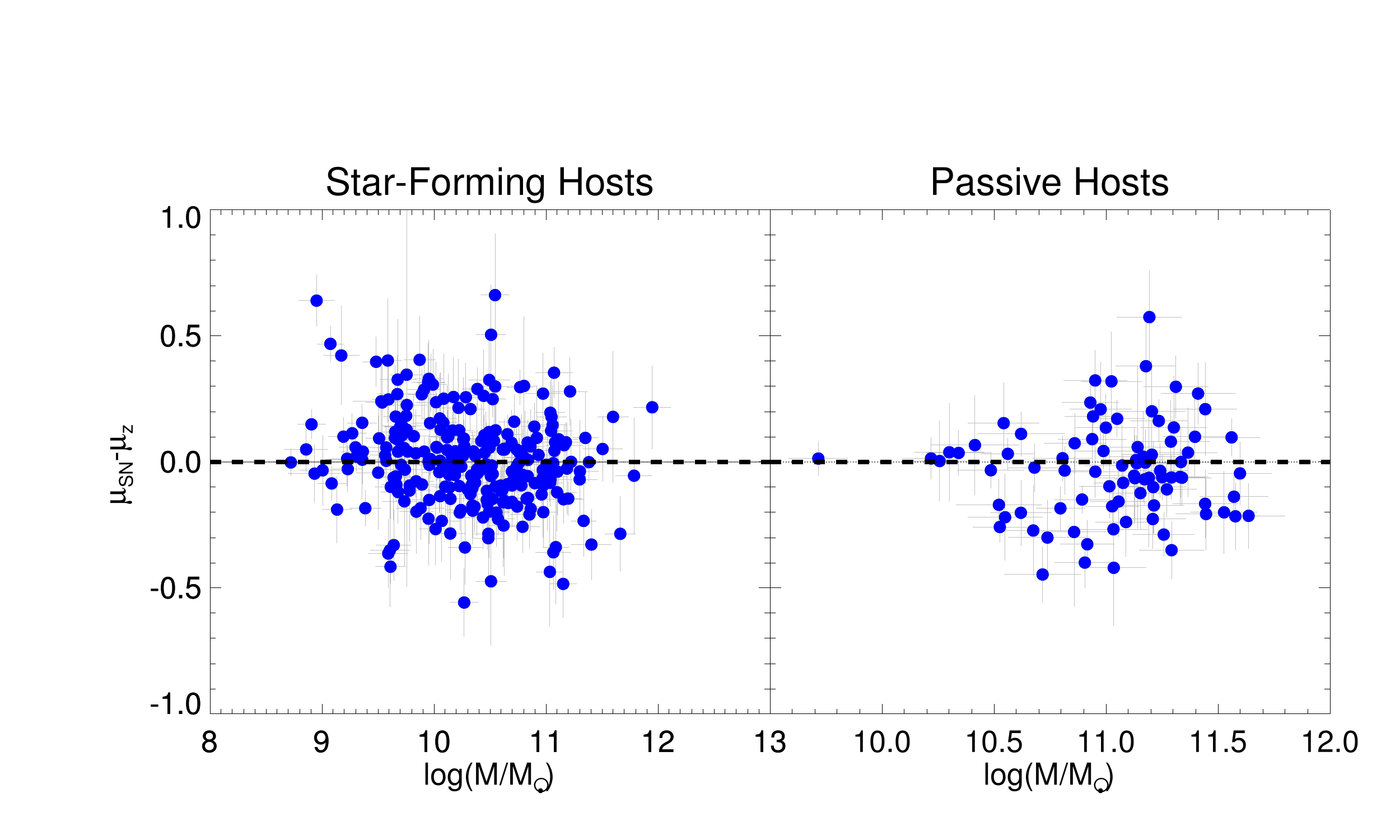}
\caption{HR as a function of host mass for the PM sample, separated into star-forming and passive galaxy groups.  LINMIX fits to the data are shown in dashed-black.  For our star-forming galaxies (left panel) we find the significance of a non-zero slope is $3.3\sigma$.  The trend with the passive galaxies (right panel) is consistent with flatness ($0.09\sigma$).}   
\label{fig:hrap}
\end{figure*} 

In the recent study by \citet{Childress14}, they predict that SNe\,Ia in star-forming hosts are a more uniform sample than those in passive hosts due to the homogeneity of young progenitors.  We expect that this uniformity would be apparent in the distribution of HRs in the sense that the HR distribution in the star-forming sample would exhibit less scatter than that of the passive sample.  A comparison of the HR distributions for the two samples reveals no statistical difference in their medians or standard deviations.  However, further analysis with a larger sample of low-mass host galaxies is necessary to make a definitive statement about the findings of \citet{Childress14}.

\section{SUMMARY AND CONCLUSIONS}

In this work we have examined the relationship between SN\,Ia Hubble residuals and derived host-galaxy properties for subsamples of SNe\,Ia from the SDSS-SNS.  Host-galaxy masses are determined using SDSS photometry as described in S14 and gas-phase metallicity and star-formation rates are derived using host-galaxy spectroscopy as detailed in Section~\ref{sec:derivprop}.  We utilize one sample of \numpm~SNe\,Ia with well-constrained host mass measurements (PM sample) and a subset of \nummzs~SNe\,Ia that also have metallicity and star-formation rate measurements from host spectra (MZS sample).  The PM sample is the largest single-survey set of SNe\,Ia and host-galaxy spectroscopic data used in a study of this type. 

To determine the relation between HR and host-galaxy properties, we perform linear fits with the LINMIX IDL routine and quote the significances of non-zero correlations. Using the PM sample, we observe with a significance of $3.6\sigma$ that more massive galaxies tend to host over-luminous SNe\,Ia after light-curve corrections, confirming what is previously reported in the literature.  This is one of the most significant detections of this effect, second only to \citet{Lampeitl10}, who also use SDSS SNe\,Ia.  We find less significant correlations between HR and metallicity ($1.4\sigma$) and HR and sSFR ($0.4\sigma$), in agreement with the results presented in previous works. We also utilize the Spearman rank test as a nonparametric measure of the correlations between HR and host-galaxy properties; we find strong evidence for a non-zero correlation ($p < 0.03$) for the HR-mass and HR-metallicity relations. The result of the HR-metallicity hypothesis test somewhat contradicts the LINMIX fit results, as it suggests there is evidence for a monotonic relation between HR and host-galaxy metallicity. This indicates that perhaps the HR-metallicity correlation is nonlinear and should be further explored using other fitting techniques.

Our large sample size also allows us to study correlations between HR and host-galaxy properties using multiple host-galaxy parameters simultaneously.  We use the multiple linear regression LINMIX package to fit for HR as a function of linear combinations of host mass, metallicity, and star-formation rate.  When using a combination of all three host parameters, no statistically significant correlation is recovered.  Similarly, no significant correlation is recovered when fitting for HR as a function of mass and metallicity, and of metallicity and SFR.  We also split our sample into two mass bins and study HR as function of metallicity and sSFR in each bin.  We find that in these mass bins, the linear trends of HR-metallicity and HR-sSFR are consistent with zero slope to within $1\sigma$.  With each multiparameter test, we find the HR correlation is consistent with flatness. Unless we are able to measure other host-galaxy properties as accurately as mass and appropriately account for the physical correlations between these host properties, then determining the true nature of this correlation will remain challenging. 

To study the effects of including photometrically-classified SNe Ia in our analysis, we divide the MZS and PM samples into spectroscopically-confirmed (Spec-Ia) and photometrically-classified (Phot-Ia) SNe.  We recompute our linear fits of HR with host-galaxy mass (metallicity, sSFR) in these subsamples; in all cases, for a respective host-galaxy property, linear fits from both subsets are in the same direction and slopes are consistent $ < 2.5\sigma$.  Using the Phot-Ia alone generally produces a fit with greater significance than that found when using the Spec-Ia alone.  The fits obtained from the Spec and Phot-Ia samples are also consistent with the larger PM and MZS samples as a whole. However, we also find that the results obtained using the Spec-Ia and Phot-Ia, particularly when comparing the HR-mass relation, could point to a striking difference between the two sets of SNe\,Ia. If we cannot assume that the PM sample is a homogeneous set of SNe Ia, or we cannot trust the purity of the photometric sample, this raises serious concerns about the usefulness of large-area surveys like DES and LSST that will observe thousands of photometrically-classified SNe Ia. As photometric typing is improved, we are confident that these Phot-Ia will be critical tools in HR-host property studies.

Throughout this analysis we determine, in several variations, correlations between HR and host-galaxy properties.  Yet we remain unsure about the physical mechanisms driving these relationships.  If progenitor age is truly responsible for the host bias, as proposed by \citet{Childress14}, and if host galaxy stellar age traces the progenitor age (which is likely true for star-forming galaxies), then a large sample of high-S/N host-galaxy spectra of a size comparable to the sample in this work would be helpful in further probing these correlations.  Obtaining such a large number of high-quality spectra will be difficult, but good S/N of the continuum is necessary to measure absorption lines and therefore infer stellar population age as was done by \citet{Johansson13}.  In this work, requiring that each host galaxy has a spectrum from SDSS or BOSS greatly reduced the size of our sample.  While the number of SNe\,Ia being discovered continues to rapidly increase, the number of host galaxies targeted for spectroscopic follow-up lags behind. We strongly advocate that current and future SN surveys strive for completeness of host galaxy spectral follow-up so that further analyses of host galaxy correlations will benefit from the increased statistics and suffer minimal bias. We are hopeful that future work using larger, higher quality datasets will contribute valuable insight into the nature of SN-host correlations and the complex combination of intrinsic and environmental features that affect SNe\,Ia.

\acknowledgements

The authors would like to thank Daniel Thomas and Ollie Steele for sharing their knowledge of {\tt GANDALF} and for their assistance with {\tt GANDALF} code modifications.

This material is based upon work supported by the National Science Foundation Graduate Research Fellowship under Grant No. DGE-1321851. Any opinion, findings, and conclusions or recommendations expressed in this material are those of the authors(s) and do not necessarily reflect the views of the National Science Foundation.

M.S. is supported by the Department of Energy grant DE-SC-0009890 and the National Science Foundation AST-1517742.  R.C.W. and J.A.F. are partially supported by the Department of Energy grant DE-SC-0009890.

Funding for the SDSS and SDSS-II has been provided by the Alfred P. Sloan Foundation, the Participating Institutions, the National Science Foundation, the U.S. Department of Energy, the National Aeronautics and Space Administration, the Japanese Monbukagakusho, the Max Planck Society, and the Higher Education Funding Council for England. The SDSS Web Site is http://www.sdss.org/.  The SDSS is managed by the Astrophysical Research Consortium for the Participating Institutions. The Participating Institutions are the American Museum of Natural History, Astrophysical Institute Potsdam, University of Basel, University of Cambridge, Case Western Reserve University, University of Chicago, Drexel University, Fermilab, the Institute for Advanced Study, the Japan Participation Group, Johns Hopkins University, the Joint Institute for Nuclear Astrophysics, the Kavli Institute for Particle Astrophysics and Cosmology, the Korean Scientist Group, the Chinese Academy of Sciences (LAMOST), Los Alamos National Laboratory, the Max-Planck-Institute for Astronomy (MPIA), the Max-Planck-Institute for Astrophysics (MPA), New Mexico State University, Ohio State University, University of Pittsburgh, University of Portsmouth, Princeton University, the United States Naval Observatory, and the University of Washington.

Funding for SDSS-III has been provided by the Alfred P. Sloan Foundation, the Participating Institutions, the National Science Foundation, and the U.S. Department of Energy. The SDSS-III web site is http://www.sdss3.org/.  SDSS-III is managed by the Astrophysical Research Consortium for the Participating Institutions of the SDSS-III Collaboration including the University of Arizona, the Brazilian Participation Group, Brookhaven National Laboratory, University of Cambridge, University of Florida, the French Participation Group, the German Participation Group, the Instituto de Astrofisica de Canarias, the Michigan State/Notre Dame/JINA Participation Group, Johns Hopkins University, Lawrence Berkeley National Laboratory, Max Planck Institute for Astrophysics, New Mexico State University, New York University, Ohio State University, Pennsylvania State University, University of Portsmouth, Princeton University, the Spanish Participation Group, University of Tokyo, University of Utah, Vanderbilt University, University of Virginia, University of Washington, and Yale University.

\appendix
\section{COMPARISON TO DR10}
\label{AppendixA}

As described in Section~\ref{sec:GalSpec}, the spectroscopic host properties used in this analysis are derived from emission-line fluxes measured using our own modified version of {\tt GANDALF}.  We detail our reasons for this re-analysis there, but note the primary motivation is to optimize the emission-line flux measurement to the redshift range of our sample.  Therefore, the host properties published in the SDSS DR10 may differ from those used in this work.

In this Appendix we use 3,787 overlapping spectra to compare results, specifically measured emission-line fluxes, $A/N$ ratios, and host-galaxy extinction.  We also show how these results would contribute to differences in derived host-properties, namely gas-phase metallicity.  For clarity, parameters derived in DR10 are denoted by the subscript ``DR10".  Some comparisons are best made using the subset of overlapping spectra with \Ha~and \Hb~$A/N > 2$, consistent with the quality requirements imposed on the host spectra in our analysis.  This $A/N$ requirement leaves 2,118 spectra in common.  

In Figure~\ref{fig:ancomp} we compare the $A/N$ values used in this work and those reported in DR10 for the four emission lines needed for the BPT diagnostic (\Ha, \Hb, \NII, \OIII).  Generally, our $A/N$ values are slightly higher than those in DR10, with the \NII~line showing the closest agreement.  This behavior is expected, as constraining the Balmer and Forbidden lines to have the same width and velocity as \Ha\ and \NII\, respectively, reduces the number of free parameters being fit (see Table~\ref{ELF}).  This effect is particularly strong at low $A/N$.  We find that 96.0\% of the spectra for which we measure $A/N>2$ in both \Ha\ and \Hb\ pass the same cuts in DR10, and that only 3.4\% of the full overlapping sample pass those cuts in DR10 but not in our sample.  The majority of the disagreement comes from just one of these two Balmer lines failing the cut (87.4\%).

This analysis and T13 (for $z < 0.45$) both use BPT diagnostics to separate star-forming galaxies from those dominated by other physical processes (i.e., AGN).  In this paper we use those galaxies that are classified as either `star-forming' or `composite' (SFC) based on this diagnostic.  We find that 6.1\% of the galaxies we classify as SFC are otherwise labelled by T13, while 8.3\% of the full overlapping sample are labelled SFC by DR10 but not in this paper.  Clearly, there is a discrepancy between the SFC classifications of the two samples.  Upon visual inspection of the spectra DR10 classifies as SFC and we do not, we observe that most of the DR10 spectra look like passive galaxies with very weak Balmer lines, which are unlikely to pass our $A/N$ cuts for inclusion in our analysis.  This confirms the need for good indicator of emission-line strength and spectral quality, such as $A/N$, to ensure a more pure sample of emission-line galaxies.

\begin{figure}[tp]
\centering
\begin{tabular}{cc}
\includegraphics[scale=0.35,angle=90]{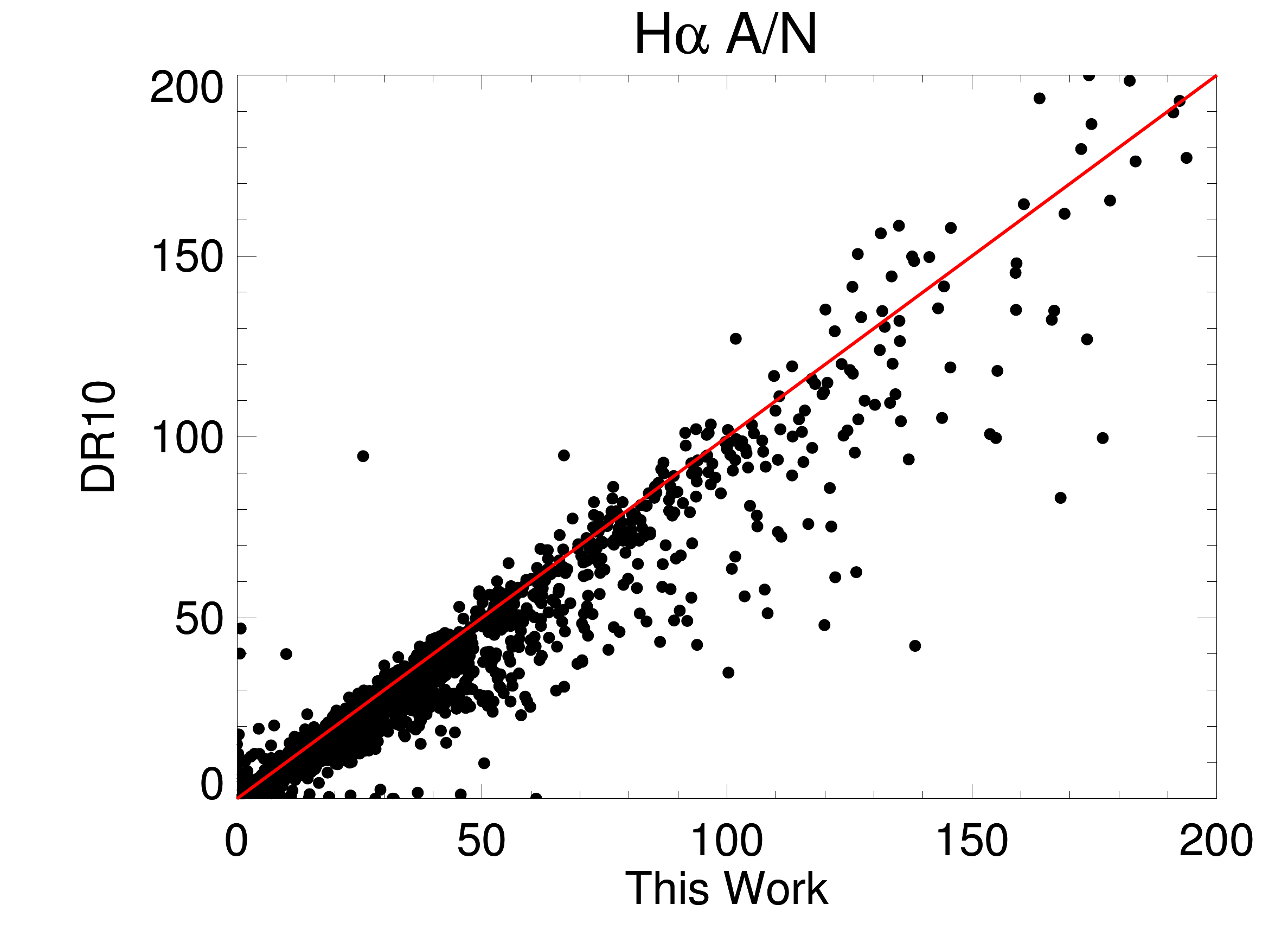}
\includegraphics[scale=0.35,angle=90]{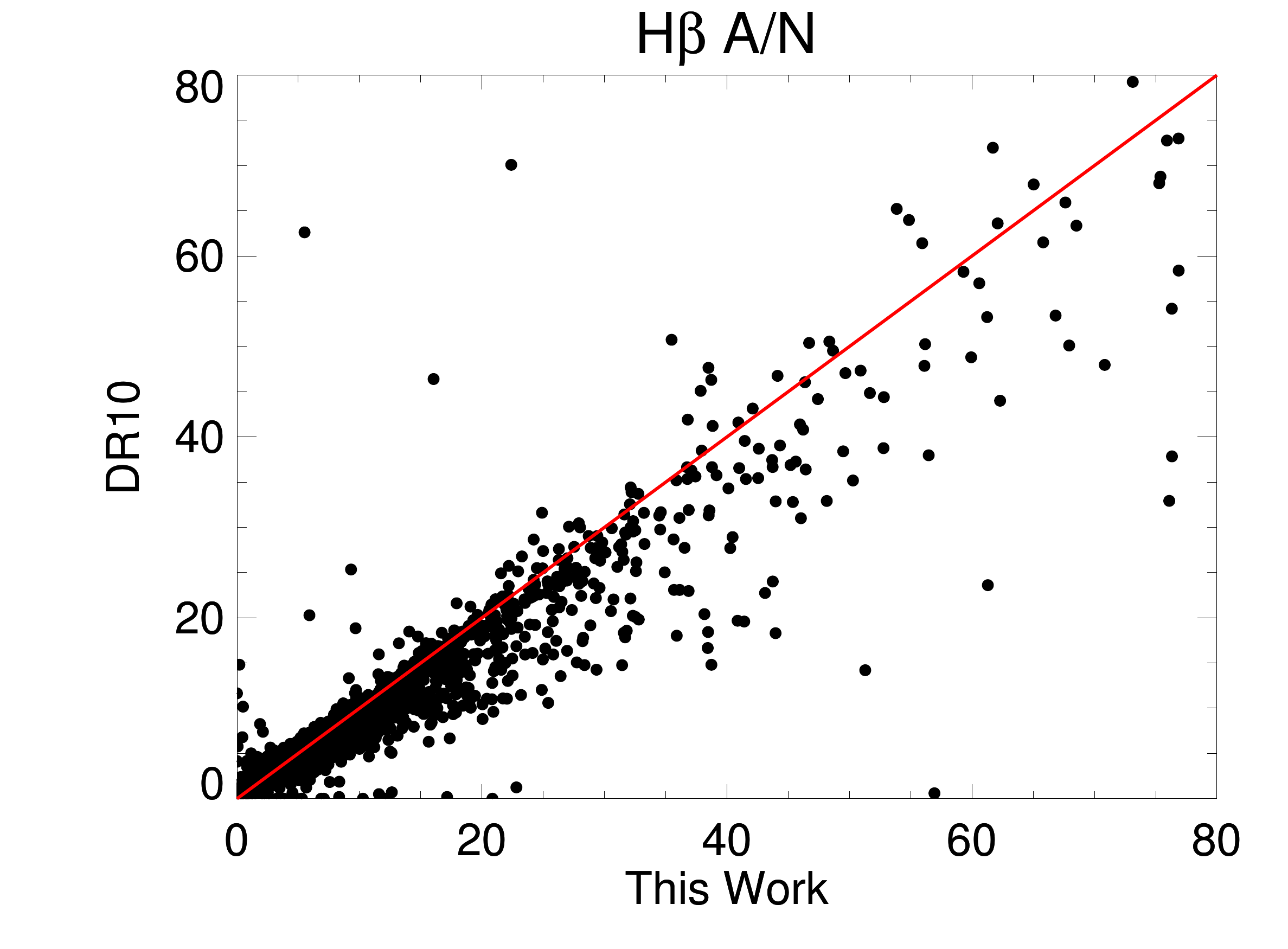} \\
\includegraphics[scale=0.35,angle=90]{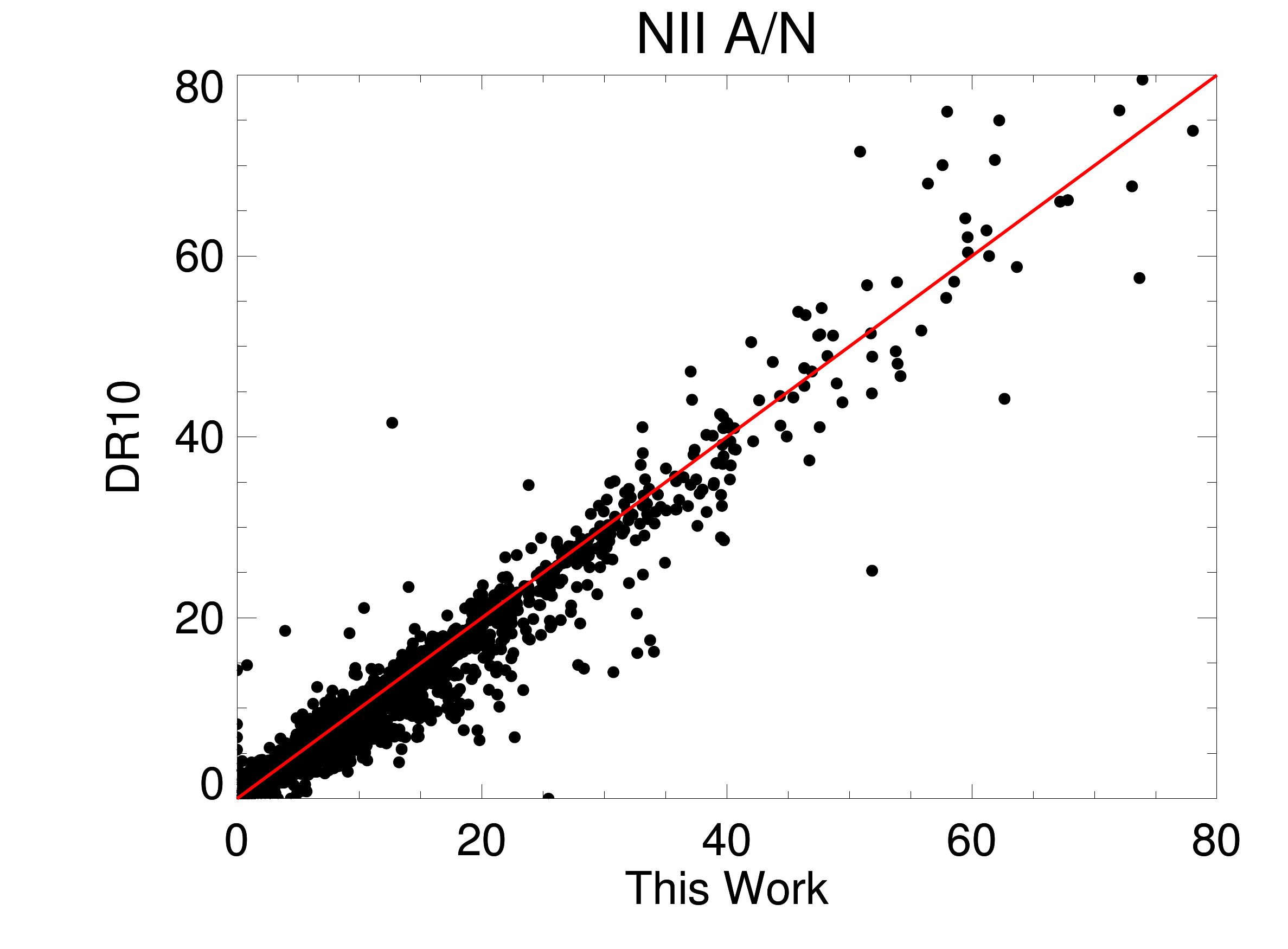} 
\includegraphics[scale=0.35,angle=90]{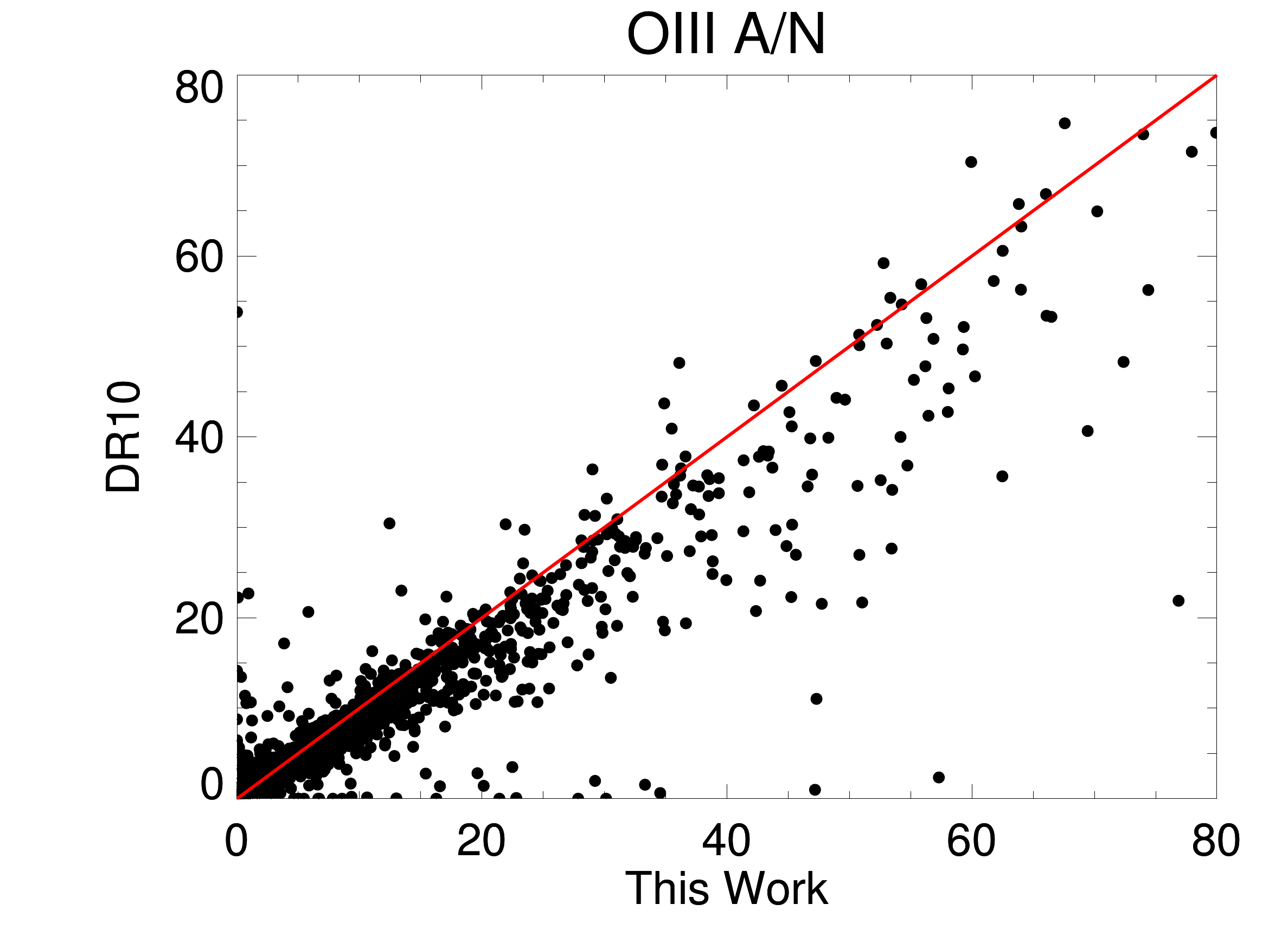}\\
\end{tabular}
\caption{Comparison of output $A/N$ values between this work and DR10.  The line $y=x$ is shown in red.  The ranges in both directions have been limited to focus on the bulk of the data; 91\%, 84\%, 91\% and 87\% of the data in the \Ha, \Hb, \NII, and \OIII\ lines are shown, respectively. For all emission lines, we find that our $A/N$ values are systematically higher than those reported in DR10.}
\label{fig:ancomp}
\end{figure}

In Figure~\ref{fig:fluxcomp} we compare the observed emission-line fluxes used for the BPT diagnostic between this work and DR10, where we have imposed \Ha\ and \Hb\ $A/N > 2$.   We make the comparison in observed flux, rather than intrinsic, as the latter quantity includes corrections for measured extinction and thus doesn't lend itself to a direct comparison.  While we can use the direct {\tt GANDALF} output parameters from our analysis, for DR10 we redden the published intrinsic fluxes via the \citet{Calzetti} law using the published $E(B-V)$ values.  We find for all four lines that our measured fluxes are on average higher than those in DR10 by $\approx10-15\%$, with no apparent dependence on flux. However, we do expect to find higher observed fluxes than DR10 due to the fact that we, unlike T13, correct the observed spectra for MW extinction before measuring fluxes.  

\begin{figure}[tp]
\centering
\begin{tabular}{cc}
\includegraphics[scale=0.3,angle=90]{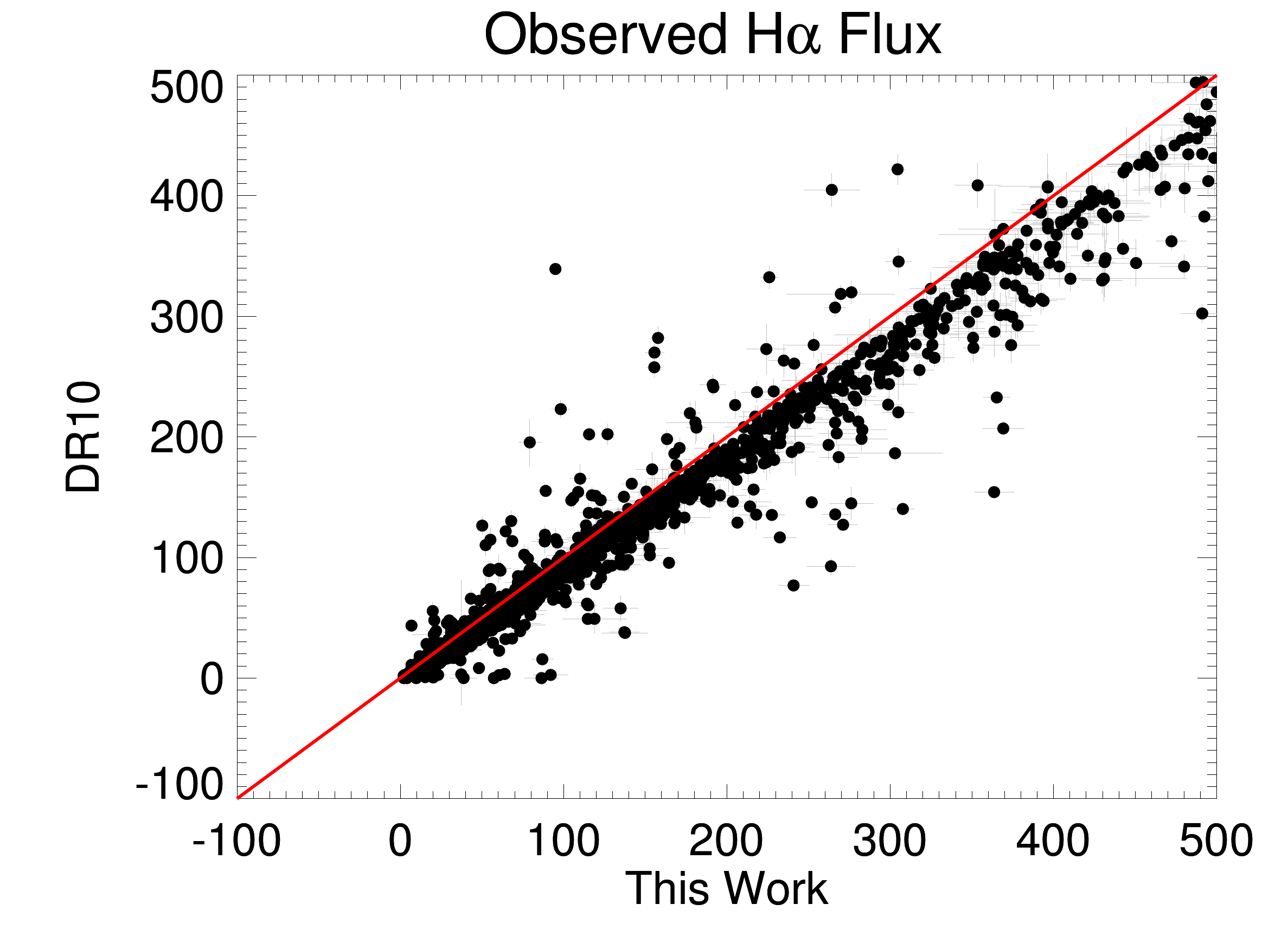}
\includegraphics[scale=0.3,angle=90]{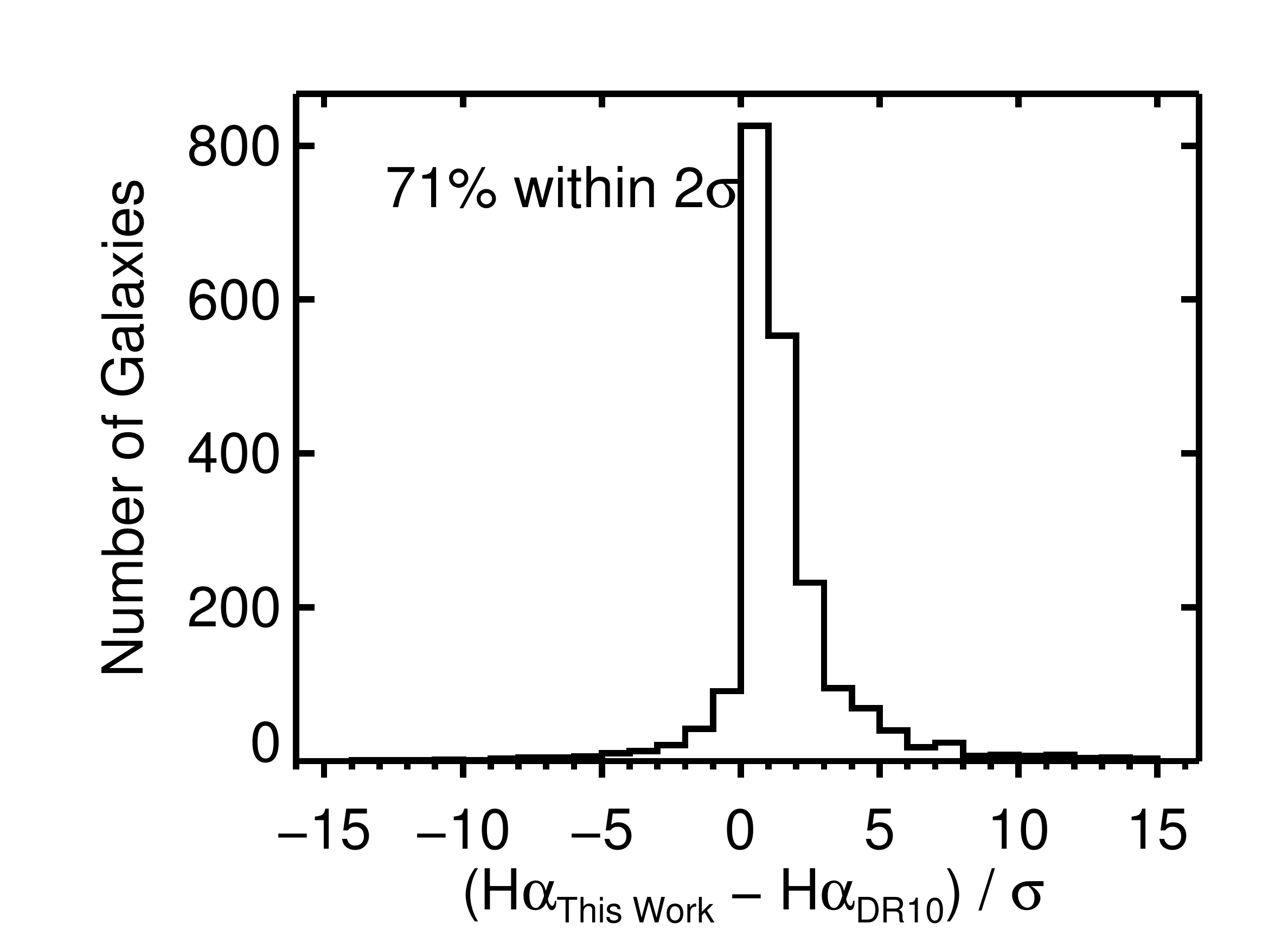} \\
\includegraphics[scale=0.3,angle=90]{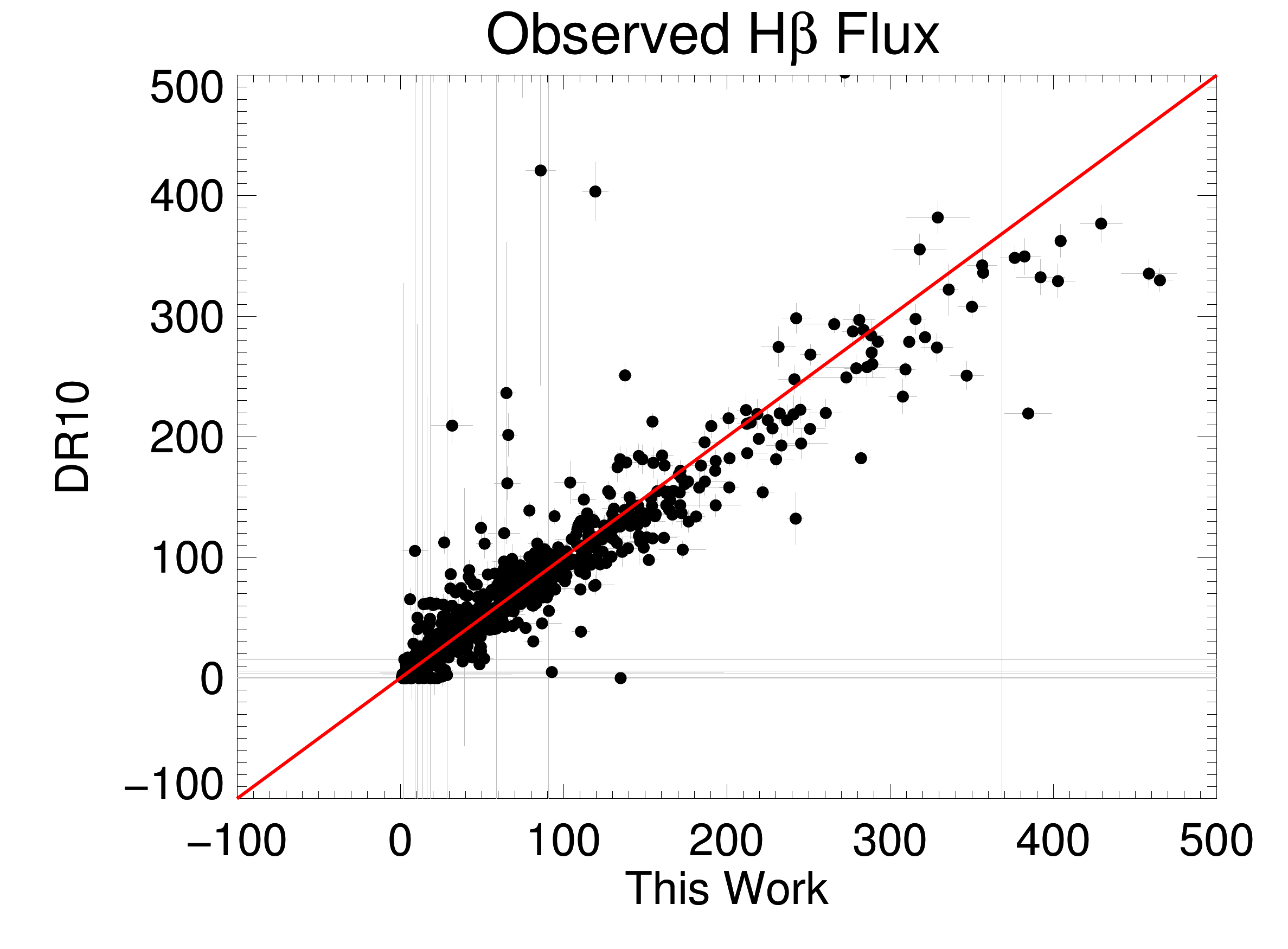}
\includegraphics[scale=0.3,angle=90]{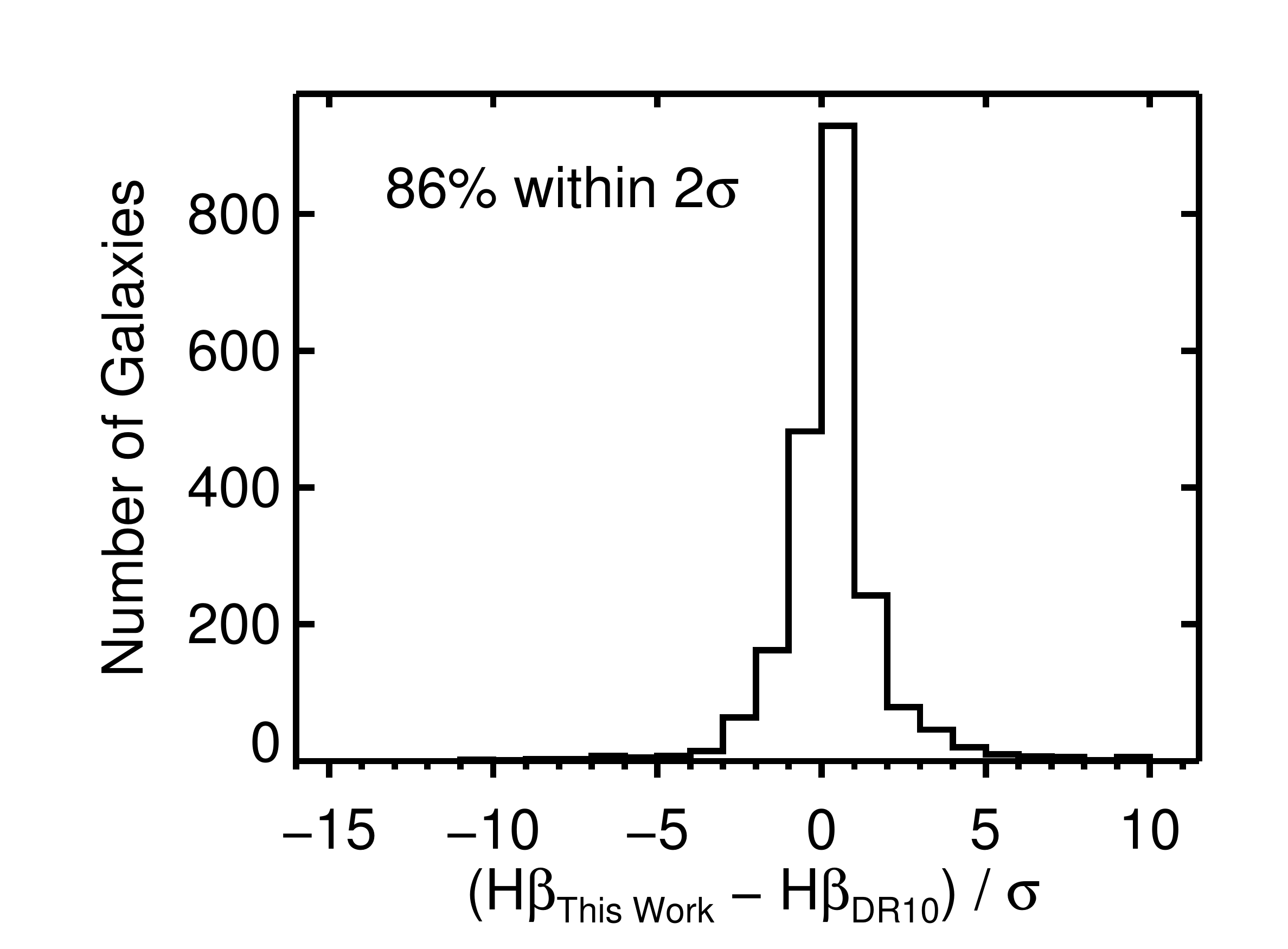} \\
\includegraphics[scale=0.3,angle=90]{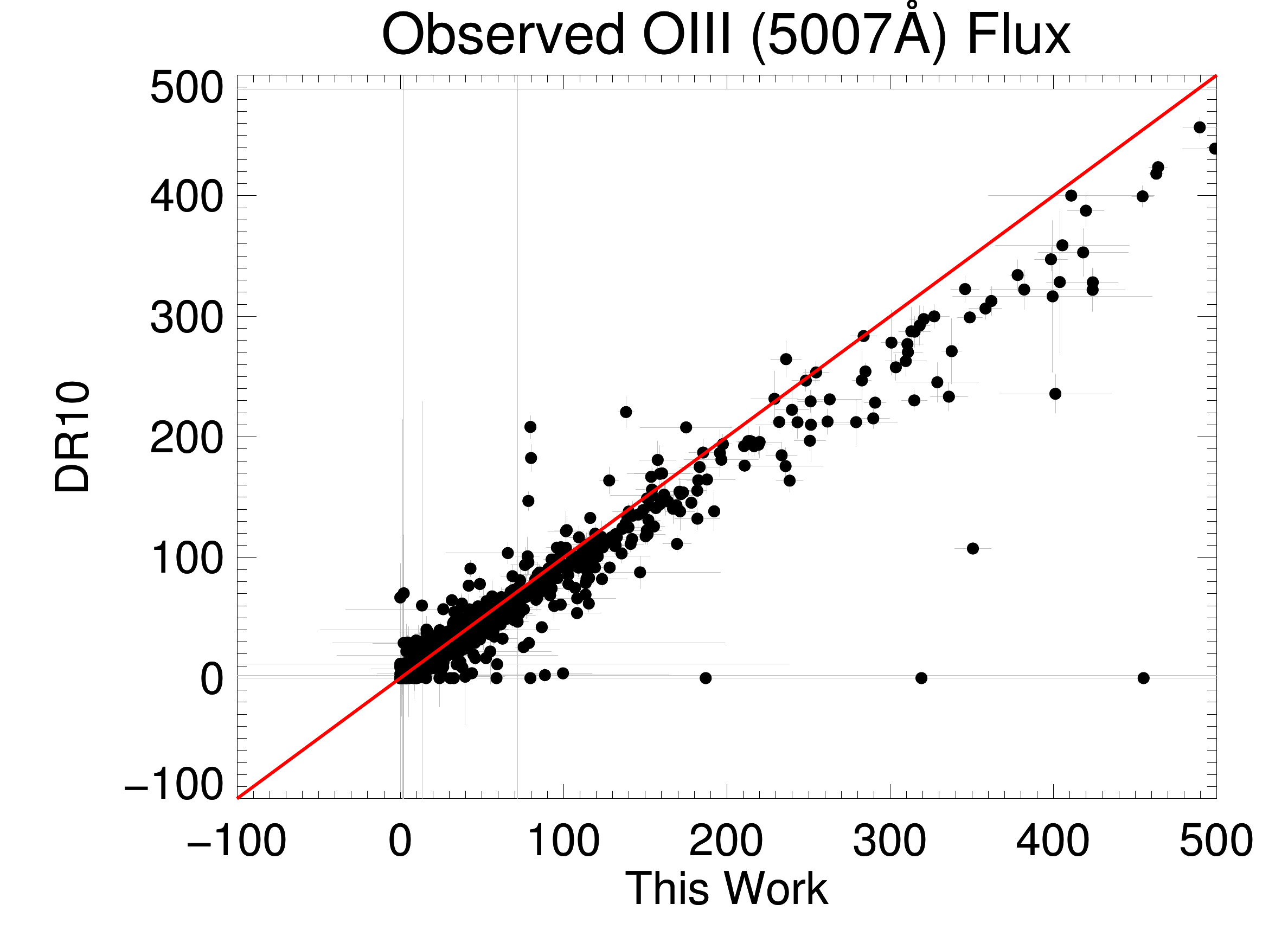}
\includegraphics[scale=0.3,angle=90]{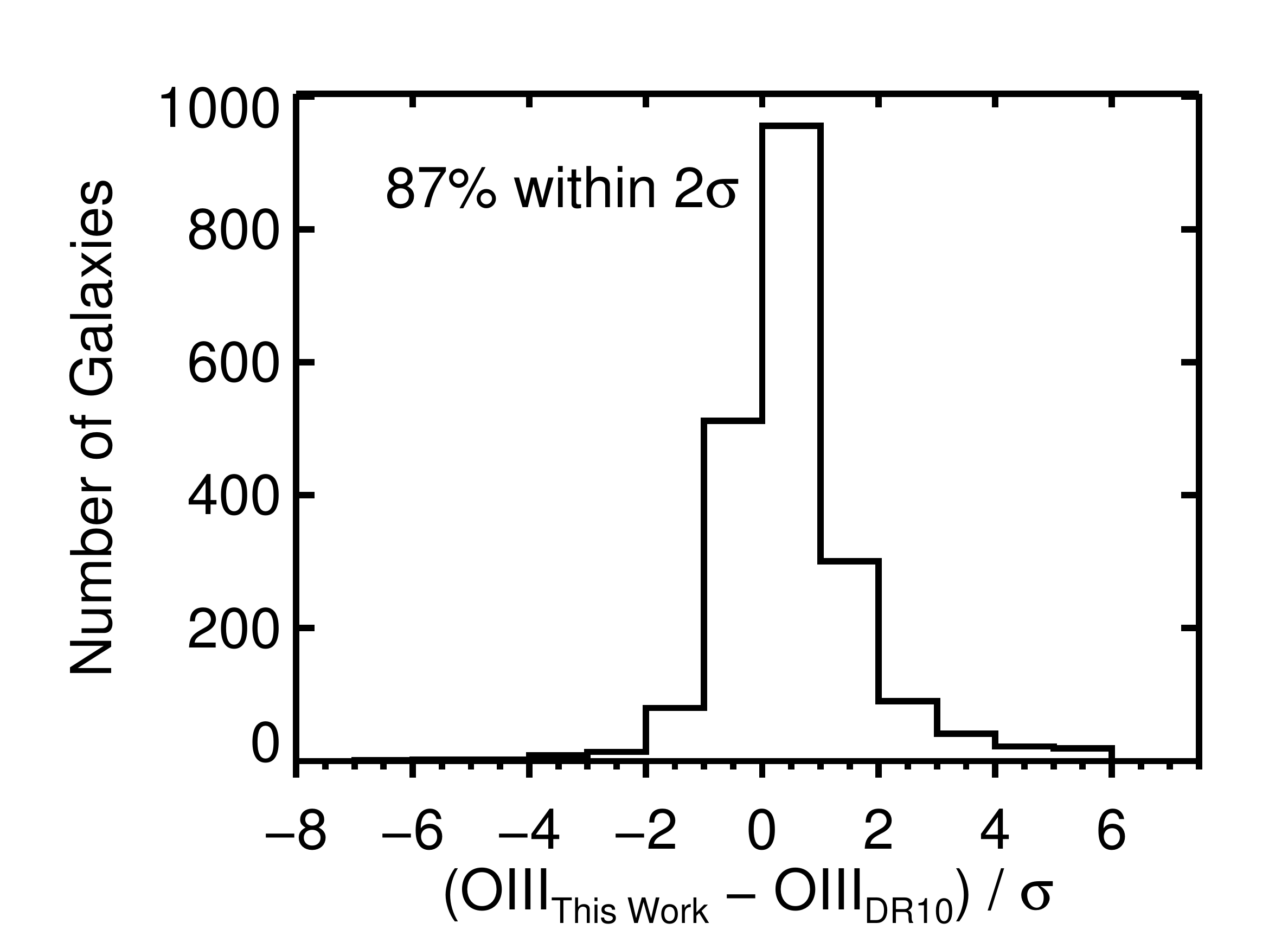} \\
\includegraphics[scale=0.3,angle=90]{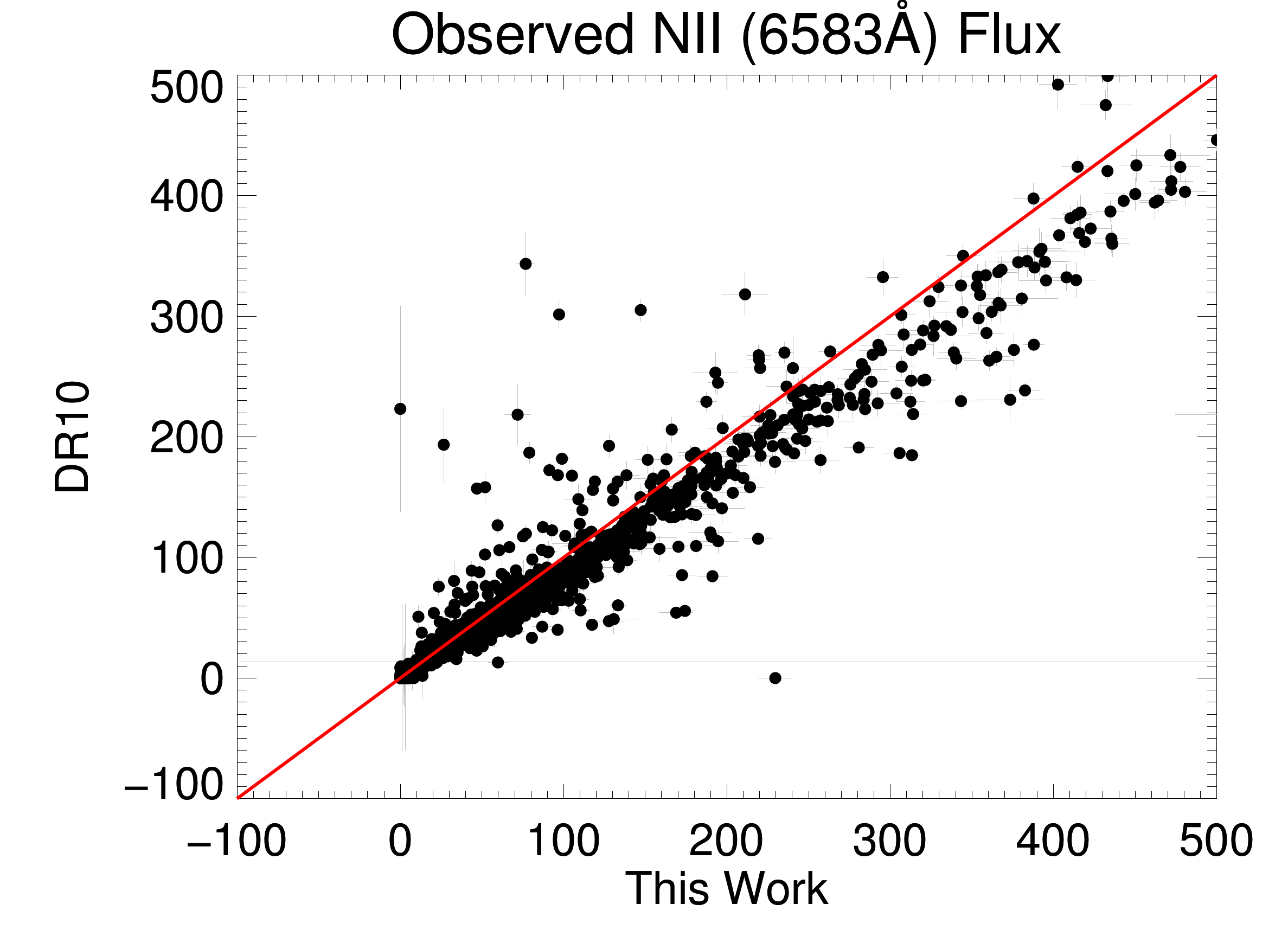}
\includegraphics[scale=0.3,angle=90]{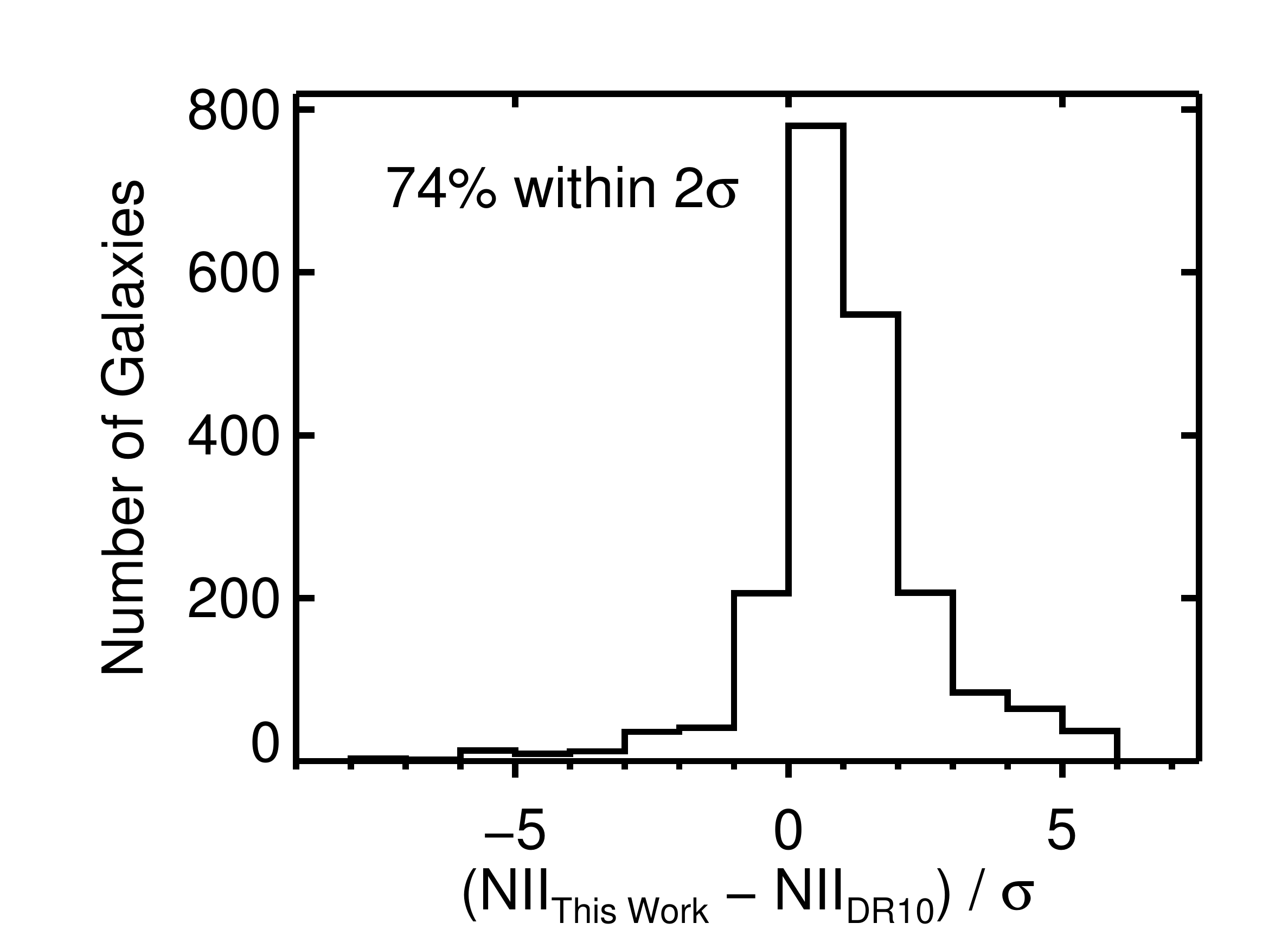} \\
\end{tabular}
\caption{\footnotesize{Comparison of the observed emission-line flux measurements between this work and DR10 where \Ha\ and \Hb\ $A/N > 2$. All figure axes have been truncated to focus on the bulk of the data; 90\%, 97\%, 96\%, and 95\% of the data points in the \Ha, \Hb, \OIII, and \NII\ lines are shown, respectively. The left columns shows a direct comparison of line flux with the line $y=x$ shown in red.  The distributions in the right columns present the difference between the DR10 measurements and those in this work.  The $\sigma$ value used is the uncertainties from each work added in quadrature.}}
\label{fig:fluxcomp}
\end{figure}

Using these observed fluxes and measured extinction, we can compute the intrinsic line fluxes necessary to estimate host-galaxy properties. While DR10 uses the extinction output by {\tt GANDALF}, measured from the continuum, we employ Case B recombination, which assumes a set ratio of intrinsic \Ha\ and \Hb\ fluxes. The difference in measured extinction values between the two methods is shown in Figure~\ref{fig:extcomp}. As expected, our decision to use Case-B recombination produces a much wider range of extinction values than what is reported in DR10. This difference in extinction values translates to a difference in intrinsic flux measurements between the two works: those reported in DR10 are systematically lower than those used here.

\begin{figure}[tp]
\centering
\includegraphics[scale=0.45,angle=90]{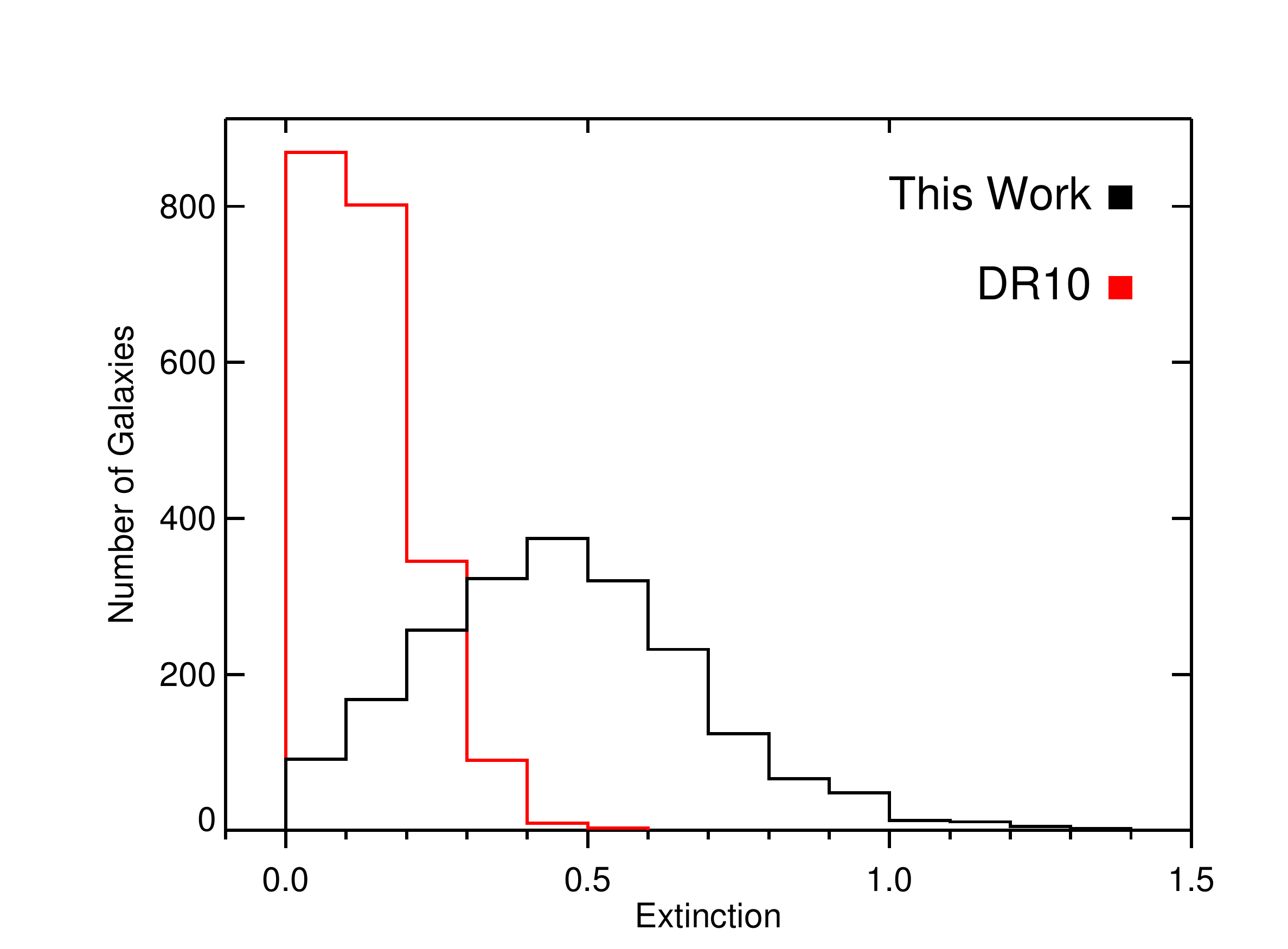}
\caption{Comparison of extinction values used in this work (black) and in DR10 (red). While this work uses Case B recombination to calculate the extinction, DR10 relies on the GANDALF output as measured using the spectral continuum fit.}
\label{fig:extcomp}
\end{figure}

This difference in intrinsic flux measurements does not seem to translate to significant differences in computed host-galaxy gas-phase metallicity. We compute the KD02 gas-phase metallicity (detailed in Section~\ref{sec:derivprop}) using the intrinsic fluxes from DR10 and this work, and present a comparison in Figure~\ref{fig:metcomp}. A physical metallicity measurement is computed for 77\% of the overlapping spectra; 54\% have a physical metallicity measurement and meet the $A/N$ requirement.  As displayed in the figure, the metallicities derived using the fluxes from this work slightly overestimate those obtained using the fluxes from DR10; yet, 98\% of the metallicities agree within $2\sigma$.

\begin{figure}[tp]
\centering
\begin{tabular}{c}
\includegraphics[scale=0.36,angle=90]{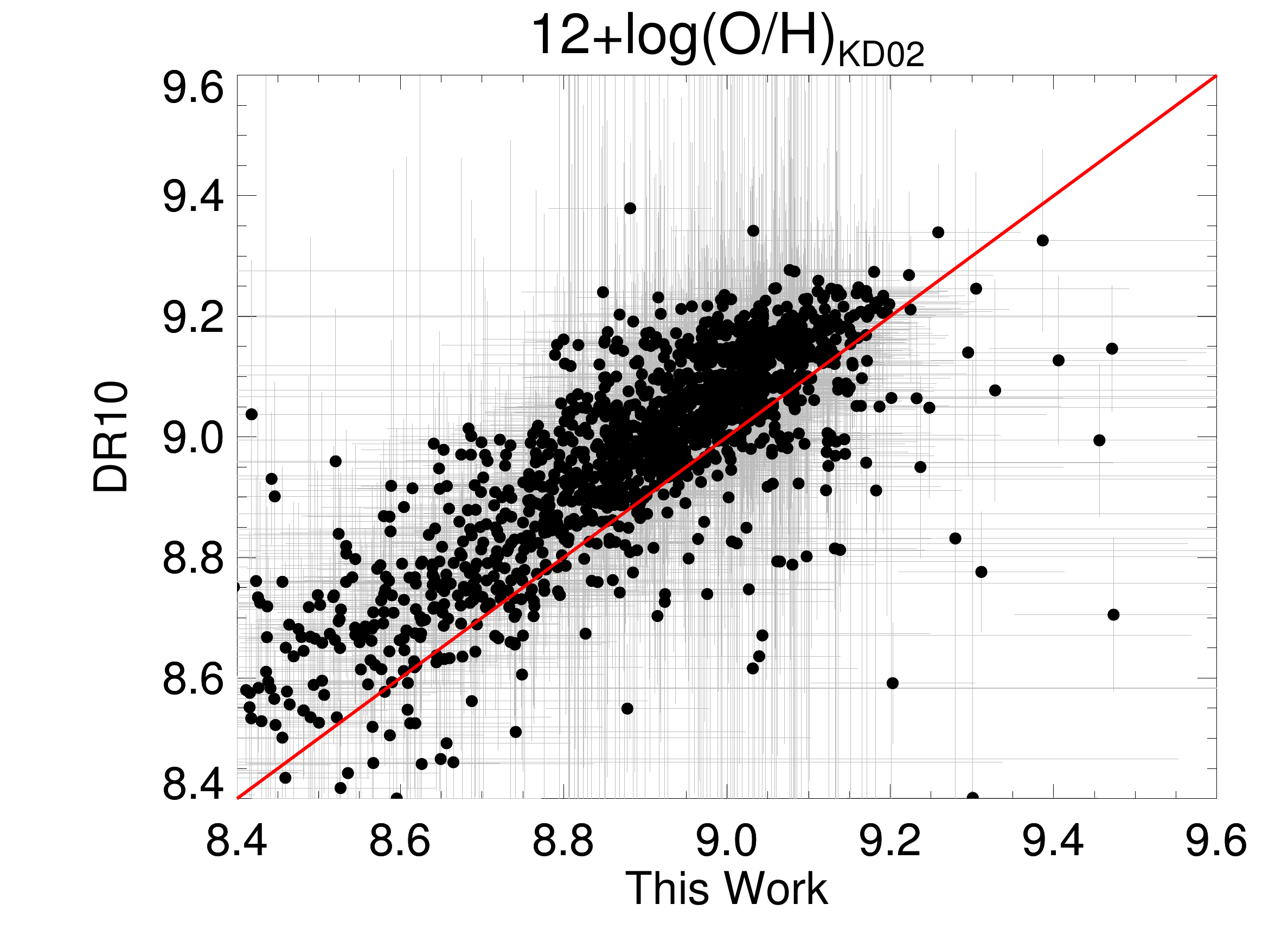} 
\includegraphics[scale=0.4,angle=90]{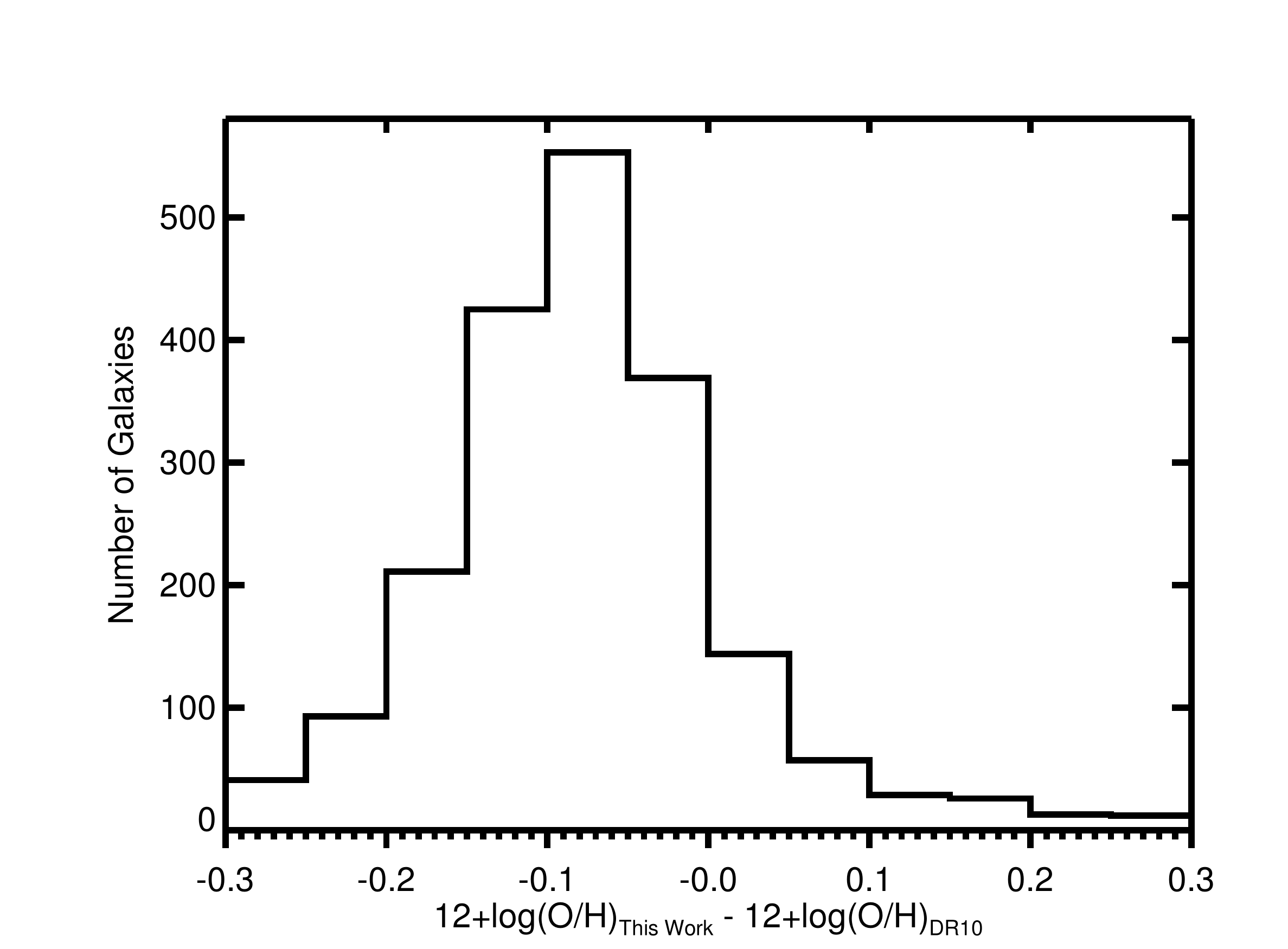}
\end{tabular}
\caption{Comparison of KD02 gas-phase metallicities derived using emission-line fluxes from this work and DR10, where \Ha\ and \Hb\ $A/N > 2$.  Figure axes have been truncated to focus on the bulk of the data; 96.7\% of the data points are shown here. A direct comparison of metallicity measurements is shown on the left (the line $y=x$ is plotted in red for comparison); the difference in metallicities is shown on the right.  The median and standard deviation of the difference, including outliers which are not displayed, are $-0.08$ \unt{dex} and $3.73$ \unt{dex}, respectively.}
\label{fig:metcomp}
\end{figure}

Because spectroscopic star-formation rate is directly proportional to intrinsic \Ha\ emission-line flux, and thus affected by choice of extinction correction, we would not expect such close agreement between the SFRs found in this work and those in DR10 (note Figure~\ref{fig:fluxcomp} only compares observed fluxes). However, we would not expect that these differences in computing SFR would significantly affect observed HR-SFR or HR-sSFR correlations.

This Appendix illustrates that the decision to optimize our analysis of emission-line spectra for our redshift range does not significantly affect the observed emission lines extracted from {\tt GANDALF}. Rather, the more important difference between this analysis and that of DR10 is the treatment of extinction. The decision to use Case-B recombination when computing extinction affects the intrinsic emission-line flux measurements, and thus creates an offset in host-property measurements. However, this is not concerning as any true correlations between SN\,Ia properties and host-galaxy properties should be observed independent of choice of extinction correction.

\section{CORRECTING FOR RESIDUAL TRENDS WITH SN COLOR}
\label{AppendixB}

An analysis of our measured HR as a function of SN\,Ia properties reveals correlations between SN\,Ia luminosity 
and the \salt\ light-curve parameters $c$ and $x_1$.  The trend with $x_1$ is not very 
strong but the trend with $c$ is significant and shows evidence that bluer SNe prefer a lower value of $\beta$, 
the slope of the color-luminosity relation (see Equation~\ref{eqn:mu}).  In Figure~\ref{fig:HRvC} we show HR as a function 
of $c$ and $x_1$ for our PM sample.  

\begin{figure}[tp]
\centering
\begin{tabular}{cc}
\includegraphics[scale=0.37,angle=90]{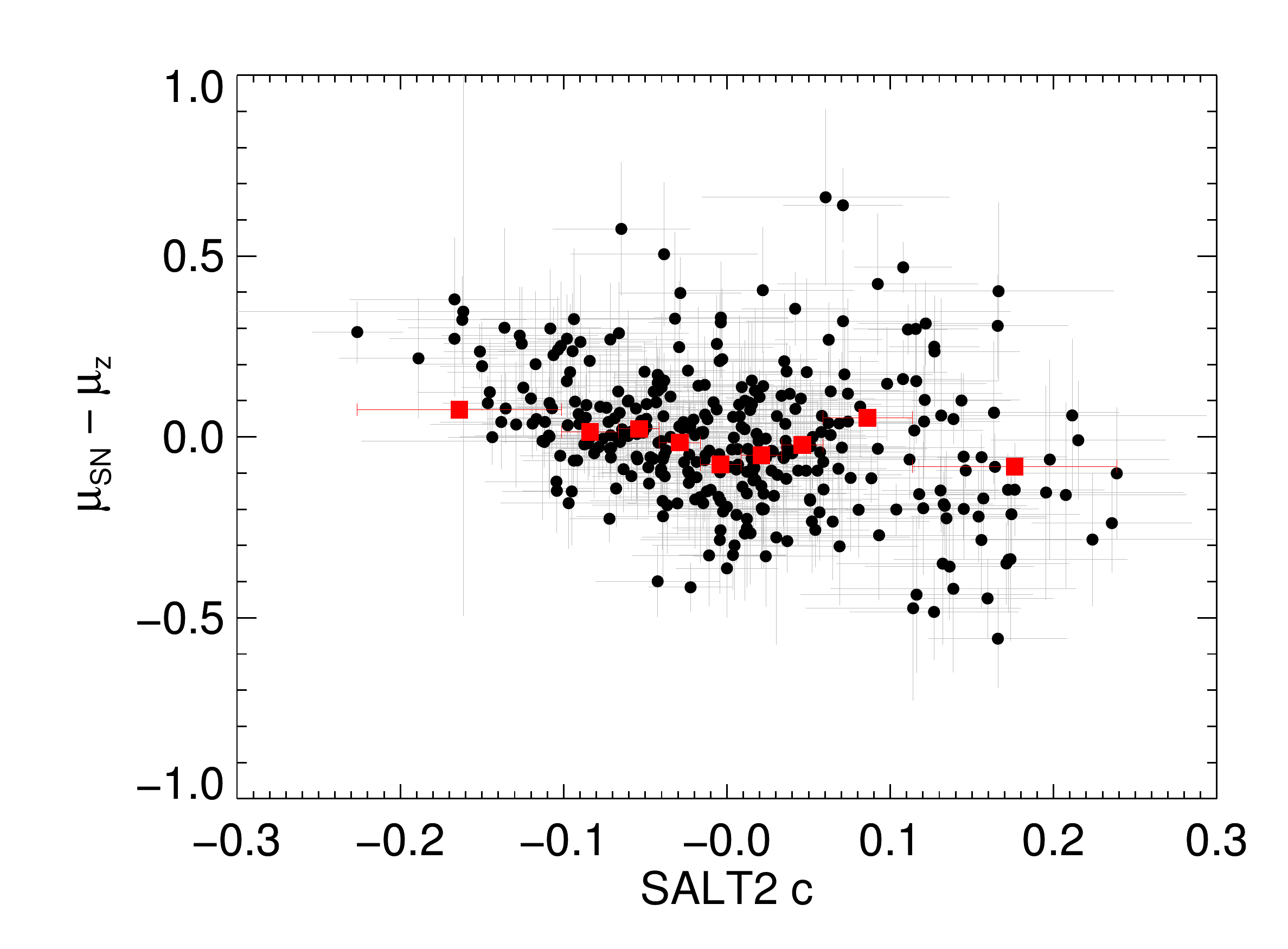}
\includegraphics[scale=0.37,angle=90]{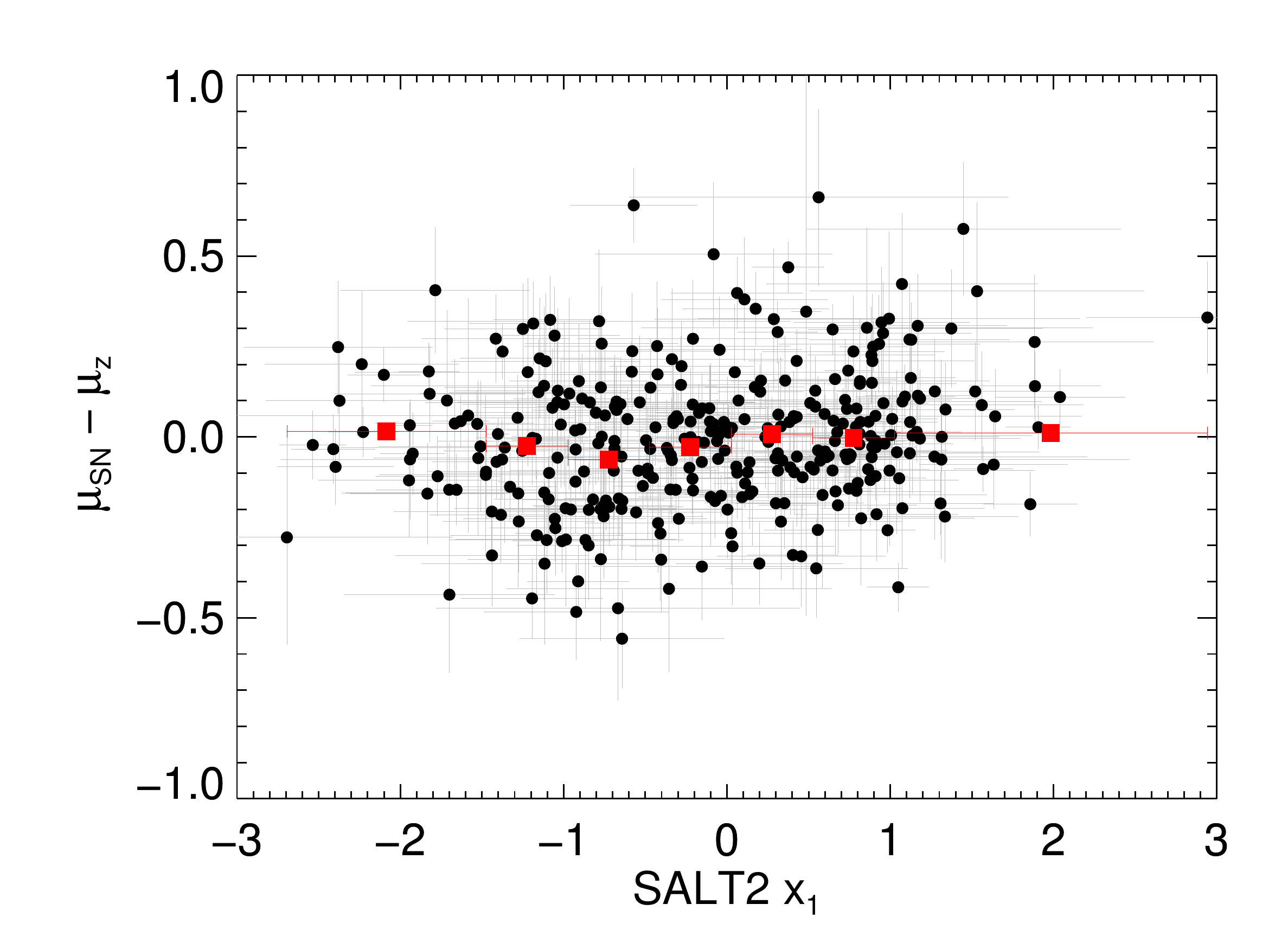}
\end{tabular}
\caption{HR as a function of {\tt SALT2} color (left panel) and stretch (right panel) for the PM sample.  Inverse-variance weighted average bins of width 0.025\unt{dex} and 0.5\unt{dex}, for $c$ and $x_{1}$ respectively, are displayed in red; each bin contains at least 30 SNe\,Ia.  }
\label{fig:HRvC}
\end{figure}

\citet{ScolnicColor} showed that
non-linear correlations between color and HR should be
expected due to the asymmetric and narrow underlying distribution of color that
correlates with luminosity. They also predict similar relations between HR and color for models of varying intrinsic scatter and reddening components; namely
one model in which intrinsic scatter is dominated by color variation \citep{Chotard} and a color-luminosity relation following a Milky Way reddening law ($\beta=4.1$) and a second model with scatter dominated by luminosity variation and a color-luminosity relation following $\beta=3.1$.  In particular, two distinct color-luminosity relations are observed for $c < 0$ and $c > 0$.  This effect is displayed in Figure~\ref{fig:HRvC}. 

To examine this color effect on our HR-host-galaxy correlations, we recompute the trends fitting for SN\,Ia color and host properties simultaneously using LINMIX.  We expect that correcting for this correlation with $c$ may weaken our host-galaxy correlations slightly; but, as discussed in \citet{ScolnicColor}, accounting for this color variation is 
not enough to explain the HR trend with host mass. When fitting for the HR-mass-color relation using the PM sample, the slope of the mass term ($-0.054\pm0.015$) is within $0.05\sigma$ of the slope recovered when fitting for the HR-mass relation only. We also recover the sSFR slope of the HR-sSFR-color relation for the MZS sample within $0.05\sigma$ ($0.015\pm0.046$) of the slope measured fitting only HR-sSFR. Interestingly, while the posterior distributions of the mass (and sSFR) and color fit coefficients are Gaussian, the distributions of the metallicity and color coefficients for the HR-metallicity-color relation are clearly skewed. Despite this skewness, we recover the metallicity coefficient ($-1.299\pm0.860$) within $1\sigma$ of the slope reported fitting the HR-metallicity relation for the MZS sample. In all cases, we find the slope of the color term to be within $1\sigma$ of $-0.705\pm0.136$.

It seems that including the HR-color correlation in our host-galaxy analysis does not have much of an effect on the observed results. This analysis, however, is only a crude estimate of these effects. LINMIX assumes a linear relation between HR and color, yet in Figure~\ref{fig:HRvC} it is apparent that the HR-color relation varies for low and high values of $c$. Future works should consider a more robust statistical treatment of this effect.

\clearpage
\LongTables
\begin{landscape}
\begin{deluxetable}{c c c l r r r l l r r r l l}
\tablecaption{Properties of SNe\,Ia and their Host Galaxies \label{alldata}}
\tabletypesize{\scriptsize}
\tablewidth{0pt}
\tablehead{
\colhead{CID}&\colhead{IAU Name\tablenotemark{a}} & \colhead{T\tablenotemark{a,b}} & \colhead{Redshift\tablenotemark{a}} &\colhead{{\tt SALT2}  $c$} &\colhead{ {\tt SALT2} $x_{1}$} &\colhead{ HR\tablenotemark{c}} & \colhead{DR8 HostID\tablenotemark{a}} & \colhead{BPT\tablenotemark{d}} & \colhead{log(M/\Msun)\tablenotemark{a,e}} & \colhead{12 + log(O/H)\tablenotemark{e}}& \colhead{log(sSFR)\tablenotemark{e}}& \colhead{gFF\tablenotemark{f}} & \colhead{Source}}
 \startdata
$703$&$\mathrm{...}$&$\mathrm{P}$&$0.298$&$-0.01\pm0.05$&$0.66\pm0.65$&$-0.15\pm0.17$&$1237663544222483004$&$1$&$9.96\pm0.13$&$8.92\pm0.05$&$-9.82\pm0.13$&$0.46$&$\mathrm{BOSS}$ \\
$762$&$\mathrm{2005eg}$&$\mathrm{S}$&$0.1914$&$-0.04\pm0.03$&$1.13\pm0.27$&$0.16\pm0.09$&$1237666338114765068$&$1$&$11.24\pm0.08$&$8.92\pm0.08$&$-10.16\pm0.09$&$0.21$&$\mathrm{SDSS}$ \\
$779$&$\mathrm{...}$&$\mathrm{P}$&$0.2381$&$0.02\pm0.04$&$0.41\pm0.39$&$-0.10\pm0.12$&$1237657069548208337$&$3$&$10.10\pm0.09$&$-999$&$-999$&$0.30$&$\mathrm{BOSS}$ \\
$822$&$\mathrm{...}$&$\mathrm{P}$&$0.2376$&$-0.09\pm0.05$&$-0.58\pm0.58$&$0.24\pm0.16$&$1237657584950379049$&$3$&$10.02\pm0.15$&$-999$&$-999$&$0.32$&$\mathrm{BOSS}$ \\
$859$&$\mathrm{...}$&$\mathrm{P}$&$0.2783$&$0.02\pm0.04$&$0.46\pm0.51$&$-0.33\pm0.14$&$1237666408438301119$&$1$&$9.64\pm0.13$&$8.81\pm0.05$&$-8.91\pm0.13$&$0.42$&$\mathrm{BOSS}$ \\
$911$&$\mathrm{...}$&$\mathrm{P}$&$0.2073$&$0.24\pm0.06$&$-0.48\pm0.73$&$-0.10\pm0.18$&$1237666407922467526$&$1$&$10.14\pm0.09$&$8.75\pm0.05$&$-9.13\pm0.09$&$0.18$&$\mathrm{BOSS}$ \\
$986$&$\mathrm{...}$&$\mathrm{P}$&$0.2806$&$0.01\pm0.06$&$-0.21\pm1.09$&$0.09\pm0.25$&$1237663463145079009$&$1$&$10.26\pm0.10$&$8.61\pm0.17$&$-9.21\pm0.12$&$0.27$&$\mathrm{BOSS}$ \\
$1008$&$\mathrm{2005il}$&$\mathrm{P}$&$0.2262$&$-0.02\pm0.04$&$0.46\pm0.48$&$-0.11\pm0.11$&$1237678617430197147$&$2$&$10.61\pm0.14$&$9.00\pm0.11$&$-11.62\pm0.31$&$0.28$&$\mathrm{BOSS}$ \\
$1032$&$\mathrm{2005ez}$&$\mathrm{S}$&$0.1298$&$0.05\pm0.04$&$-2.54\pm0.20$&$-0.02\pm0.10$&$1237666302164664434$&$2$&$10.68\pm0.07$&$9.10\pm0.06$&$-11.38\pm0.09$&$0.46$&$\mathrm{SDSS}$ \\
$1112$&$\mathrm{2005fg}$&$\mathrm{S}$&$0.2576$&$-0.04\pm0.05$&$-0.53\pm0.71$&$0.10\pm0.18$&$1237663478724428434$&$10$&$11.35\pm0.06$&$8.93\pm0.09$&$-9.42\pm0.07$&$0.13$&$\mathrm{SDSS}$ \\
$1119$&$\mathrm{2005fc}$&$\mathrm{S}$&$0.2978$&$-0.14\pm0.06$&$0.86\pm1.38$&$0.30\pm0.28$&$1237663458851619714$&$0$&$10.80\pm0.07$&$8.99\pm0.02$&$-9.25\pm0.07$&$0.29$&$\mathrm{BOSS}$ \\
$1241$&$\mathrm{2005ff}$&$\mathrm{S}$&$0.0898$&$0.05\pm0.02$&$-0.54\pm0.08$&$-0.09\pm0.06$&$1237656567586226517$&$2$&$10.70\pm0.11$&$8.78\pm0.05$&$-9.33\pm0.11$&$0.13$&$\mathrm{BOSS}$ \\
$1253$&$\mathrm{2005fd}$&$\mathrm{S}$&$0.2631$&$-0.10\pm0.04$&$-0.93\pm0.47$&$-0.12\pm0.14$&$1237663457779384632$&$2$&$11.15\pm0.09$&$9.15\pm0.05$&$-11.55\pm0.12$&$0.22$&$\mathrm{BOSS}$ \\
$1354$&$\mathrm{...}$&$\mathrm{P}$&$0.2494$&$0.20\pm0.08$&$-1.12\pm1.22$&$-0.15\pm0.30$&$1237663784195129684$&$1$&$10.63\pm0.08$&$8.88\pm0.03$&$-8.72\pm0.09$&$0.21$&$\mathrm{BOSS}$ \\
$1371$&$\mathrm{2005fh}$&$\mathrm{S}$&$0.1193$&$-0.10\pm0.02$&$0.79\pm0.10$&$-0.15\pm0.06$&$1237663277923106978$&$3$&$10.89\pm0.08$&$-999$&$-999$&$0.45$&$\mathrm{SDSS}$ \\
$1415$&$\mathrm{...}$&$\mathrm{P}$&$0.2119$&$0.17\pm0.04$&$0.92\pm0.50$&$-0.21\pm0.13$&$1237663716016980100$&$0$&$11.64\pm0.13$&$9.08\pm0.07$&$-11.34\pm0.17$&$0.29$&$\mathrm{SDSS}$ \\
$1658$&$\mathrm{...}$&$\mathrm{P}$&$0.2773$&$0.00\pm0.05$&$0.43\pm0.47$&$0.06\pm0.15$&$1237657191977845356$&$1$&$9.73\pm0.12$&$8.57\pm0.95$&$-9.51\pm0.18$&$0.27$&$\mathrm{BOSS}$ \\
$1794$&$\mathrm{2005fj}$&$\mathrm{S}$&$0.1419$&$0.03\pm0.03$&$1.17\pm0.32$&$0.11\pm0.08$&$1237663542603809147$&$3$&$9.27\pm0.08$&$-999$&$-999$&$0.21$&$\mathrm{BOSS}$ \\
$1979$&$\mathrm{...}$&$\mathrm{P}$&$0.2869$&$0.01\pm0.06$&$-1.28\pm1.08$&$-0.16\pm0.26$&$1237678617406604390$&$1$&$9.74\pm0.17$&$8.78\pm0.16$&$-10.09\pm0.21$&$0.33$&$\mathrm{BOSS}$ \\
$2017$&$\mathrm{2005fo}$&$\mathrm{S}$&$0.2631$&$-0.11\pm0.04$&$1.37\pm0.56$&$0.30\pm0.16$&$1237663479793714269$&$1$&$10.55\pm0.08$&$9.09\pm0.03$&$-9.73\pm0.09$&$0.32$&$\mathrm{BOSS}$ \\

$2081$&$\mathrm{...}$&$\mathrm{P}$&$0.2517$&$-0.10\pm0.05$&$-0.43\pm0.82$&$0.25\pm0.18$&$1237660024493834637$&$1$&$10.09\pm0.09$&$8.86\pm0.07$&$-9.43\pm0.10$&$0.53$&$\mathrm{SDSS}$ \\
$2149$&$\mathrm{...}$&$\mathrm{P}$&$0.2956$&$-0.09\pm0.06$&$0.29\pm0.73$&$0.33\pm0.20$&$1237666338652487684$&$3$&$10.49\pm0.16$&$-999$&$-999$&$0.35$&$\mathrm{BOSS}$ \\
$2330$&$\mathrm{2005fp}$&$\mathrm{S}$&$0.2132$&$0.02\pm0.06$&$-1.79\pm0.58$&$0.41\pm0.17$&$1237678434328183252$&$1$&$9.87\pm0.10$&$9.14\pm0.05$&$-10.42\pm0.13$&$0.28$&$\mathrm{BOSS}$ \\
$2372$&$\mathrm{2005ft}$&$\mathrm{S}$&$0.1805$&$0.03\pm0.03$&$0.31\pm0.22$&$-0.09\pm0.08$&$1237657070091108996$&$1$&$10.60\pm0.08$&$9.02\pm0.05$&$-9.90\pm0.08$&$0.31$&$\mathrm{BOSS}$ \\
$2440$&$\mathrm{2005fu}$&$\mathrm{S}$&$0.1911$&$-0.08\pm0.03$&$0.43\pm0.29$&$0.21\pm0.09$&$1237678617436487971$&$1$&$10.32\pm0.08$&$8.86\pm0.02$&$-8.82\pm0.08$&$0.28$&$\mathrm{BOSS}$ \\
$2532$&$\mathrm{...}$&$\mathrm{P}$&$0.2689$&$0.00\pm0.05$&$0.89\pm0.63$&$0.21\pm0.19$&$1237663783672676591$&$20$&$11.44\pm0.10$&$9.06\pm0.13$&$-11.42\pm0.19$&$0.14$&$\mathrm{BOSS}$ \\
$2561$&$\mathrm{2005fv}$&$\mathrm{S}$&$0.1181$&$0.04\pm0.03$&$-0.08\pm0.11$&$0.04\pm0.06$&$1237678437019287600$&$1$&$10.76\pm0.06$&$8.78\pm0.07$&$-10.36\pm0.07$&$0.20$&$\mathrm{SDSS}$ \\
$2639$&$\mathrm{...}$&$\mathrm{P}$&$0.2163$&$0.00\pm0.03$&$0.40\pm0.28$&$-0.33\pm0.10$&$1237663544219926794$&$0$&$10.92\pm0.07$&$9.55\pm0.05$&$-14.40\pm0.61$&$0.39$&$\mathrm{BOSS}$ \\
$2766$&$\mathrm{...}$&$\mathrm{P}$&$0.1499$&$-0.05\pm0.03$&$-0.05\pm0.40$&$-0.06\pm0.09$&$1237666300019802272$&$0$&$11.25\pm0.11$&$8.98\pm0.05$&$-10.57\pm0.12$&$0.34$&$\mathrm{SDSS}$ \\
$2789$&$\mathrm{2005fx}$&$\mathrm{S}$&$0.2905$&$-0.11\pm0.05$&$-0.77\pm0.55$&$0.00\pm0.17$&$1237663444906017256$&$3$&$11.22\pm0.17$&$-999$&$-999$&$0.25$&$\mathrm{BOSS}$ \\
 & & & & & &  & ... & & & & & &  \\
\enddata
	\tablenotetext{a}{As specified in S14.}
	\tablenotetext{b}{Denotes if the SN\,Ia is spectroscopically confirmed (S) or photometrically typed (P).} 
	\tablenotetext{c}{Uncertainties on HR do not include the intrinsic $\sim 0.1$ mag scatter.}
	\tablenotetext{d}{BPT diagnostic flag that indicates a star-forming (1), composite (2) galaxy, or AGN (0). Star-forming (10) and composite (20) hosts as determined by the BPT diagnostic where some line fluxes are measured to be zero are also included. In some cases we cannot measure the necessary line fluxes for the BPT diagnostic (3).}
	\tablenotetext{e}{Measurement errors on derived host-galaxy properties do not include systematic uncertainties; -999 indicates no measurement could be made.}
	\tablenotetext{f}{$g$-band fiber fraction.}
\end{deluxetable}
\clearpage
\end{landscape}
\end{document}